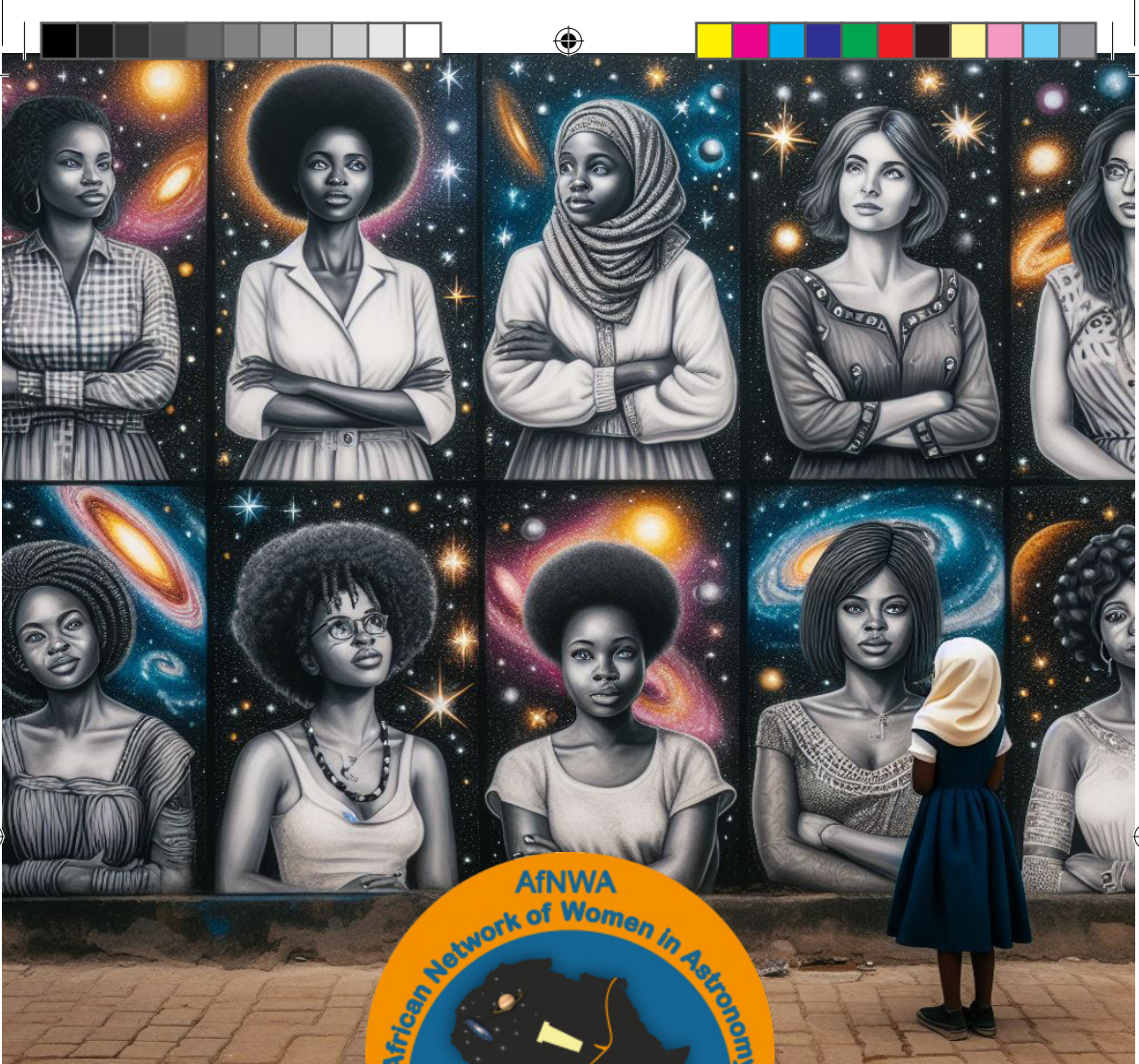

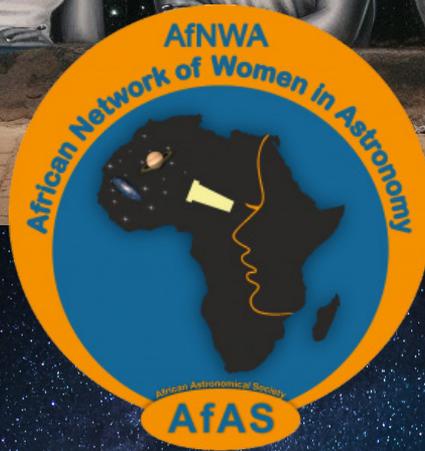

# INSPIRING STORIES FROM WOMEN
## IN ASTRONOMY IN AFRICA

**Edition 1**

# EDITORIAL TEAM

## AfNWA board

**Dr. Priscilla Muheki**
Main Editor
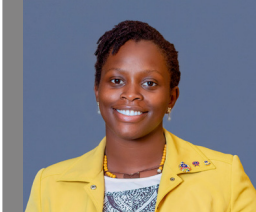

**Prof. Mirjana Pović**
Co-editor, AfNWA Chair
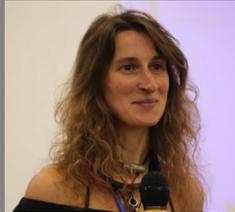

**Prof. Somaya Saad**
Co-editor
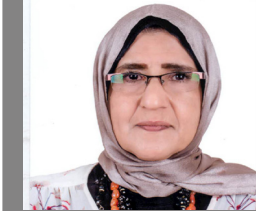

**Ms. Salma Sylla Mbaye**
Co-editor
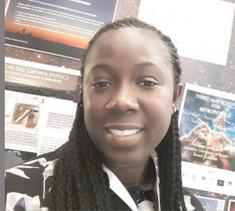

**Prof. Vanessa McBride**
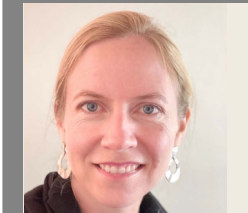

**Dr Naomi Asabre Frimpong**
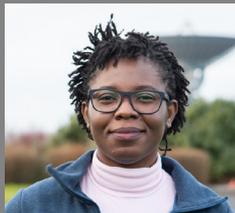

**Dr Meryem Guennoun**
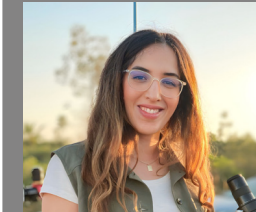

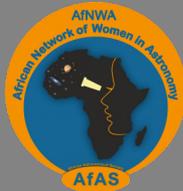

## African Science Stars team

**Graphic designer:** Thinavhuyo Desmond Mudimeli

**Science Communications Intern:** Odirile Mamba

**General Manager:** Mutshidzi Mclloyd Nelwamondo

**Publisher:** Madambi Rambuda

Subscriptions • info@sciencestars.co.za

*Inspiring stories from women in astronomy in Africa*
is published by Science Stars (Pty) Ltd

**Date of Publication:** 09 August 2024

African Science Stars is an initiative under the African Astronomical Society and funded by the Department of Science, Technology and Innovation.

1 Bridgeway,
Century City,
Cape Town,
South Africa, 7441
+27 21 830 5200
info@sciencestars.co.za
www.africansciencestars.com

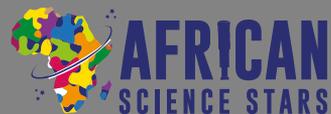

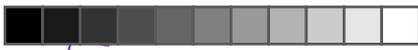

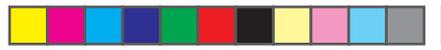
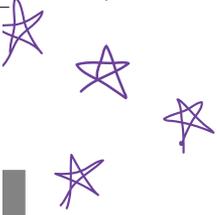

*This book is inspired by the stories of women who managed to fulfill their dreams despite the challenges and become astronomers and scientists.
It is dedicated to all girls in Africa and around the world to support them in their life journey to become what they want to be, making their dreams a reality.*

*We are inmensely grateful to all women who shared their inspiring stories with us in this book. Thank you.*

*AfNWA Board*

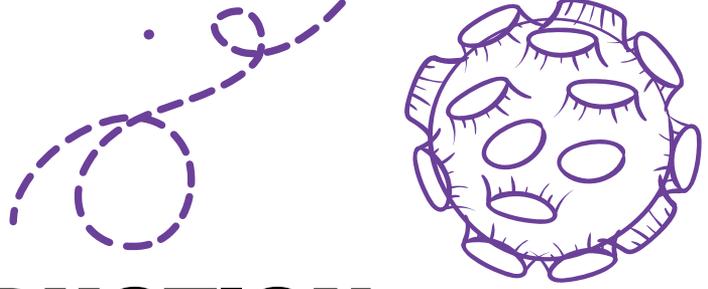

# INTRODUCTION

In preparation for the International Astronomical Union (IAU) General Assembly (GA) 2024, the first GA to be held in Africa (6-15 August 2024, Cape Town, South Africa), the African Network of Women in Astronomy (AfNWA) embarked on a visionary project: the creation of an inspiring storytelling book that showcases the remarkable journeys of professional female astronomers in Africa. This book is not merely a collection of biographies; it is a tapestry of resilience, passion, and scientific excellence woven through the lives of women who have ventured into the cosmos from the African continent.

The primary aim of this book is twofold. Firstly, it seeks to bring greater visibility to women astronomers in Africa, highlighting their groundbreaking research and the personal stories that have shaped their careers. By shining a light on their achievements and awards, we hope to acknowledge their contributions to the field of astronomy and underscore the importance of diversity in science.

Secondly, this book aspires to inspire and empower the next generation of scientists, particularly young women and girls across Africa. Through the personal narratives and professional achievements of these trailblazing astronomers and students in astronomy, we aim to spark curiosity, foster a love for science, and demonstrate that the sky is not the limit but just the beginning for those who dare to dream.

As you delve into the stories within these pages, you will encounter a rich array of experiences and insights that reflect the unique challenges and triumphs women face in astronomy. From overcoming societal barriers to making groundbreaking discoveries, these women have carved paths that others can follow, proving that with determination and passion, the stars are within reach for everyone.

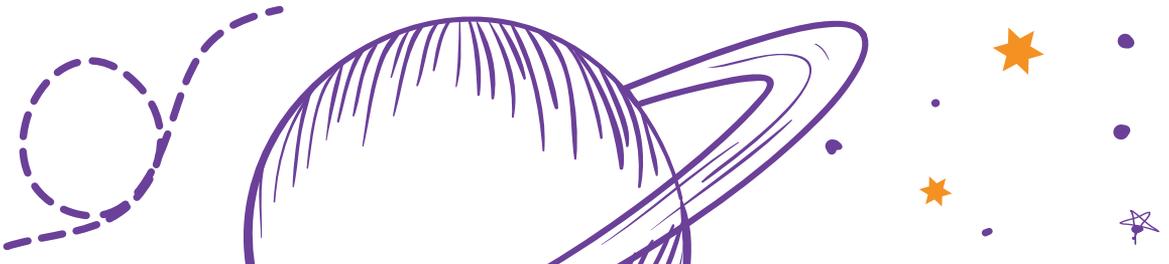

# About AfNWA

The African Network of Women in Astronomy (AfNWA) is an initiative that aims to connect women (or individuals who identify as such) working in astronomy and related fields in Africa. It was established in September 2020 as one of the committees of the African Astronomical Society (AfAS). With AfNWA we want to ensure the future participation of girls and women at all levels in the development of astronomy and science across Africa. Our main objectives are to improve the status of women in science in Africa and to use astronomy to inspire more girls to do Science, Technology, Engineering and Mathematics (STEM).

AfNWA was established in September 2020 and launched publicly in January 2021, with strong support from the AfAS, becoming one of its committees. The Network currently has ~ 150 members from over 30 countries, with ~ 80%/20% female/male participation. Nearly 80% of all members are early-career researchers, master and PhD students. This means that having AfNWA now is timely to offer the right support to all young researchers and make sure that we do not lose them throughout their scientific career.

The initiative behind AfNWA started when Prof. Mirjana Pović received the 2019 Inspiring Science Award from Nature Research in collaboration with Esteé Lauder, and decided to use the financial part of the award to create AfNWA. AfNWA was then initially created by Prof. Mirjana Pović (Institute of Space and Geospatial Science in Ethiopia), Prof. Vanessa McBride (at that time Office of Astronomy for Development in South Africa, now International-al Council for Science in France), Dr. Priscilla Muheki (Mbarara University of Science and Technology in Uganda), Prof. Carolina Ödman-Govender (University of the Western Cape in South Africa), Prof. Somaya Saad (National Astronomy and Geophysics Research Institute for Research in Astronomy and Geophysics in Egypt), and Prof. Nana Ama Brown (University of Ghana in Ghana).

Later, Ms. Salma Sylla Mbaye (Cheikh Anta Diop University in Senegal), Dr. Naomi Asabre Frimpong (Ghana Space Science and Technology Institute in Ghana) and Dr. Meryem Guennoun (Oukaimeden Observatory in Morocco), also joined the AfNWA Board. This book is part of the activities carried out by AfNWA, among many others, such as the training of early-career women researchers, outreach and education activities focusing on girls, and the AfNWA awards for women in astronomy in Africa, among others.

*With AfNWA we want to ensure the participation of women at all levels in the important developments in astronomy in Africa for the benefit of society as a whole, and that through astronomy we can inspire and empower many more girls to contribute in the future to making our world a better place for all its children and inhabitants.*

# CONTENT

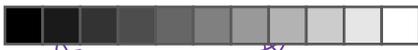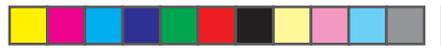

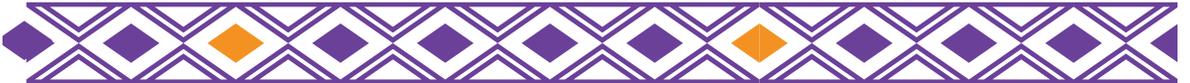

# A tribute to  ▬▬▬  1973 - 2022
# Prof. Carolina Ödman Govender

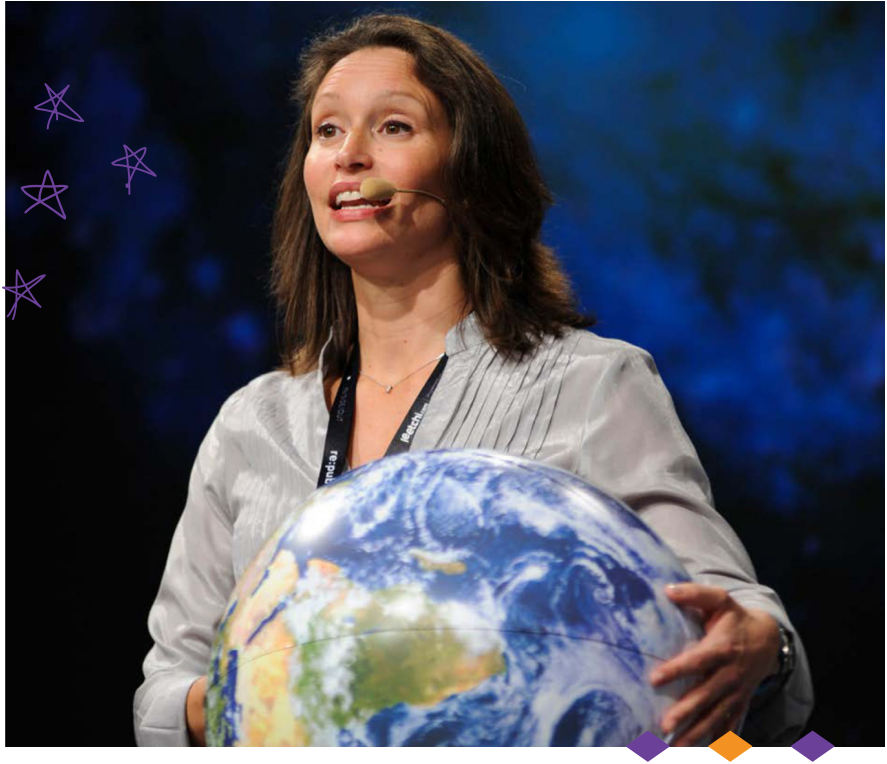

**C**arolina was someone special, with a brilliant mind and a heart full of compassion. Her unwavering commitment to see others grow was unmatched.

One of her most admirable qualities was her ability to bring out the best in others. She always recognized the strengths in everyone and encouraged them to strive for greater heights. Her enthusiasm, heart melting smile and warm presence can only leave us missing her the more.

Prof Ödman was an Associate Professor in Astrophysics at the University of the Western Cape (UWC) and the Associate Director for Development and Outreach at the Inter-University Institute for Data Intensive Astronomy (IDIA). This research institute tackles the challenges of Big Data in science. In this position, she played an enormous role in science communication and engagement. The impact of her work included engaging with the media, inspir-ing and mentoring physics and astronomy students, community outreach and offering leadership in public forums on science issues that affect the public.

Prof. Ödman was very much interested and worked on the translation of scientific terms into indigenous languages. She believed that the science community needed to be sensitive and make science available in other languages, which was an opportunity to decolonize science, so that astronomy could be introduced to high school learners in some schools in Cape Town. This multidisciplinary research project involved postgraduate students at UWC and the University's Xhosa Department.

During her career, Prof Ödman-Govender received numerous accolades, including the IAU Prize, called the Special Executive Committee Award for Astronomy Outreach, Development and Education recognized for her pioneering work in astronomy outreach, development and education. The award was presented at the 2018 XXX IAU General Assembly in Vienna, Austria. Award of the International Science Council, 2020 for her concept of developing a 'Science for Development' course to equip science graduates with a broad science perspective on development challenges. Then in 2021, the National Science and Technology Forum's (NSTF) awarded Prof Ödman-Govender one of the (dubbed the 'Science Oscars') for her work in science communication and outreach.

Ödman-Govender, Carolina was passionate about the role of science in addressing societal and global challenges, and amongst her many accolades, Prof Ödman received the International Astronomical Union Special Executive Committee Award for Astronomy Outreach, Development and Education. Another notable achievement was that she was one of the founders of Astronomy for Africa and played a key role in Universe Awareness (UNAWE), which was an outreach programme for children to foster an interest in astronomy.

Prof Ödman-Govender received the Special Posthumous Recognition for Contribution to Science Diplomacy accolade.

Professor Carolina Odman Govender was a founding member of the African Network for Women in Astronomy (AfNWA). During the many challenges and effects of the Covid-19 pandemic since the beginning of 2020, and despite her suffering, Professor Odman made a great effort to provide a number of remote training programs through AfNWA. She served AfNWA and the African Astronomical Society with all pride and grace until the end. And in recognition for all this service, the AfNWA board decided to name the early career award for women in astronomy after her and the first of these awards was awarded at the AFAS 2023 conference.

Prof. Ödman was a tireless champion for astronomy development in Africa. A little more than a year after she passed away, Professor Carolina Ödman-Govender was also honored by the Science Forum South Africa (SFSA). As part of a tribute, Prof Ödman-Govender's friends and colleagues successfully motivated the International Astronomical Union to name an asteroid in her honor and this was done in February. 2023.

Prof. Ödman had set an example that would continue to be an inspiration for all of us, she touched so many lives and her legacy will live on.

# Boutheïna Kerkeni

**Tunisia**

**Professor, Universities La Manouba and Tunis el Manar**

> *Success is no accident. It is hard work, perseverance, learning, studying, sacrifice, and most of all, love of what you are doing or learning to do."* — Pelé

Since my secondary studies in Tunis, I have been fascinated by all aspects of Physics, and Space Science, and my dream was to become an astrophysicist. I remember always asking my parents about several life phenomena and objects in the Universe and realized that the answers to all my questions were Physics. I was most fortunate to have passionate Physics teachers who introduced me to complex scientific situations and elucidated them with simple and elegant theories. As a result, this has attracted my eagerness for a deeper insight into the subject, and Physics was my choice to follow later at the University. I was passionate about physics and started helping some pupils among my relatives with their Physics studies and answered several of their why questions.

At Tunis el Manar University, where I pursued my undergraduate studies, I was always eager to understand the research activities of my professors and started reading their research papers. I spent most of my time in the library reading books and understanding the different assumptions, methods, and their implications over time. This developed into a much deeper scientific interest in the subject. At that time, we did not have access to the internet and so our time was focused and entirely dedicated to reading encyclopedias, books, magazines to extract the information required. In this aspect, I encourage young students to be organised, have clear ideas about their scientific dreams, and work continuously with enthusiasm and eagerness to achieve their goals.

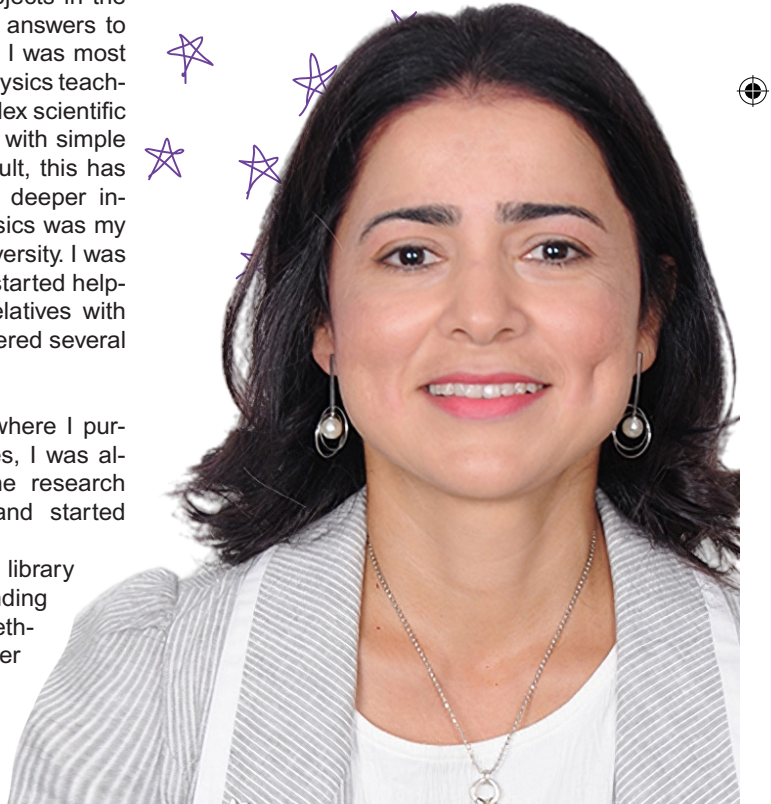

After my graduation in Fundamental Physics with honours, I moved to Paris VI University for a master's degree in atomic and molecular physics as I wanted to study the tiniest quantum interactions up to how galaxies form. My decision to study Astrophysics at the Observatory of Paris/Paris VI University with a gained CNRS-PhD scholarship had been driven by my desire to understand the complex, rich, and fascinating Universe that we are all a part of. During my PhD, I fostered several collaborations with astronomers involved in the GAIA mission. This led to fruitful findings as I was able to use my knowledge in the computation of collisional parameters and quantum chemical rates in astrophysical models to solve the master equation. This interaction task helped me gather invaluable insights by using quantum chemical data in astrophysical models and for the first time give an interpretation of the observed polarisation spectra near the Solar limb.

My interest in astrophysics research has further triggered my eagerness to transfer the knowledge to others and I seized the opportunity to teach life sciences undergraduate students at University of Cergy Pontoise (France) where some of our laboratory experiments on H2 formation at ice model surfaces were being held.

Why study the routes of H2 formation? The answer is that molecular hydrogen is the most abundant molecule in the Universe. It is the first one to form and survive photo-dissociation in tenuous environments. Furthermore, molecular hydrogen, either in its neutral or ionized form, controls much of the chemistry in the region of gas and dust between stars known as the interstellar medium (ISM).

The diffusion and recombination of H atoms on carbonaceous grains is a crucial path towards the gas-phase formation of H2 under the physical conditions of the interstellar medium. The astrophysically important catalytic formation of molecular hydrogen stimulated research over 40 years ago both theoretically and experimentally.

Despite significant advances, the characterization of several mechanisms for ISM chemistry remains largely incomplete. This observation can be made both for processes in the gas phase and for the elementary mechanisms involved at the interface between gas and interstellar dust grains. I pursued research through my postdoctoral visits to (Oxford University and later University College London) on the topic of H2 formation on graphene and several models of dust grains detected via their infra-red signature in the interstellar medium. I have developed a general method for quantum reaction dynamics to predict state-to-state cross-sections and rate constants.

My research continued to be computational-theoretical based with a particular focus on the quantum chemical (ab initio/DFT), molecular dynamics (MD), and chemical Kinetics modelling of molecular collisions and chemical reactions both in the gas phase and at hybrid interfaces. The goal of my research is to derive fundamental and molecule-specific data/parameters, like IR spectra (to compare two observations), reaction rates, and diffusion barriers. These can then be included in astrochemical models that simulate specific species evolution, track their complexity over realistic typical timescales, and constrain the properties of the underlying sources.

After my postdoctoral training, I returned to Tunisia (end of 2009) as an academic. I have organised my own research group, opened professional training (through short visits to my collaborators' laboratories), and opportunities to my students in astrophysical sciences. On returning to the University of Gafsa in 2009, I began the difficult job of building up a research group as well as preparing lecture courses which impacted negatively on my research output for a short time. However, I continued to foster several collaborations through international conferences with different groups across the world. I organised workshops and conferences and invited visiting researchers. I participated in short visits to the collaborator's laboratories. I transferred my expertise in the usage of various computational chemistry software through organised tutorials.

My current projects focus on producing credible routes by which the simplest and abundant molecules like water, ammonia, carbon monoxide, hydrogen cyanide, etc can be transformed into more complex species at fast enough rates over the 1-to-50-million-year lifetimes of molecular clouds. Abiogenesis or the origin of life is one of the most challenging questions that can be posed today.

In essence, organic molecules on the early Earth could have had either terrestrial (synthesis driven by lightning, vulcanism, ultra-violet light or cosmic rays) or extraterrestrial (formation on interstellar dust clouds and delivered to Earth by meteorites) origin. Many complex organic molecules are known to exist in molecular clouds and circumstellar envelopes, and also in comets and meteorites so the extraterrestrial origin is a possible scenario for a prebiotic chemistry startup.

Radio-astronomy has enabled the identification of over 256 species not just in our backyard, interstellar clouds in the neighborhood of our Sun, but throughout our Galaxy, the Milky Way, in nearby galaxies, and some distant quasars. Powerful new equipment like the Atacama Large Sub-Millimeter Array (ALMA) and the James Webb Space Telescope promise to deliver more insight into the composition of molecular clouds and maybe even of water-ice grains.

My main interest is to tackle challenging and controversial scientific projects and to find adequate solutions. I intend to be one of the top researchers capable of adding a plus in my specific field of investigation. I am very motivated about my work and always willing to learn new techniques in this unstoppable and fast-growing technology world. Furthermore, I am passionate about applying new techniques to solve problems differently but more efficiently. This allows me to thrive and always be active and happy in my life. I always aim at reaching a solid standing in the international scientific community which will bring me a step closer to my ideal long-term goal of building the foundations of developing contacts and establishing a permanent strong collaborating network in Africa and beyond the Mediterranean.

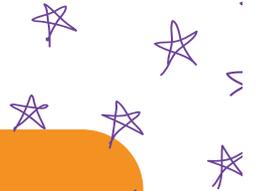

**Awards**

1. Senior Mujeres por Africa Visiting Fellowship at Instituto de Astrofisica de Canarias, Spain (2023).

2. Senior AfOx (Africa in Oxford) Visiting Fellowship at University of Oxford (2020)

3. The PRESTIGE FP7 Marie Curie-Actions, IFPEN, France (2016)

# Gloria Raharimbolamena

Madagascar

University of Bristol - United Kingdom

"*Embrace the decisions you make and the outcomes that follow.*"

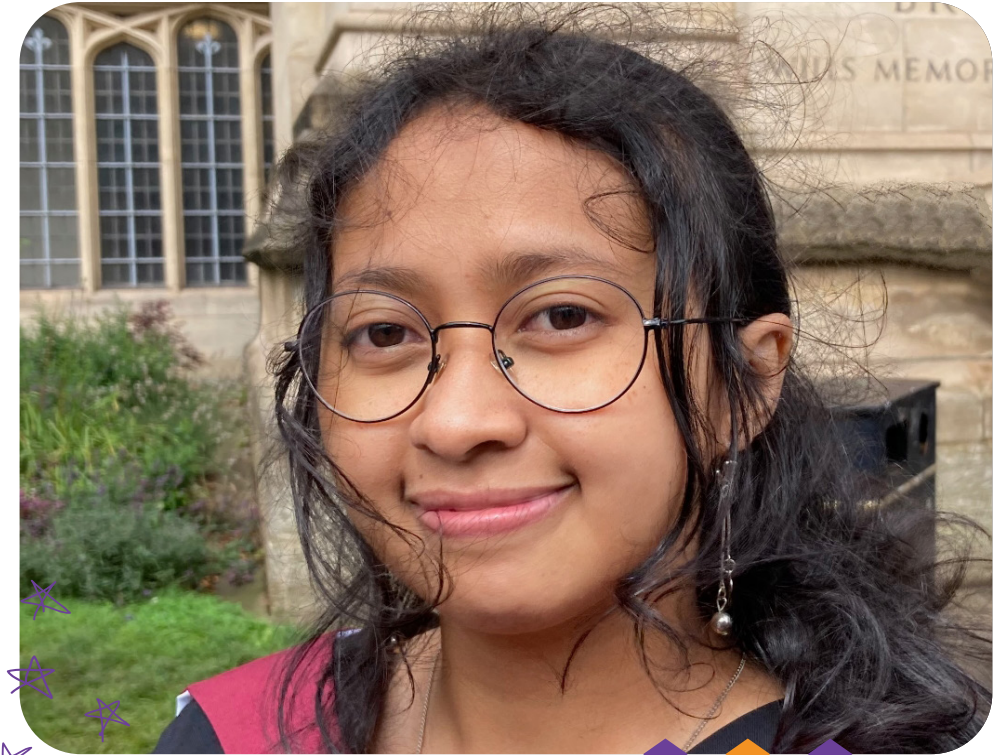

I am not the first nor the last Malagasy woman who studied astronomy, but I can proudly say that I am one of them. I grew up in Madagascar, a country with natural wonders yet considered one of the poorest in the world. My life was simple, though my knowledge of technology was limited. I did not know yet what astronomy meant during my childhood, but enjoying the night sky with my naked eyes has always been one of my favourite leisure activities. As far back as I can remember, I have always had a passion for science. It helps us understand the world around us and contributes to expanding our knowledge. Physics and mathematics were my favourite subjects at school, even if I was not good at them, but having a great teacher really helped.

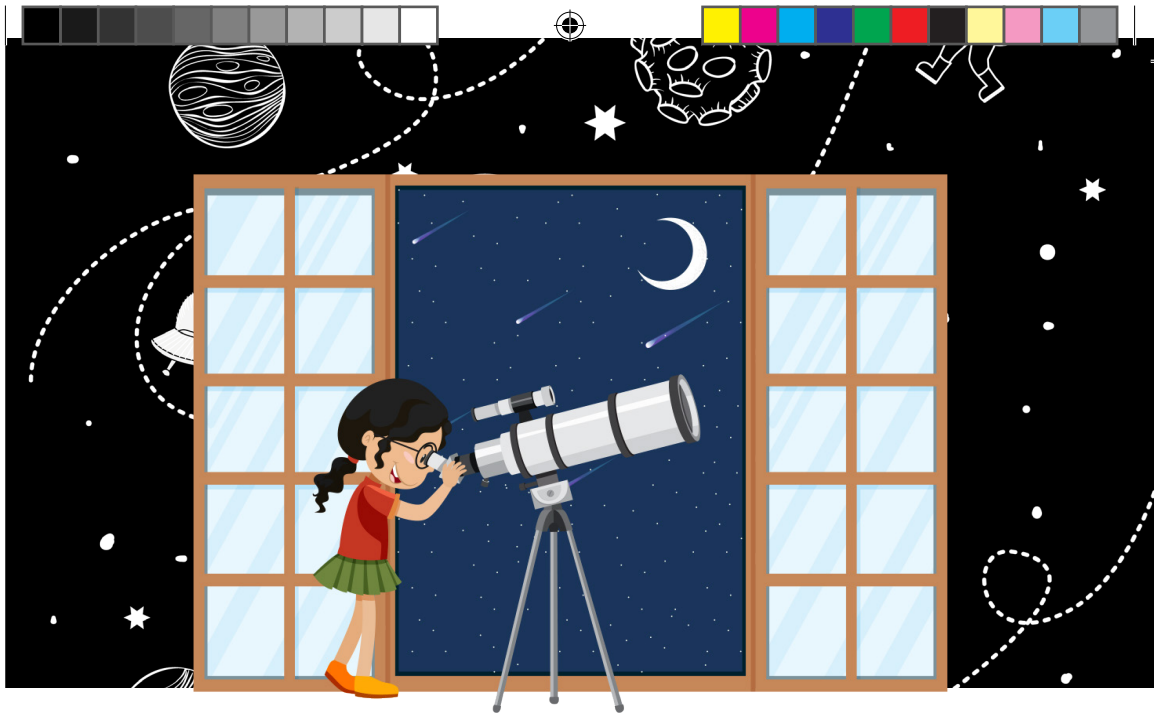

My first planned career was never as an astronomer or astrophysicist; I wanted to become a building engineer. Unfortunately, my parents could not afford a private engineering school, and the public school for this subject is further away from my city, requiring a bigger budget. Consequently, I chose to do physics at the nearest public university with the intention of branching into nuclear physics later. My first year at the University was challenging. Even if I liked the subject, I failed most of my exams and had to re-take them. It was excruciating and I was mentally broken. I could have changed my academic course back then, but I could not think of a better option. I could not afford to give up too early, as I just started. Then, I worked hard to improve my marks and finally passed.

Astrophysics and astronomy were only introduced to me during my last year of undergraduate study. When I applied for Physics, I didn't even know the existence and potential openings for the Astronomy field in Madagascar. However, I was fascinated by the subject and decided to follow that path for my master's degree. Astrophysics was a recent subject; you can only do a master's degree in Madagascar. My Astrophysics journey started at this point. I was part of the fourth cohort of Astrophysics courses at the University of Antananarivo and was the only one graduating from my cohort.

My first summer school has been the most exciting part of my odyssey. In 2018, I was given the opportunity to attend the International School for Young Astronomers in Egypt; I could never have imagined that I would be selected among several international students on my first-ever application. This school has opened a path for my future career. Many summer schools and workshops followed, namely the African School of Physics in Namibia and the JEDI Workshop in Nosy Be in the same year, 2018, allowing me to improve my knowledge and skills. Through the Advancing Theoretical Astrophysics, a summer school in Amsterdam in 2019, I developed a passion for black holes and their surroundings, which is now my current research topic.

After completing my master's degree in Madagascar in January 2020, the next step of my study was to apply for a PhD abroad. Sadly, our degree is not accredited internationally, which is compelling me to pursue a second MSc degree abroad. Through the DARA (Development in Africa with Radio Astronomy) Advanced Training Programme, I was granted a one-year scholarship to study for an MSc by Research degree in Astrophysics with Dr Andrew Young and the late Prof Mark Birkinshaw at the University of Bristol (UK). My research consisted of modelling the radio-to-X-ray emission of jets and hotspots using the Julia programming language.

However, due to the pandemic lockdown and some problems with my funding and English test, my starting date was postponed multiple times and extended to almost two years. That was particularly challenging, as keeping my motivation waiting for so long was not easy. There was a time I was thinking about finding another opportunity and giving up on Bristol. Then, I was selected for a 3-month internship programme in SISSA (Trieste-Italy), but unfortunately, it clashed with my starting date in Bristol. Fearful of missing this significant opportunity, I have been holding onto Bristol until the end and gave up on SISSA. To be honest, I wish I could have done both, but it was just unfortunate.

Luckily, at the end of January 2022, I could finally travel and move to Bristol. Adapting to a new country after the pandemic was another challenge I had to face. It was also my first time away from home for a long time. Difficulty in communicating, lack of confidence, and lack of fluency in the English language posed a barrier initially. But my determination to follow my passion, with the help of my surroundings, acted as a bridge that enabled me to overcome this challenge. After passing my viva and doing some corrections, I graduated happily in July 2023.

Meanwhile, after submitting my MSc thesis, I started a PhD in Bristol directly with the same supervisor. Finishing my MSc by Research degree and securing a PhD scholarship at the University of Bristol have been my most significant achievements so far. I am currently a 2nd-year PhD student, and my research mainly focuses on studying the emission mechanisms of jets from Active Galactic Nuclei. My biggest dream is to have my own research group in my home country one day, but for now, I will keep focusing on my studies, finishing my PhD, and securing a nice post-doc to gain more expertise.

Apart from doing research, I have also been involved in several outreach programmes through the Malagasy Astronomical Society, formerly known as Malagasy Astronomy and Space Science. Previously, I used to be the graphic designer of the association, and now I am the representative of the students studying abroad. Even if far away from home, I stay in touch with the Malagasy Astronomy community and continue contributing to the development of Astronomy in Madagascar as much as I can.

When looking back on time, I never thought I would get this far in my studies, but opting for a degree in astrophysics has genuinely been the most significant choice I have ever made. What I enjoy most about it is the exploration. Failure and challenges are part of the journey, but once you have explored and faced them, they contribute to the intrinsic beauty of your research. So, never give up; hold onto your dream and keep up the good work.

# Naomi Asabre Frimpong

**Ghana**

Ghana Space Science and Technology Institute, Ghana/ African Astronomical Society

*" I want to see Africa lit up! - that is my dream for Africa."*

As a young girl in Ghana, I was mesmerised by the constellations in Ghana's vast starlit skies. With eyes full of wonder and a heart brimming with curiosity, I would spend countless nights gazing at the heavens, yearning to unravel their mysteries. My enduring fascination with the night sky has been the guiding light on my remarkable journey through the realm of astronomy.

My journey into science and astronomy began with my academic studies in Ghana, where I obtained my BSc in Chemistry from the University of Cape Coast, delving into the phytochemical and antimicrobial analysis of Psychotria ankasensis leaves, a study about the chemistry of plants. I then advanced to earn my MPhil in Nuclear and Radiochemistry at the School of Nuclear and Allied Sciences, University of Ghana, focusing my thesis on the hydrochemistry and stable isotope analysis of the Densu Wetlands in Ghana, which was a study about water and how it evolves over time. Though seemingly distant from the study of the cosmos, these scholarly endeavours laid the foundation stones for my future in astronomy, as my gaze never strayed from the stars.

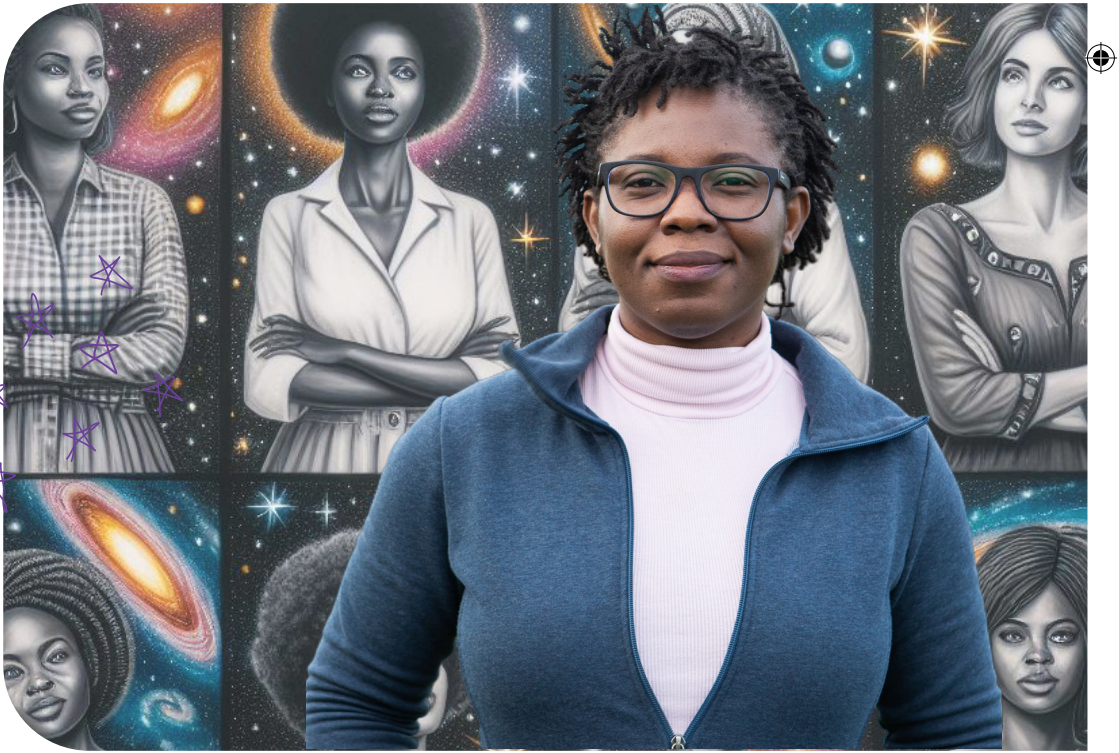

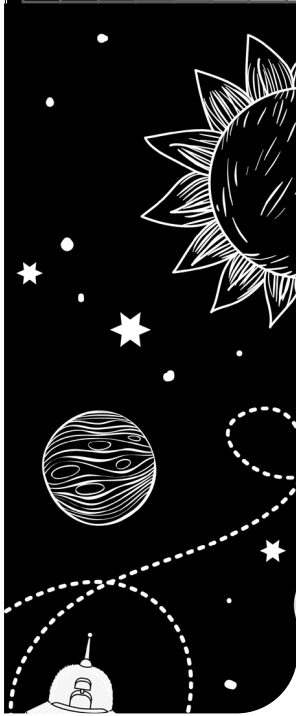
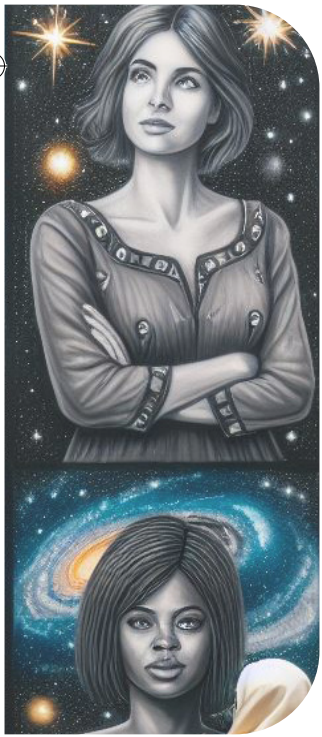

A defining moment in my career emerged when I joined the Ghana Atomic Energy Commission, home to the newly established Ghana Space Science and Technology Institute, as an assistant research scientist. I was thrust into a pioneering project that saw a 32-meter telecommunication dish converted into a radio telescope, marking the inception of the Ghana Radio Astronomy Observatory. This project was more than a mere technical feat; it was a portal to the international astronomical community and a beacon of hope for Ghana's and Africa's growing role in space science.

My passion for astronomy deepened through collaborations with esteemed astrophysicists and astronomers who recognized and fostered my growing interest. Their support encouraged me to expand my knowledge and skills by participating in the basic astronomy training program organized by the Development of Africa with Radio Astronomy (DARA) project. Additionally, I participated in an intensive three-week astronomy training program for young African scientists in India, facilitated by the Inter-University Centre for Astronomy and Astrophysics (IUCAA) through the Indian High Commission. These enlightening experiences opened my eyes to the vastness of astronomy and the exciting possibility of intertwining my love for chemistry with my celestial passion. This newfound knowledge propelled me towards pursuing a PhD in astronomy and astrophysics at the University of Manchester's Jodrell Bank Centre for Astrophysics, funded by the Newton Fund through DARA.

My research delved into the majestic world of massive young stars; colossal entities capable of overshadowing our sun by up to 500 times. Their unique evolutionary processes captivated me, prompting an in-depth investigation into their evolution and formation within dark molecular clouds. By harnessing the power of the Atacama Large Millimetre Array (ALMA) telescope, I probed the intricate atmospheric conditions of these young stars, analysing complex organic compounds like methanol and methyl cyanide to decipher the mysteries of their formation and evolution.

While working on my PhD, I had the amazing opportunity to volunteer as a producer, presenter, and editor for the Jodcast, an astronomy podcast created every two months by PhD students and postdoctoral researchers from the Jodrell Bank Centre for Astrophysics. As I neared the end of my PhD journey, I was honoured to step up as the executive producer of the Jodcast for two years. This role was incredibly rewarding, and being part of this project was a lot of fun.

Navigating through my PhD was a testament to resilience, presenting a formidable challenge that refined my work ethic and profoundly shaped my outlook on life. Transitioning from a chemistry background to a domain dominated by physics was daunting, yet it was an invigorating experience that highlighted the critical importance of pursuing one's passions, regardless of the hurdles encountered.

Beyond the confines of research, I have embraced the mantle of science communicator, eager to share the universe's wonders as the Head of Science Communication and Outreach at the Ghana Space Science and Technology Institute. Additionally, as the current National Outreach Coordinator (NOC) for Ghana for the Office of Astronomy Outreach (OAO), I have spearheaded and contributed to various astronomy outreach and education initiatives, including science exhibitions, tours, school visits, mentorship and training programs, and workshops. These efforts aim to foster a love for astronomy and space science within Ghana. Currently serving as the Vice-President of the African Astronomical Society, a Pan-African Professional Society of Astronomers which is committed to advancing astronomy and human capacity across Africa.

My work as a mentor in the Space4Women mentorship program by the United Nations Office for Outer Space Affairs (UNOOSA) in 2023 further underscores my dedication to empowering the next generation of scientists, particularly young women, and girls. My passion for mentorship stems from the invaluable guidance of exceptional teachers and lecturers, the motivation provided by remarkable mentors, and the unwavering support of wonderful friends and family throughout my journey.

While my unbridled curiosity, combined with my perseverance, has led me to become the first female astronomer in Ghana, I certainly hope I won't be the last. I hope that my journey will serve as a beacon of encouragement, especially for girls and women aspiring to delve into the sciences. My message is straightforward: 'exert your best effort so that your presence and contributions cannot be overlooked'. As I look to the future, my aspirations are as boundless as the universe I study. I aim to be a source of inspiration, not only for my daughters but for all the young dreamers daring to reach for the stars. My plans include continuing my research, fostering international collaborations, and passionately engaging in outreach and science communication.

Reflecting on my journey from being a starstruck girl beneath Ghana's night sky to a scientist exploring the cosmos, I am reminded of the power of dreams and the importance of perseverance. My story is a testament to the beauty of following one's passion, the significance of mentorship, and the boundless opportunities that science offers to change our world. Let this story inspire you to reach for the stars, for in the vast expanse of the universe, there is room for every dream to soar.

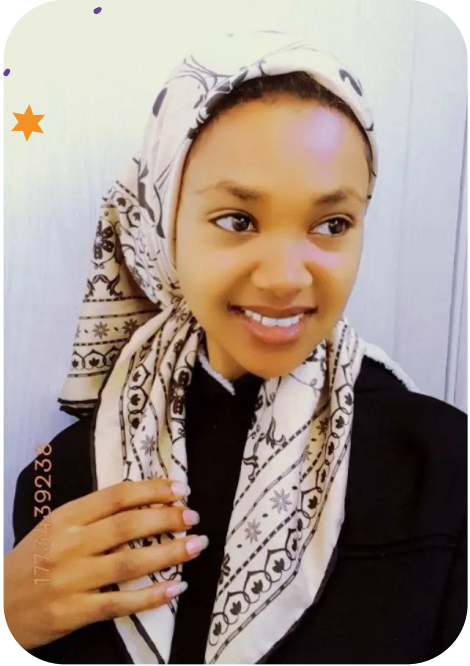

# Mehbuba Ahmed Mohammed

**Ethiopia**

**Space Science and Geospatial Institute (SSGI), Ethiopia**

> *Embrace challenges with unwavering determination, for they are stepping stones to growth and breakthroughs. Let curiosity guide you, and inspire others to reach for the stars.*

### A Journey Through the Dark Skies

Once upon a time, in the enchanting village of Dessie, Ethiopia, there lived a very young girl named Mehbuba. She was captivated by the mesmerizing night sky that spread its velvety blanket overhead.

Despite growing up in a village with limited access to tools and education about any astronomical objects, Mehbuba's fascination with the brightest stars in the dark sky fueled her curiosity about the universe. As Mehbuba grew older, her passion for astronomy remained undimmed. She dreamed of studying astronomy and becoming an astronomer.

After completing her studies in physics, she joined an astronomy institute in her home country to pursue a master's degree in astronomy and astrophysics. Her focus was on studying active galaxies, a captivating area of research in the vastness of space. She recently completed her MSc studies and has now joined the Astronomy and Astrophysics Department at the Space Science and Geophysical Institute (SSGI) in Ethiopia.

This exciting new chapter in her academic and professional journey allows her to further explore her passion for the mysteries of the universe and contribute to the advancement of scientific knowledge in Ethiopia. Throughout her journey, Mehbuba encountered obstacles and challenges. Despite facing the adversity of being a victim of the conflict, she refused to let those challenges define her or hold her back. Instead, Mehbuba chose to confront each obstacle head-on, demonstrating an unwavering determination to overcome the difficulties in her With unwavering determination and the support of mentors, like-minded friends, and individuals, she overcame each hurdle that came her way.

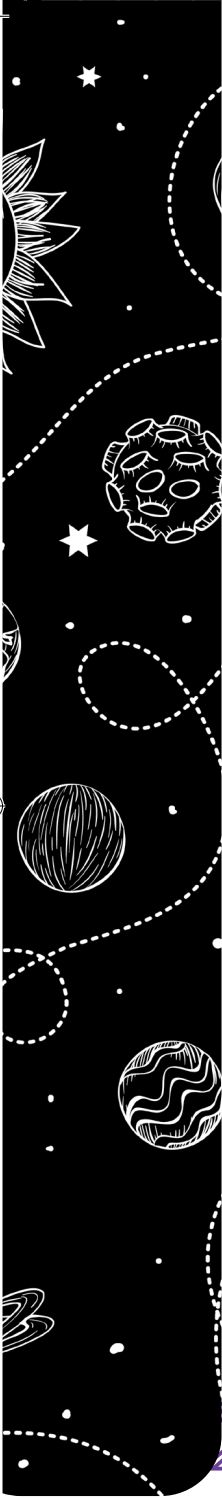

Her dreams went beyond personal success. Mehbuba was deeply committed to sharing her love for astronomy with others, especially young people. Her dedication to advancing the understanding of the universe and her efforts in promoting astronomy education left a lasting impact. She knew that her journey was far from over. She founded an outreach programme group that brought different activities to the schools, allowing young students to marvel at the wonders of the universe and inspiring them to become future astronomers.

In the realm of science, Mehbuba found joy and fulfillment. The thrill of learning, discovering, and pushing the boundaries of knowledge in the field of astronomy brought her immense happiness. She believed in the power of curiosity, perseverance, and embracing challenges as stepping stones to growth and breakthroughs.

Mehbuba's biggest dream is to become a change-maker scientist. She aspires to make significant contributions to the field, leaving an indelible mark on the scientific community. Her ultimate goal is to inspire future generations of scientists, particularly girls and women, to pursue their dreams and break the barriers that society might impose.

As Mehbuba looked to the future, she envisioned a world where astronomy education would be accessible to all, nurturing a new generation of curious minds and expanding our collective knowledge of the cosmos. And so, Mehbuba's story continues, filled with determination, passion, and a deep love for the mysteries of the universe.

**Awards**

1. The 2022 Ethiopian Physical Society North America (EPS-NA) Scholarship award at a graduate student level

2. The American Physical Society Women in physics (APS Wip) Group Grant, 2021

3. The 2019 Ethiopian Physical society North America (EPS-NA) Scholarship award at an undergraduate student level

# Alaa Salah Afifi

**Egypt**

**Astro Tech Hub**
**Physics specialist /space mentor**

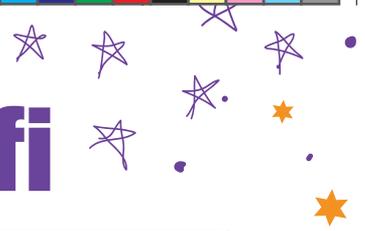

*In the vast expanse of the cosmos, where stars twinkle like diamonds scattered across a velvet sky, there exists a passion that burns brighter than the fiercest supernova. It is the passion of Alaa Salah, a young astronomer whose journey into the world of space and science is as inspiring as it is remarkable."*

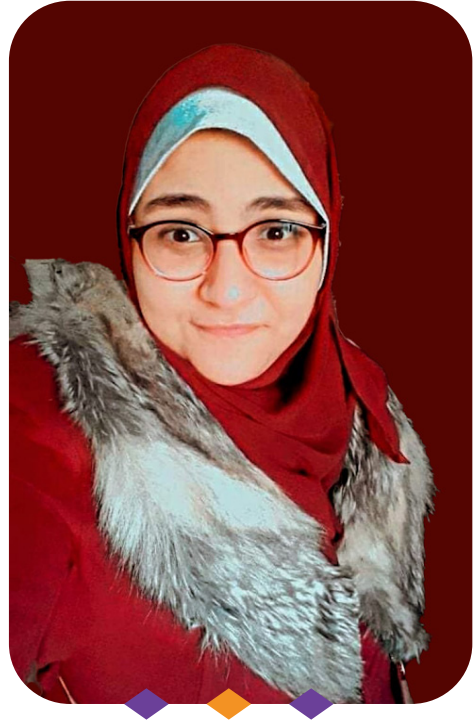

Alaa's story begins with a dream, a dream fueled by the teachings of their father, a physics teacher whose passion for science ignited a fire within them. From a young age, Alaa saw science not just as a subject of study, but as a magical force that could transform the ordinary into the extraordinary. They watched in awe as their father turned the complexities of physics into simple, understandable concepts, much like a magician turning dust into diamonds.

With this inspiration guiding their way, Alaa embarked on a journey of discovery, pursuing a bachelor's degree in astronomy. But the path to their dreams was not without its challenges. Upon completing their degree, Alaa faced the daunting task of entering the space industry without a master's or PhD degree. Undeterred, they took matters into their own hands, enrolling in numerous courses in space science and programming, and immersing themselves in the world of professionals.

Through sheer determination and hard work, Alaa secured a spot in the space industry, a feat that seemed impossible at first. Their journey taught them an invaluable lesson: the importance of perseverance. Like the thorns that adorn a rose, life's challenges may be sharp and painful, but they are also what make the journey beautiful and rewarding.

As a space mentor at the Space Generation Advisory Council and a Physics Specialist at Astro Tech Hub, Alaa's passion for science shines brightly. They conduct sessions and present at conferences, sharing their knowledge and insights on the role of radio waves in communication, connectivity, and space exploration. Their dedication to education is evident in their efforts to educate others on the importance of radio waves and promote understanding of key concepts in space science and technology.

In the realm of education and knowledge-sharing, Alaa Salah shines as a beacon of enlightenment. Their journey has taken them to various platforms where they have shared their expertise and insights with eager minds.

At the Space FUTA Club, Federal University of Technology Akure, Alaa was an invited speaker, where they conducted a one-hour session on the role of radio waves in communication, connectivity, and space exploration. Their session not only educated attendees on the importance of radio waves but also provided valuable insights that sparked curiosity and promoted a deeper understanding of key concepts in space science and technology. Engaging with students, Alaa was able to inspire a new generation of space enthusiasts and future scientists.

Another notable speaking engagement for Alaa was at the Virtual Conference on Condensed Matter Physics (ISCMP2023). Here, they presented a thought-provoking session titled "What if our Cell Brain is a Space-Traveling Compass?" This session, which lasted an hour, was a collaborative effort with professors and doctors from around the world. Together, they presented a review article and delved into the implications and theories related to the role of our cell brain in space travel. This session not only showcased Alaa's depth of knowledge but also their ability to collaborate with experts from diverse backgrounds to explore cutting-edge concepts in science.

But for Alaa, science is more than just a subject of study; it is a mirror that reflects not just the world around us, but also ourselves. They believe that in the journey of scientific exploration, it is not just the destination that matters, but the path we walk. Each step we take, and each challenge we overcome, adds depth and meaning to our journey, making it truly unforgettable.

Looking to the future, Alaa's dreams are as vast as the universe itself. They aspire to become an astronaut, to float among the stars and witness the wonders of outer space firsthand. But beyond their personal aspirations, Alaa's heart is set on helping others achieve their dreams. They dream of becoming a lecturer in astronomy, inspiring the next generation of scientists to reach for the stars.

Alaa's story is a testament to the power of perseverance, passion, and unwavering belief in the pursuit of one's dreams. It is a story of hope, of resilience, and of the boundless possibilities that lie within each of us. And as they continue on their journey, Alaa remains a shining example of what it means to reach for the stars.

**Awards**

1. Moons of our Solar System Badge Issued by The Open University

# Meryem Guennoun

**Morocco**

**Oukaïmeden Observatory / Laboratory of High Energy Physics and Astrophysics**

> *First female astronomer in Morocco"*

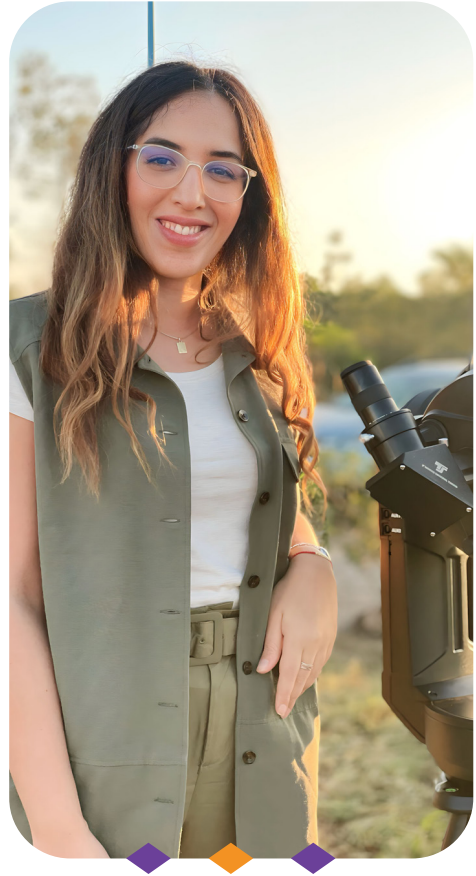

During my formative years in Beni Mellal and Kelaa Sraghnas, small cities that could easily be mistaken for villages, my fascination with astronomy began between the ages of 3 and 5. I found myself drawn to celestial wonders through pictures and my own sketches (see pictures below).

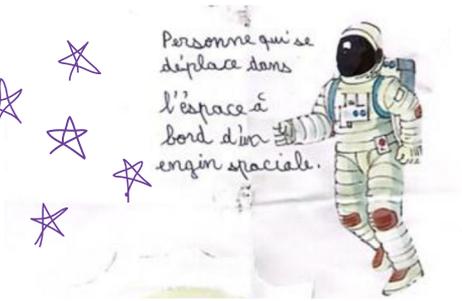

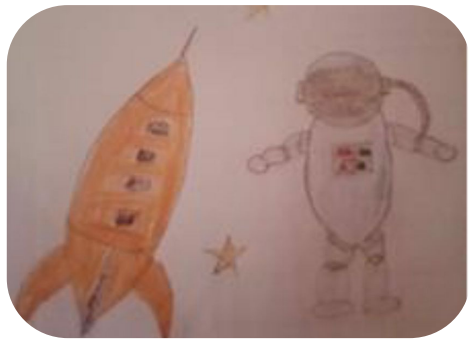

However, growing up in 1990 in Kelaa Sraghna, I unwittingly nurtured a limiting belief. The notion that astronomy was reserved for ambitious careers in America seeped into my consciousness, dampening my childhood dreams. Over time, my celestial aspirations faded from my conscious thoughts, creating a mental distance even to this day.

As the years unfolded, I excelled academically, standing among the top students in my class. Yet, the trajectory of my career remained uncertain. In a society where options for someone like me often gravitated towards medicine or engineering, I found myself certain about what I didn't want, which was medicine. The realm of engineering beckoned, but the specific path remained elusive. Choosing to keep my options open, I pursued a major in MIPC (Mathematics, Computer Science, Physics, and Chemistry) and eventually earned a scientific bachelor's degree in physics for engineering.

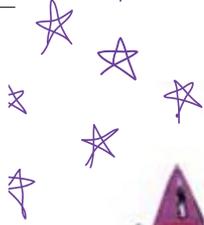
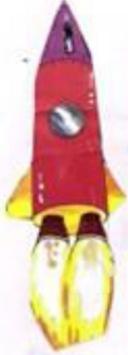
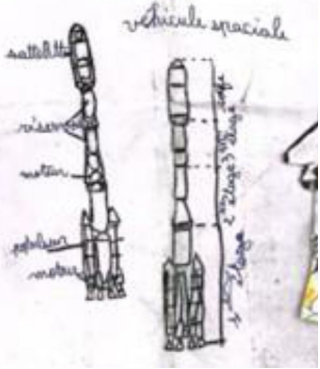
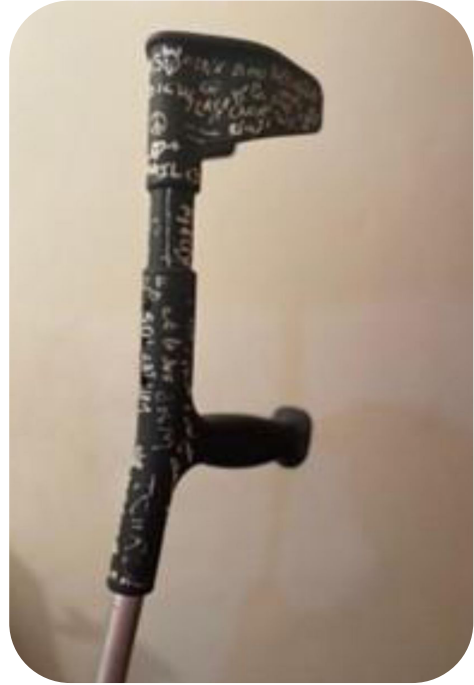

The turning point in my journey arrived during a summer overshadowed by illness. Stricken with joint rheumatism, I found myself unable to walk for a month. Concerned about my future, my best friend took the initiative to apply to every master's program in physics at the Faculty of Science in Marrakech on my behalf. Supported by medication and a walking cane (see picture below), I gradually regained my strength, just in time to face the oral exams for master's degrees.

It was during this challenging period that I stumbled upon a revelation. The master's program in High Energy Physics and Astrophysics, a hidden gem in the Moroccan educational landscape, materialized before me. The mere existence of such an option in our country left me in awe and profoundly changed the course of my journey.

Entering the realm of astrophysics and preparing for my PhD., I encountered a new challenge, an external limiting belief. A disheartening fact loomed over me: no other woman had successfully defended their PhD. in this field. Unbeknownst to me initially, the weight of this statistic manifested in discouraging remarks, not necessarily ill-intentioned, but nonetheless casting a shadow on my aspirations. Strangely, it was the voices of those around me, particularly younger individuals, that echoed skepticism about the compatibility of astronomy with a woman's capabilities. The whispers became audible: "Astronomy is hard for girls; perhaps you should consider an alternative career…" These comments, seemingly innocuous, became a constant companion to my struggles as I navigated the challenges of my PhD.

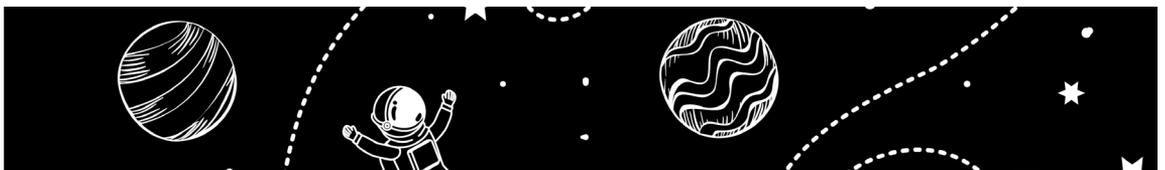

Amidst the discouraging whispers, I found solace in the unwavering support of my two supervisors and some other teachers at our university. Fortunately, these mentors did not subscribe to the limiting beliefs that circled around me. Their encouragement served as a beacon of light during moments of darkness. While the road seemed arduous, I clung to the belief that women can excel in any field if they wholeheartedly pursue their passions. This conviction became my anchor in turbulent times.

The remarks, instead of breaking my spirit, became fuel for my determination. Each comment, intended or not, served as motivation to defy expectations and prove that astronomy knows no gender barriers. There were moments when the journey seemed insurmountable, but I persisted, holding onto my core belief that women can conquer any career path with resilience and dedication.

In my astronomical pursuits, I specialize in unraveling the mysteries of meteor trajectories and orbit calculations, a captivating journey that extends to identifying the parent bodies of observed meteors, revealing asteroids as their celestial origins. Beyond these celestial connections, my work represents a groundbreaking shift in meteor science. I challenged and reshaped the conventional criteria for defining meteor showers, exposing their inherent limitations and advocating for a more robust and nuanced approach.

In a triumphant moment, I defended my PhD. The culmination of years of dedication and perseverance marked not only a personal victory but also a historic moment for Moroccan astronomy. During the official ceremony, my teachers proudly declared that I was the first female astronomer in Morocco.

Fast forward four years, and I am delighted to witness a shift in the narrative. Three more women have successfully defended their PhDs in astrophysics within our lab. The echo of my journey has, hopefully, served as a source of hope and inspiration for these young women. It is my fervent wish that my story becomes a testament to the fact that dreams in science, or any field, are within reach for every aspiring girl, encouraging a future where more women carve their paths in pursuit of their passions.

Looking ahead, my most cherished dream is to continue my journey in astrophysics, seamlessly balancing the pursuit of cutting-edge research with a fulfilling career in teaching. I envision a future where the realms of academia and inspiration converge, allowing me to impart knowledge and passion to the next generation of scientists in Africa. A particular focus of my aspirations lies in empowering young minds, especially girls, to fearlessly pursue scientific careers. My wish is to be a guiding light, encouraging them to defy societal expectations, and embrace the limitless possibilities within the realms of science.

# Yara Herminia Simango

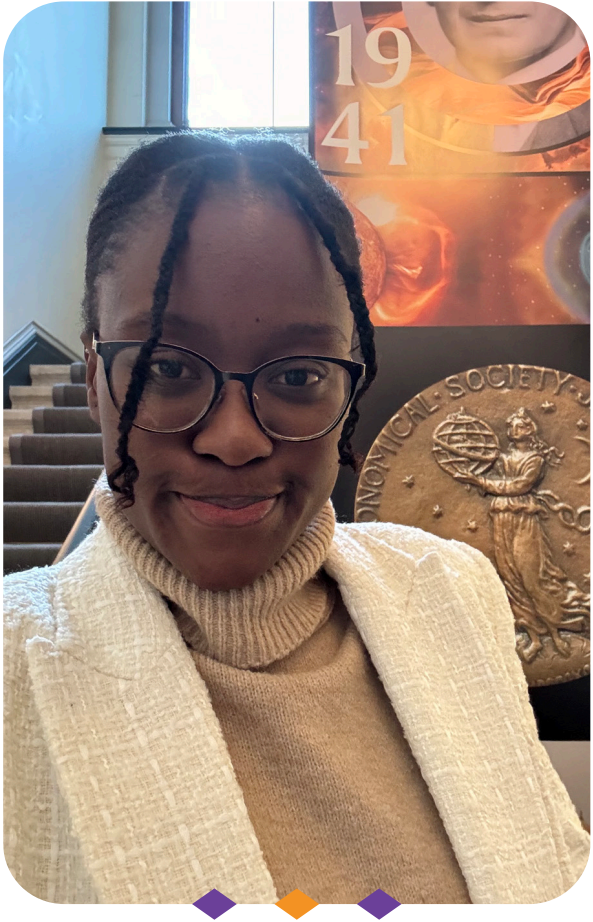

**Mozambique**

**PhD student, University of Bristol and University of Cape Town**

> *The fact that Astronomy is not done in Mozambique is an opportunity to start. We start from where we are with what we have, everywhere in the world people started from somewhere."*

I am Yara Hermínia Simango, 28 years old and I am from Mozambique.

I grew up in one of the neighbourhoods on the outskirts of the city of Maputo, the capital of Mozambique. I grew up as an ordinary child. I went to a public primary school close to my house, for reasons of practicality and safety. Where I come from, children are placed in schools close to home. I was always very outgoing, I interacted with everyone, my teachers knew me for being a smart girl and everyone else at school knew me for being part of the traditional dance group.

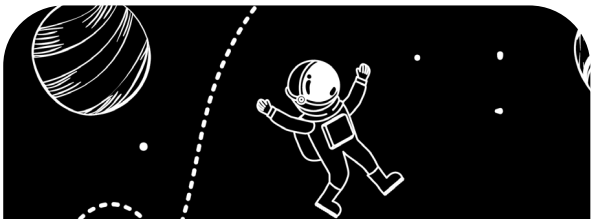

When I went to secondary school, my desire to be a scientist began to emerge. At that stage of my life, I began to have access to some television programs and films where scientists were people of great social prestige. The scientists in the films that I used to watch always came up with solutions that helped a very large group of people and that seemed fascinating to me. At the same time, I was doing better in basic science subjects at school, I mean physics, chemistry, mathematics, and biology.

One of the most difficult decisions of my school journey was deciding on which course to apply for at university. I had little access to information about the courses taught by the faculty of science. Furthermore, in my community, courses at the Faculty of Science are undervalued compared to those at the Faculty of Engineering. Few people understand the real applications of science and because of this, whenever I spoke about my desire to take a science course at university I was extremely discouraged. The claim has always been "There are no jobs for what you want to do" or "You are so smart, you could do engineering, agronomy, medical science, or education". I ended up having the courage to apply for faculty of science courses; I applied for computer science as my first option and for Meteorology as my second option. In fact, computer science was just because the only person I knew who had managed to study at that college was taking this course. He was a friend who I admired a lot, the lack of role models within my small social bubble was a barrier at that stage. I am glad I was not admitted to my first option, because I really had a strong inclination to physics. Being admitted to the Faculty of Science was one of the greatest achievements of my life until that time. I had been admitted to the most prestigious university in Mozambique; on average five thousand people compete for 40 places on a course.

I then did my bachelor's degree in meteorology at Eduardo Mondlane University. When I was in my second year I had an introductory course in astrophysics. I had a great professor. He used to talk a lot about astronomy and the projects that were coming to Africa in this field. He also used to talk a lot about the SKA project. I was completely fascinated, and I decided that I wanted to be an astronomer. The same professor used to organise a lot of activities related to astronomy (outreach and educational activities) within the physics department, so I started to be involved in all activities. Right after I finished my bachelor's degree, I was admitted to the Development in Africa with Radio Astronomy (DARA) basic training, which is where I gained basic skills.

After some unsuccessful applications for a master's in astrophysics, I was admitted to the University of Leeds and received a scholarship from the DARA project. When I was ready to go to the UK in early 2020, the pandemic came, and I could not go. As time passed, the pandemic situation only worsened. It was a very desperate moment in my life. At the same time, I was approved for a one-year grassroots leadership academy that amplifies women's talent, raises gender equality, and promotes sustainable transformation, through a program called CHANGE from Girl Move Academy which I turned down in order to go for my master's. I then decided to do a four-month post-graduate course in teaching methodologies for physics at Pedagogic University. At this point, the plan was to finish the course and become a physics teacher.

In 2021 the pandemic calmed down; I was able to travel to the UK to do my long-awaited master's degree in astrophysics. At the end of my master's degree, I applied for some PhD positions and was admitted to a joint PhD program between the University of Bristol and the University of Cape Town, which I started in 2023. The place I applied for at the University of Bristol was filled by someone else, in fact, the first response I received was that I had not been admitted to the program, it was very frustrating. However, my supervisor considered it perfect for the Cotutelle program, which was a possibility. She proposed it to the department, and it was accepted. At this stage of my life, two people were very important, my master's supervisor Prof. Melvin Hoare, who taught me everything I needed to be able to conduct research and achieve results, and my current supervisor Dr. Natasha Maddox, who believed in my potential, and proposed the university to open an extra position for me.

Professionally I am trying to develop leadership skills. Together with some Mozambican astronomers we are establishing a solid community of astronomers actively working in the field based in Mozambique. I am currently the president of the Mozambican Astronomical Society.

My master's research was in massive star formation; characterizing the nature of ultra-compact HII regions from the southern Galactic plane, found in the CORNISH-South survey. In my PhD I am working on a project whose goal is to do mass modelling first for nearby, well-resolved galaxies, then extend the technique to galaxies at greater distances.

I believe science is a very powerful tool to achieve Sustainable development goals, especially in underdeveloped countries. My advice to the girls (especially African) reading my story, even if doing science seems to be very difficult, if it is your dream, break the stereotypes and move forward. You can have a huge impact on the development of our continent. And what I like the most about doing science is the fact that I travel to different places to attend conferences, without even paying for it.

In conclusion, the biggest challenge of my journey was the language. Coming from a Portuguese speaking African country, learning English was not an easy task.

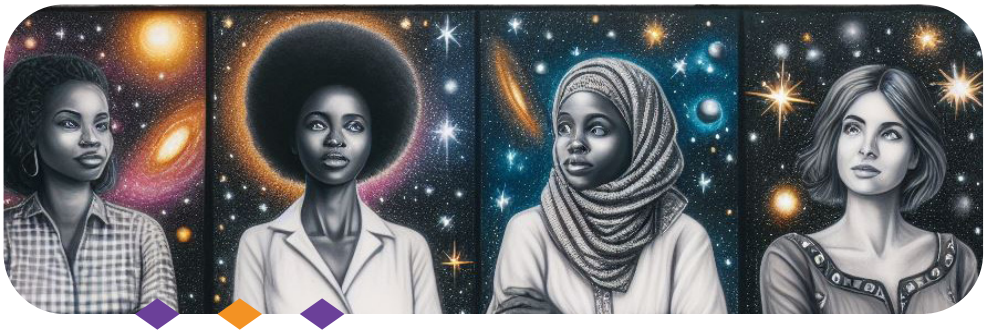

# Valentine Nyirahafashimana

**Rwanda**

**East African Institute of Fundamental Research (EAIFR)/Universiti Putra Malaysia**

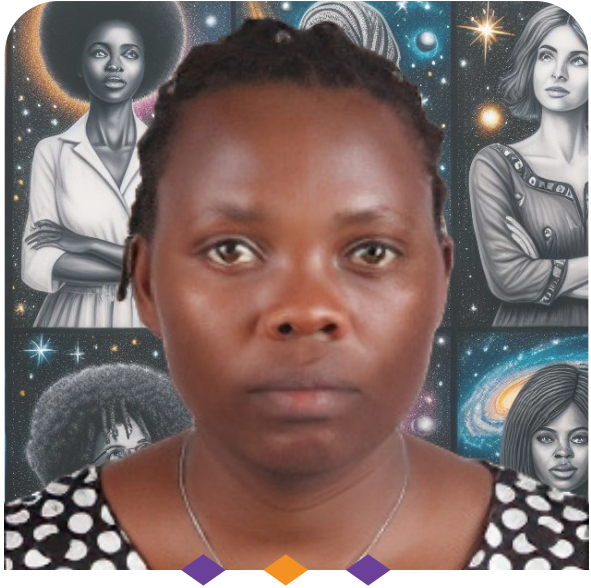

> *"I am a scholar who seeks to learn and improve myself to serve society and solve problems."*

**M**s. Valentine Nyirahafashimana is a 34-year-old Rwandan with a strong passion for science, particularly Physics. Her academic journey began with certificates in Physics, Chemistry, and Mathematics as her major subjects in G.S.Saint Joseph Nyamasheke. She pursued her interest in Physics at the University of Rwanda, where she earned a second class upper division Bachelor's degree in Science with an option in Fundamental Physics in August 2017. Her final year project focused on "The Deuterium Fluoride Carbon Dioxide laser – Characteristics and its military applications".

She taught Physics at a prestigious high school, a Technical and Vocational Education Training School in Rwanda, and at Kigali Independent University Polytechnic Institute in Civil Engineering, Electronics & Telecommunication and Electrical Technology. In these roles, she actively participated in various team activities, including organizing science competitions, workshops, and coaching sessions, with a special focus on supporting female students in STEM. Her dedication was recognized when she attended ASP2018 in Namibia and received a certificate.

Subsequently, she pursued a Master of Science in Physics in High Energy Physics and Astroparticles at EAIFR-ICTP; an international hub of research excellence in Rwanda in 2021, supported by the Or-ganisation for Women in Science for the Developing World (OWSD) fellowship. Her master's research focused on Multi-wavelength studies of strong gravitational lensing phenomena which involved investigating the bending of light around massive objects, such as galaxies or galaxy clusters, to understand the properties of both the lens and the source. This can provide valuable insights into the distribution of matter in the universe and contribute to our understanding of dark matter, dark energy, and the large-scale structure of the cosmos. The research was under the supervision of Dr Lucia Marchetti from the University of Cape Town, South Africa. This experience broadened her expertise in the field of Astronomy.

- Ms. Valentine Nyirahafashimana is currently a PhD. student candidate at Universiti Putra Malaysia, specializing in mathematical physics and engineering. Her research, titled "Advancing Quantum Error Correction: A Comprehensive Investigation into Innovative Geometric Approaches, Non-Stabilizer Groups as QEC Codes, and the Surface Code," reflects her focus on cutting-edge topics in quantum error correction. Overall, Ms. Nyirahafashimana's academic and research background demonstrates her commitment to advancing quantum error correction and her expertise in high-energy physics and astroparticles.

Science is crucial for understanding the world around us, solving problems, and advancing human knowledge. It enables technological progress, medical breakthroughs, and a deeper comprehension of our place in the universe.

Scientists find joy in discovery, exploration, and knowledge contribution. Young women in astronomy enjoy collaborative efforts, overcoming challenges, and contributing to cosmic understanding. Professional networking, breaking gender stereotypes, and participating in outreach programs foster a sense of belonging. Achieving research milestones, maintaining work-life balance, and continuous learning contribute to overall joy.

Therefore, she encourages curiosity and a passion for learning from a young age, especially among ladies, advocates for equal opportunities in STEM fields, and provides support and mentorship to girls and women interested in pursuing careers in science.

Her challenges included funding constraints, complex data analysis, and competition for resources. The next was the journey from a master's in astronomy to a PhD. in Quantum Information Science. This made her realise the importance of persistence, collaboration, and staying updated on advancements. The stimulating intellectual journey, international exposure, collaboration, networking, teaching, and advocacy opportunities in this interdisciplinary field have kept her going.

As an astronomer, she aspires to contribute significantly to our understanding of the universe, whether through groundbreaking research, discoveries, or advancements in technology.

She plans to keep on with the ongoing research projects, and collaborations, and stay engaged in the scientific community. Also, she wants to be involved in mentoring the next generation of scientists and promoting science education.

# Adaeze Lorreta Ibik

**Nigeria**

**PhD student, University of Toronto, Canada**

*My passion for understanding the Universe drives me to break barriers in astronomy, from teaching to pursuing a PhD, while advocating for greater representation of African women in STEM. With determination and support, I strive to inspire curiosity and make science accessible to all."*

My name is Adaeze Lorreta Ibik. I'm a PhD student studying Astronomy and Astrophysics at the University of Toronto in Canada. I come from Enugwu-Ukwu in Anambra State, located in the South-Eastern part of Nigeria. I got my bachelor's and master's degrees in physics from the University of Nigeria, Nsukka. Before starting my PhD in Toronto, I worked as a Physics teacher for 5 years. During my time in Toronto, I've also been a Teaching Assistant for an Astronomy course and a science communicator of different flavours.

My interest in Astronomy began during my final year of undergraduate studies when I took a course in astronomy. I was fascinated by how much we could learn about the universe and our environment. I joined an amateur astronomy club, and my interest grew further when I attended the Pan-African School for Emerging Astronomers (PASEA) school in 2013. The mix of international and local instructors inspired me, and I realized that I could be a scientist too. After my first PASEA, I focused on facilitating basic space science outreach programs in primary and secondary schools, collaborating with the Center for Basic Space Science in Nsukka Nigeria. Because of my passion for science communication, I agreed to join the PASEA instructor team, where I had the opportunity to conduct teaching sessions at the Zambia 2022 PASEA school.

During my National Youth Service year after graduation, I continued basic science outreach in secondary schools in the South-Western part of Nigeria. I pursued my master's degree in physics with a major in Astronomy. In 2017, I was selected to attend the International School for Young Astronomers (ISYA) in Ethiopia, which gave me exposure to the research component of astronomy. At that moment, I faced a crucial decision regarding my future amidst the array of opportunities before me. Though I had harbored a desire to become a scientist since high school, it was only after being inspired by PASEA that I seriously considered pursuing a career as a Lecturer/Professor in astronomy.

Despite hailing from a humble background, where such career paths were uncommon among my peers, I persisted in my dream, even as I struggled to finance my master's program and afford basic necessities, let alone application fees. However, fueled by determination and guided by my Christian faith, I held onto the belief that my story would one day take a different turn, as I trusted in God's guidance and remained steadfast in speaking positivity into my life.

After several failed attempts to get admission for my PhD. abroad, I finally received a fully funded scholarship from the University of Toronto, Department of Astronomy, in 2019. This became possible because I received waivers for application fees, financial help from friends, and the generous support of the Dunlap Institute and one of the astronomy professors to whom I had applied. Just before starting my PhD, I joined the Helping Hand Network (an educational non-profit organization) based in Liberia to conduct a teacher workshop for science teachers in Liberia co-sponsored by the International Astronomical Union. This was a transformative experience for my teaching career as I had to teach astronomy to teachers and not students.

As a PhD student, I engaged in lots of outreach initiatives organized by the Dunlap Institute of Astronomical Instrumentation and Science Rendezvous at the University of Toronto across the Greater Toronto Area in Ontario Canada. My current research focuses on studying bright explosions in the universe, such as fast radio bursts (FRBs) and luminous supernovae (SLSNe). FRBs are short energetic mysterious lightening coming from distant galaxies in the Universe. Recent discoveries have shown that some come from different types of galaxies beyond our Milky Way, but we still don't understand their origin or what causes them.

Part of my research is to use radio imaging data to find their progenitors and the cause of the explosion if possible. SLSNe on the other hand accompanies the death of a massive star or binary star system. We have found a number of these explosions and are now using them to trace the history of their progenitors a few decades before they exploded. I use data from telescopes of different wavelengths (colors of light) to carry out my analysis. This is an important aspect of my work because collecting light of diverse colors from an astronomical object or event provides valuable insights into their characteristics and behaviors, like piecing together a puzzle.

I love science because I enjoy understanding how things work. As an aspiring astronomer, I am amazed at the magnificent Universe and the creator–God. I'm passionate about making science accessible to everyone, especially girls in Africa. My goal is to contribute to increasing the representation of African women in STEM through various means such as outreach programming, online content creation, mentoring, etc. I recognize that every career, particularly for women, presents challenges. Nonetheless, we have the power to contribute significantly, as our unique perspectives are vital for driving technological advancements.

Having encountered obstacles firsthand, I recognize the significance of having a support network. I've been fortunate to have supportive individuals and organizations who have assisted me on my journey. Despite the demands of balancing mother-hood and academia, I'm determined to continue my research and become a science communicator. I also know some people who are successfully doing this with many children, and they remain my distant support system.

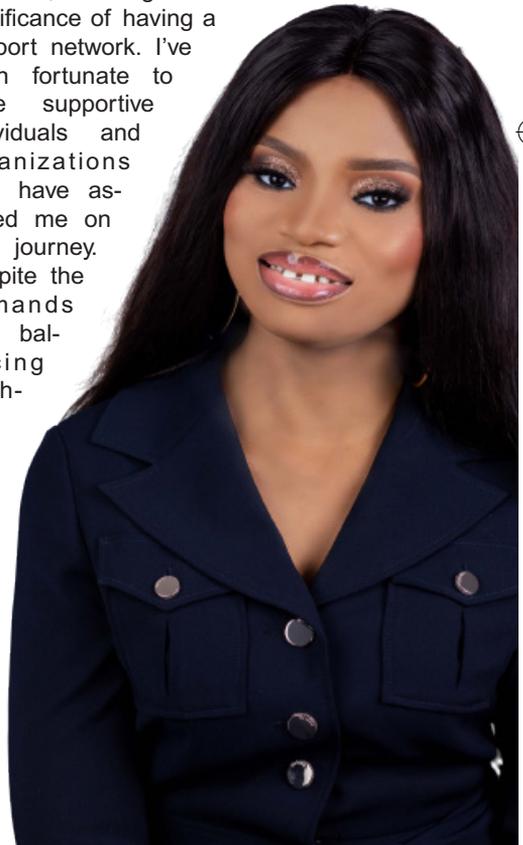

With a heart full of hope and determination, I strive to make science accessible to all, especially young African girls, inspiring them to pursue their dreams in STEM fields. As I continue my quest to unlock the mysteries of the cosmos, my story serves as a beacon of hope for aspiring scientists everywhere and a testament to the power of dreams and the resilience of the human spirit. With each discovery and every challenge overcome, I am proof that with perseverance and belief in oneself, anything is possible.

My final word to every young African girl out there is this. Embrace your curiosity and passion for science, believe in your abilities, and never let anyone make you feel less of who you are. Your unique perspective and contributions are invaluable in shaping the future of science and technology. Pursue your dreams fearlessly, knowing that you have the power to make a difference in the world. Even if you are afraid, DO IT AFRAID!

**Award**

1. Dundie 'Changing the World' Award 2022

# Amanda A. Sickafoose

" *This too shall pass."*

USA

Planetary Science Institute / SALT

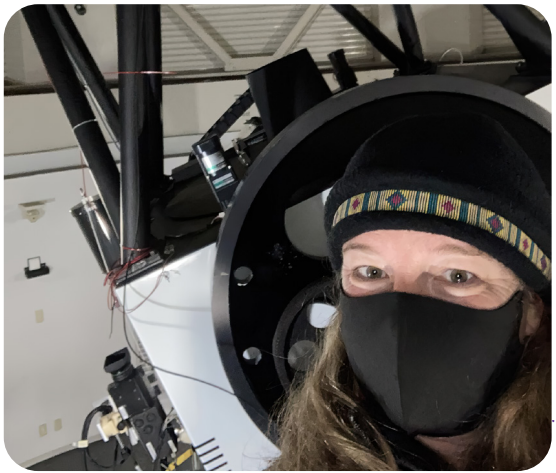

**W**hen I was only six years old and asked what I wanted to be when I grew up, I said " A female astronaut and the first person to visit the planet Uranus". I didn't understand the path to reach that goal. What I understood was that only very smart and persistent people became what I thought was the coolest profession on (or off) the planet. Throughout school, I competed with my older brother and excelled in math and physics – I think now it was because I appreciated both a challenge and not following stereotypes.

I followed in my late uncle's footsteps to attend the U.S. Air Force Academy, continuing to compete and push myself while majoring in Physics and flying glider planes, going through survival training, and even spending a summer with an F-16 squadron at Aviano Air Base, Italy. After two years in the military, I realized that it wasn't the career path for me: I wasn't going to become an astronaut by flying planes because of my eyesight, so I should concentrate on being a scientist. I left the Air Force for the more mundane life of completing a BSc with a double major in Physics and Mathematics, minoring in Astronomy.

The weather at my school in Ohio was not the best, but I learned the basics of observing the sky. I also made great connections by spending summers in programs at Observatories in the western U.S.. I went on to complete an MSc and PhD in Astrophysical, Planetary, and Atmospheric Sciences. When selecting a graduate school, I followed the advice of my mentors to choose a location where I would enjoy living for many years and to pick an interesting topic. That advice has served me well: I went on to work on a variety of topics and to live in very nice places. Over time, I realized that the realities of life as an astronaut were not for me and that I would rather continue to do astronomy research. Little did I know where that would lead!

My thesis dealt with laboratory experiments of levitating lunar dust, to better understand astronauts' observations on the Moon. I also continued to collaborate with the observational astronomers I met during my undergraduate days, becoming part of the discovery of hundreds of objects in the outer Solar System's new "Kuiper Belt". My laboratory and observational work came together in developing instruments that were specialized to observe stellar occultations, which occur when a distant star passes behind a foreground body as observed at a specific time and place on the Earth. That experience led me to building an instrument for a telescope on the summit of Mauna Kea, Hawai'i – what a privilege to spend time in that sacred place.

After becoming a research scientist, I noticed an opportunity to work at the Southern African Large Telescope (SALT). There are not many chances to spend time at 10-m class telescopes, or to live in exciting cities like Cape Town. So, I moved halfway around the world in 2008 and have never looked back. I've gone on to manage a technical team to support existing astronomical instruments and build new ones, to study distant bodies in the Solar System like Pluto and the Centaurs Chiron and Chariklo, and to collaborate with other astronomers on observations of stars, asteroids, and exoplanets. Chasing the shadows of stellar occultations has taken me all around the world, to use telescopes in Australia, Chile, Japan, Namibia, New Zealand, South Africa, and the U.S. More importantly, I've worked with wonderful people as teachers, advisors, colleagues, students, and collaborators.

I will likely never physically visit a planet other than the Earth. However, through telescopes, I have visited more than we even knew existed when I was six years old. There are so many things yet to discover, in so many different areas of research. I am grateful to be a part of the amazing group of people working to advance human knowledge. I never knew that a life in astronomy could be such an adventure!

**Awards**

1. Asteroid 9525 Amandasickafoose

# Betelehem Bilata Woldeyes

**Ethiopia**

**PhD student, Instituto de Astrofísica de Andalucía (IAA-CSIC), Spain**

" ———————— *God is good!"*

My name is Betelehem Bilata Woldeyes, and I am from Ethiopia. I am currently pursuing a PhD in Astronomy and Astrophysics at the Instituto de Astrofísica de Andalucía (IAA-CSIC) in the Extragalactic Astronomy department, located in Granada, Spain. I have been undertaking this program since December 2021.

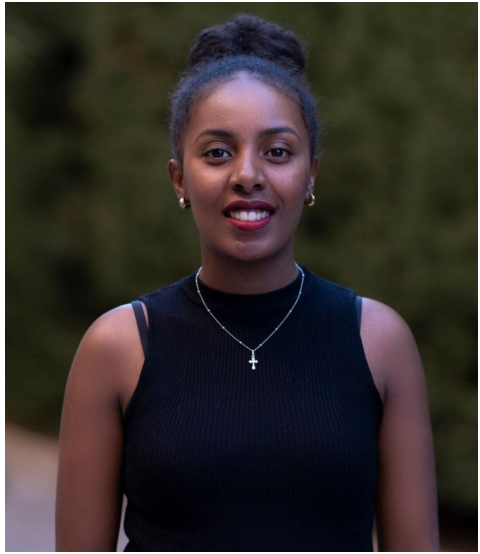

Growing up, I never imagined I would become an astronomer, considering it seemed unlikely to have a career in such a field in my country. However, I've always been fascinated by the universe and eager to explore the cosmos beyond our planet Earth. After completing my Bachelor's degree in Physics, God provided me an opportunity to pursue my Master's degree. It was during this time, that I learned about the Space Science and Geospatial Institute (formerly known as Ethiopian Space Science and Technology Institute), Entoto Observatory and Research Center, and their Master's program in Astronomy and Astrophysics in conjunction with the Addis Ababa University. Recognizing it as a remarkable opportunity to embark on my journey towards becoming an Astronomer, I eagerly joined the program. Since then, my passion for the field has continued to flourish, inspiring me to pursue further research opportunities and pursue a PhD program. Today, I am actively engaged in advancing our understanding of the cosmos through my doctoral studies.

I obtained my Master's degree (MSc) in Astronomy and Astrophysics from Addis Ababa University, Entoto Observatory and Research Center in Ethiopia in 2019. Prior to that, I completed my Bachelor's degree in Physics at Debre Berhan University (DBU) in Ethiopia in 2016. My journey in academia began as a graduate assistant at DBU in August 2016, where I had the opportunity to contribute my skills and knowledge for over a year. Following the completion of my MSc degree, I transitioned into a role as a lecturer in the Physics Department at DBU, delivering various courses to students.

In 2020, I embarked on a new chapter in my career by joining the Space Science and Geospatial Institute, Entoto Observatory and Research Center (SSGI-EORC) as an associate researcher in the Astronomy and Astrophysics Research and Development Department. During my tenure, I actively participated in a range of research projects in the field of Extragalactic Astronomy, focusing particularly on active galactic nuclei (AGN). Additionally, I contributed my expertise to a potential astronomical site selection project.

Passionate about promoting Astronomy and advocating for women's participation in science, I engaged in various outreach programs and initiatives. I participated in organizing national and international conferences, meetings, and astronomical events aimed at raising awareness and fostering interest in Astronomy. From May 2020 - February 2022, I had the honor of serving as the chair of the International Astronomical Union - Office of Astronomy for Education's (IAU-OAE) National Astronomy Education Coordinators (NAEC) of Ethiopia. Additionally, I am actively involved in a collaborative project called 'SciGirls – Empowering Girls in Science through Astronomy'. This project aims to empower girls in secondary schools with diverse socioeconomic backgrounds, ultimately bridging the gender gap in STEM fields through Astronomy.

As a person involved in science, I see science as a journey of exploration, discovery, and enlightenment. From a young age, I was captivated by the mysteries of the natural world, always curious about how things work, and eager to explore the unknown. As I grew older, my passion for science only deepened, and I came to realize the profound significance it holds in shaping our understanding of the universe and improving the quality of human life. From unraveling the mysteries of the cosmos to advancing human civilization by playing a crucial role in driving technological innovation, and broadening our thinking on becoming a critical thinker, science has the power to transform our understanding of the world and shape the course of history.

What I enjoy most about science is its ability to inspire wonder, curiosity, and intellectual growth, as well as its capacity to bring people together in pursuit of a common goal. It is not just about finding answers; it is about asking questions, challenging assumptions, and pushing the boundaries of knowledge which fosters a spirit of collaboration and cooperation, transcending cultural, geographical, and ideological boundaries. As I continue on my own scientific journey, I am excited to see where it will take me and what new wonders I will uncover along the way.

My research interests center around the formation and evolution of galaxies and the properties of the intergalactic medium. My doctoral work focuses on studying the effects of the gravitational interaction between galaxies in gravitationally bound systems and the intergalactic gas evolution and changes due to interactions and its hydrodynamics. In particular, one of the components we are studying is the Intragroup Light (IGL), an extended low-surface-brightness component formed by stars stripped from their host galaxies within galaxy groups, a component formed as a result of a gravitational interaction. We aim to investigate the classification technique of IGL and its correlation with the properties of the host galaxy groups, shedding light on its formation and evolution over cosmic time.

While being part of this journey is undeniably exhilarating and awe-inspiring, it is not without its challenges. Along the way, we may encounter various difficulties that test our strength. One of the challenges I faced in my journey was grappling with persistent self-doubt and a fear of failure. However, as I progressed in my professional journey and met different people in the field, I came to realize that we don't always know everything and become perfect; rather, we learn and progress on the way.

And also, if we don't push ourselves to take the first step and try, we won't even learn if we're capable of doing it. With God's help and embracing this mindset allowed me to overcome my doubts and fears and take the courageous step forward. Moreover, the unwavering support and encouragement from my family, friends, and colleagues played a pivotal role in bolstering my confidence and propelling me forward on my path as a researcher.

In the end, my advice to the people who read my story, especially women, is to become your own biggest cheerleader, as this self-encouragement may not always come from others. Learn to uplift yourself throughout your journey, and simultaneously embrace the opportunity to support, encourage, and inspire others in your community, in particular women.

# Costecia Ifeoma Onah

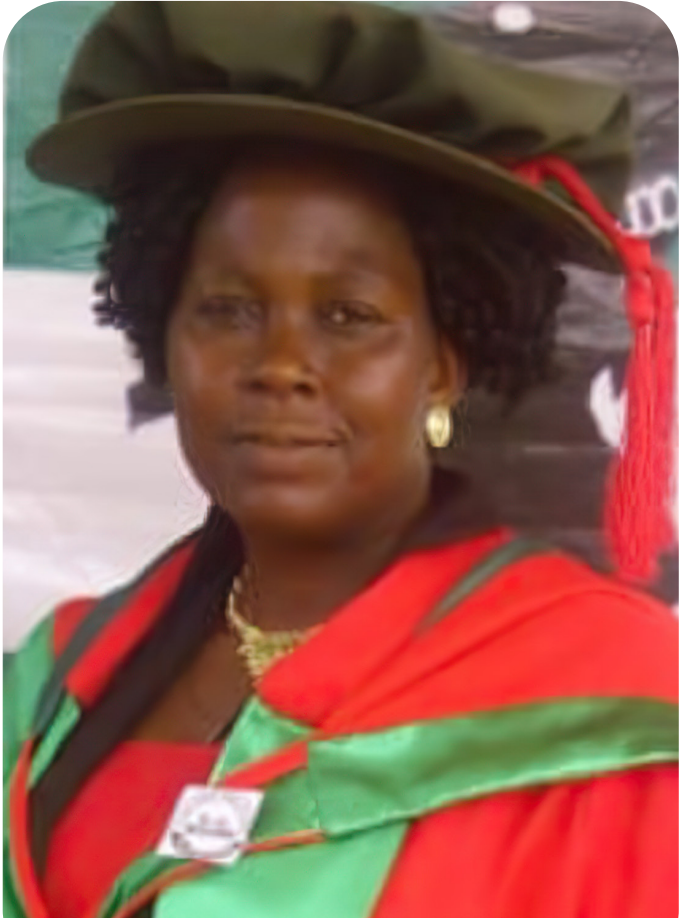

Nigeria

Federal University of Technology, Owerri

> *My biggest dream is to be a Professor of Astronomy and Astrophysics, and travel around the world"*

**M**y Professional background started as a student in the Department of Physics and Astronomy, at the University of Nigeria Nsukka. As a Physics student, I developed much interest in the study of Astronomy and Astrophysics. These gave me much joy since through this study, I was able to know much about the Universe I am staying in.

My research work as an Astronomer is mainly on the Evolution of Radio Source Components and the quasar/galaxy Unification Scheme. I did a lot of research work here and achieved much in my findings which led to many publications. I developed an interest in studying heavenly bodies, Extragalactic objects, and many more. This study brought about my enthusiasm for construction of some instruments that helped me in the investigations of these heavenly bodies and the Extragalactic objects in space. All these made my publications so far. Currently, I am working on cosmic ray emissions, a very interesting topic, by applying data from different Neutron monitoring stations.

In one of my research works, I analytically developed a model to explain the unification of the Extragalactic Radio Sources (EGRSs) via evolution based on observed temporal evolution, relativistic beaming, source orientation, and asymmetries in the EGRSs and also developed a theoretical model that could unambiguously explain the temporal evolution in extragalactic radio sources. I also analytically examined the observed radio properties of different samples of EGRSs sources' structural asymmetry and used a relativistic beaming and source orientation model to explain any observed structural asymmetry in the radio sources.

Most of my works have led me to where I am today as an Astronomer. I have attended many conferences and workshops where I met many powerful scientists in the field of Astronomy. I always advise young ones especially girls to develop an interest in studying sciences because I have tested it and seen that it is good and if I made it, they too will make it.

Along the way, I have had many challenges confronting my motion as an Astronomer including having access to research institutions. This is mostly due to a lack of funds for travel and research visits to other institutions abroad. Therefore, I depended a lot on the Internet to do most of my work.

This field of study has granted me many opportunities. Currently, I am a Senior lecturer in a Federal Institution, handling many students both undergraduate and postgraduates, thereby breeding many graduates at different levels. My biggest dream is to be a Professor of Astronomy and Astrophysics, and plan to travel around the world in the course of this study.

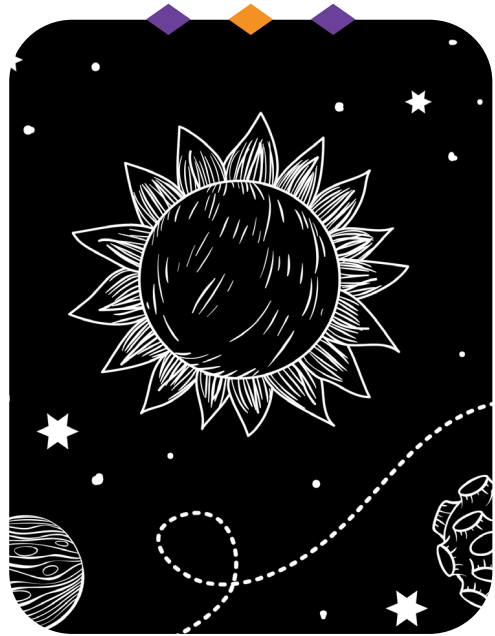

# Tshilengo Vhuthu Miranda

**South Africa**

**Msc student, University of Venda**

*"It only takes a village girl to dream of the stars; your background does not determine your future, but your passion and perseverance do."*

As I consider my experience in the field of astronomy, I'm drawn to sharing the core of my goals, struggles, and passions that have shaped my journey so far. I am currently enrolled in a master's studies in Astronomy at the University of Venda. My research interests are closely linked to radio astronomy, specifically, the study of transient and variable sources through the advanced MeerKAT telescope. I find this area of astronomy fascinating because it provides a special lens through which we can see the dynamic aspect of the universe.

2019 was the beginning of my career as an astronomer when I attended The National Astrophysics and Space Science Programme- North-West University (NASSP-NWU) winter school. There, the vast field of astronomical observations opened up for me, inspiring me to learn more about this subject. This realization inspired me to seek an honours degree and, subsequently, to start a Master's program in Radio Astronomy at the University of Venda. It fills me with pride to be the first officially enrolled master's student studying Astronomy at the University of Venda. Most importantly, I want to thank my supervisors, Professors Patrick Woudt and Eric Maluta, for constantly supporting me.

I see science not only as a profession but as a thorough knowledge of all that exists in our universe. Science has had a huge impact on my life, from the Big Bang theory's explanation of the universe's origins to the countless professions it supports, like astronomy. My advice to young girls and other women is to never doubt their skills because of this conviction. You have what it takes to be a great leader in any subject, but especially in fields like science where women's presence is vital yet still somewhat low.

The practical application of theories to solve cosmic mysteries is the tangible component of science that most interest me. However, I have faced some challenges along the way. Choosing to focus on Astronomy at the University of Venda came with its own set of difficulties, chief among them being the isolation of being the only student in this discipline. Patience and flexibility were qualities that were essential in overcoming the logistical and geographic obstacles to getting help.

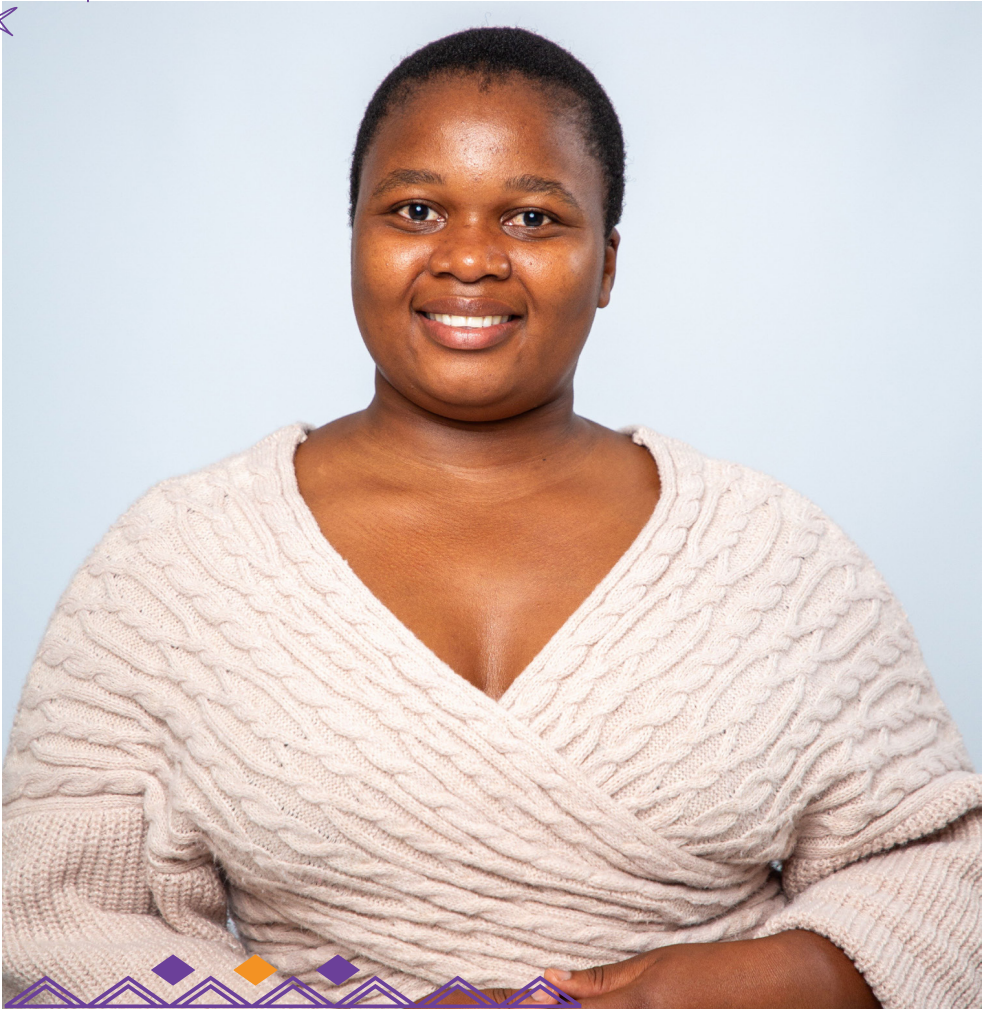

My goals go beyond what I can do personally. To increase public awareness and accessibility to this important field, my ambition is to develop a strong astronomy program at the University of Venda. To spread awareness of the wonders of astronomy worldwide, my future goals are based on both outreach and academics.

By telling my experience, I want to encourage not only a love of science but also a conviction that everyone can dream big and break through obstacles. I'm confident that if we work hard and are determined, we can all have a big influence and open doors for astronomy and other fields to future generations.

# Mayssa El Yazidi

**Tunisia**

**Planetary scientist**, Catholic University of Sacro Cuore

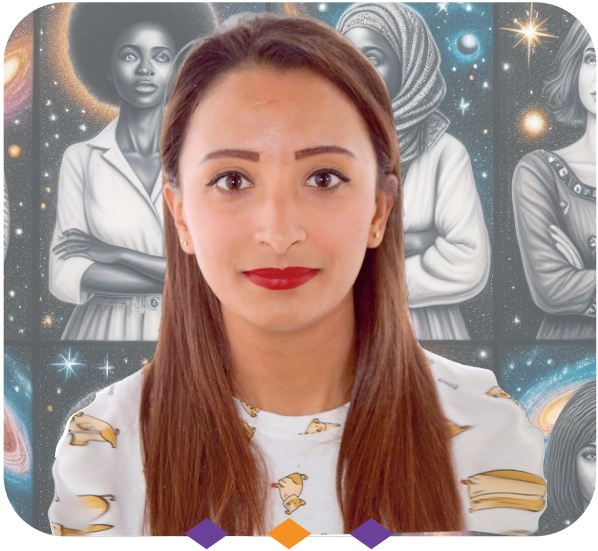

> *Remember to look up at the stars and not down at your feet. Try to make sense of what you see and wonder about what makes the universe exist. Be curious. "* — Stephen Hawking

Since I was a child, I was always attracted by the desire to discover the Universe, and to learn more about the planets and the Moon which is relatively closer to us. This desire drew my path and oriented my interest to study Planetary science. In 2011, when I successfully completed my baccalaureate degree, I wanted to do my university studies in Astronomy or astrophysics, and at that time, I discovered that in the Republic of Tunisia, there were no university courses in Astronomy, and not in closer fields. I decided to study geology but apply it on a planetary scale.

In 2011, I was accepted at the Faculty of Sciences of Tunis, University Tunis El Manar, to study Geology courses, the courses were interesting and very important, but my brain and soul were always connected to the universe, and between the stars. At this moment, came to my mind the idea of studying geology, but not the geology of the Earth, the geology at an extraterrestrial scale, which we call nowadays, "Planetology" or "Planetary Science.

I discussed the idea with one of my university professors, Prof.Dr. Slim Shimi Najet, who warmly welcomed the idea and she accepted to be my supervisor. She believed in me; she supported me and she gave me all the courage I needed to continue my path. Since 2011, Prof.Dr. Slim Shimi Najet and I worked together, and she supervised my Bachelor research project titled "Comparative geology of the celestial bodies of the solar system" in 2014, and continuously my master research project titled "Introduction to the study of the geology of Mars" that I successfully defended in 2016.

- Technically, my university degrees are in geology since planetology is not classed within the Tunisian universities as a university course, but my research project is in Planetary Science. Every single success in my career I sincerely address it to Prof.Dr. Slim Shimi Najet! She is the one who made me the person I am today.

In 2023, I completed my PhD in Sciences, technologies, and Measurements for Space from the University of Padova, with emphasis on geological mapping, and structural and geomorphological analysis for some areas of interest on Mars and Mercury. I became the first female in the Tunisia Republic to hold a thesis in planetary science, particularly the study of the geology and the structural features of Mars and Mercury.

My PhD project was funded by the Fondazione Cassa di Risparmio di Padova e Rovigo Doctoral Scholarship, so I pursued my research at the Centre of Studies and Activities for Space CISAS "G. Colombo", Universities of Padova. I faced some challenges during my PhD at the University of Padova, related to the gender balance, and the presence of female in such a field of science, which is mostly considered a "Masculine field". I also faced some challenges related to my North African origin; some people consider African students as not good enough and do not have enough astronomy background. These challenges gave me in fact more courage to continue my career and be excellent at what I was doing.

I was able to receive one of Italy's most selective scholarship programs, including three years of funding from the Fondazione Cassa di Risparmio di Padova e Rovigo Doctoral Scholarship, reserved for foreign candidates, and one year of funding from the Centre of Studies and Activities for Space CISAS "G. Colombo", to support the definition of Ganymede's observations for the study of the libration. Additionally, I have been invited at ESA- ESTEC to work with HRSC archival Data on the frame of ESA Archival Research Visitor Programme from September to December 2022, in order to increase the scientific return from HRSC instrument and I was the only student from the University of Padova that was accepted for such a program.

My PhD work was concerned with the geological mapping and the study of the tectonic features on the western of the Eminescu (H-09) quadrangle of Mercury. I used available basemaps derived from the NASA MESSENGER Mercury Dual Imaging System (MDIS) images, to produce a 1:3M regional geological map. The preliminary analysis of this work shows an intriguing morphology related to endogenic and exogenic processes on the surface of Mercury, where intensive tectonic and cratering features constitute the main geological events that provided the heterogeneity of terrains. The tectonic events were probably driven by global cooling, however, we found both compressive and tensional tectonic features on the surface.

The work is still in progress. This map will be the first geological product for this region with such a scale, and this will allow us to determine the absolute ages of the units to classify the terrains in chronological order and provide a complete geological and morphological analysis for the selected area. During this work, I was able to propose 38 targets (i.e., Terrains, structures, craters, hollow deposits, and volcanic features) to be covered by the SIMBIO-SYS instrument in order to support the ESA/JAXA BepiColombo mission to Mercury, and contributing to the investigation and the understanding of Mercury.

These studies are new for Tunisian Universities, especially since Planetary sciences is one of domains that is still absent within the Universities in Tunisia. My dream is to continue my research and go back to Tunisia to implement this science inside the Tu-nisian University allowing students to study and explore our universe and other planets.

I feel I have a huge responsibility on the top of my shoulders. I feel responsible for my country, for my university and my people. I aim to continue my research in the exploration of Mars and Mercury, and other planetary bodies. Currently, I'm looking for a Postdoc to pursue my research.

**Awards**

1. Astronomy at home IAU Award

2. International Astronomical Union, 1st-Prize for the best outstanding Online Events, Office for Astronomy Outreach (OAO), International Astronomical Union

# Marian Selorm Sapah

Ghana

University of Ghana, Accra-Ghana

*"Ghana's first trained Cosmochemist"*

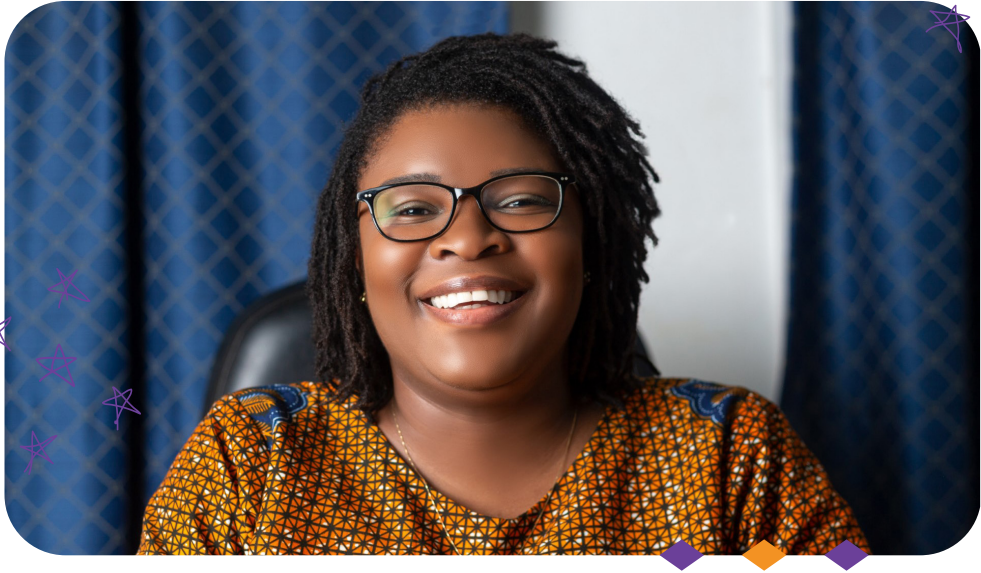

I am Dr. Marian Selorm Sapah a Senior Lecturer at the Department of Earth Science, University of Ghana. I graduated from the University of Ghana, with a BSc honors in Geology and from The Australia National University, with a PhD in Earth Chemistry, with a specialization in Cosmochemistry. This currently makes me Gha-na's only trained Cosmochemist.

Cosmochemistry is an astronomical science that applies the tools and principles of chemistry to better understand the origin, age, formation, and evolution processes of extraterrestrial materials such as meteorites, asteroids, cosmic dust, and other matter in the universe, through laboratory experiments. Cosmochemistry is a fascinating multi-disciplinary field that overlaps with Astronomy through the advancement of space technology that allows for telescopic data, spacecraft instrument data, and laboratory data to complement each other. My main research interests in the field of Cosmochemistry are focused on understanding early Solar System processes through the elemental and isotopic studies of Meteorites.

Since I can remember, I've always been curious and wondered what was beyond the skies, why the universe exists and how it came to be in existence. Even though answers had been difficult to come by for a long time, I did not discard these questions. After my undergraduate degree, I was looking for opportunities in higher education and was offered a six-week internship at the Institute of the Study of the Earth's Interior (ISEI), Okayama University, Misasa Japan. The institute is well known in the field of cosmochemistry and at the time was involved in the Hayabusa sample return mission. While there, I was party to scientific experiments and discussions involved in the handling of the returned samples.

I was truly fascinated and it was the first time I had come across tangible efforts seeking to answer some of the nagging questions I had about our place in the universe. I knew then that I wanted to know more, and this was an opportunity to be involved in something profound and entirely alien to me. I therefore took up a project in cosmochemistry a few months after the internship when the institute offered me a PhD position. Albeit with some reservation since nobody back home in Ghana knew or understood what I was talking about when I told them what I wanted to do for my PhD. I did it anyway, not allowing societal, cultural and professional differences, stereotypes, discrimination and prejudices deter me.

Science to me is a form of creativity that allows me to test my ideas through experimentation, find answers, and use those answers to solve everyday problems. I basically want to understand why things are the way they are or behave the way they do, and Science helps me with that. I believe there is a mystery in all things waiting to be uncovered through Science. Science drives imagination and curiosity which leads to innovation and innovation is crucial for human and sustainable development. This is particularly true in the areas of Planetary and Space Science and Astronomy.

The world has been looking to space for answers for some time now, especially when it comes to Earth's resources, communication, and climate problems. Understanding the Earth's place in the universe and the impact of the universe and its systems on Earth will help us become better stewards of the Earth. It will help us better understand, manage, and preserve the Earth. Other exciting benefits of the study of planets and outer space include the advancement in technology as a result of spin-off technologies from space technology. Typical examples of such spin-off technologies include; GPS, wireless, and LED lighting.

For Africa to be able to take part in global Planetary and Space Science endeavors and its benefits, there is the need to develop human resources in the field. Unfortunately, human resources in Planetary and Space Science and Astronomy in Africa is inadequate and even worse so, for women and girls in the field. Teaching and research in Planetary and Space Science and Astronomy plays an important contributing role in this endeavor, which is where I come in as a teacher and researcher to help teach, mentor, and build capacity in my home country Ghana and on the continent of Africa.

Apart from teaching and scientific research, I also have a special interest in creating awareness, interest and promoting Planetary and Space Science in Ghana and Africa. To this end, I have carried out outreach activities at basic schools in Ghana, given talks at both local and international conferences on the topic, and written articles targeted at basic and high school students as well as the general public. I am also a founding member of the Africa Initiative for Planetary and Space Science (AFIPS), an initiative that seeks to elevate Planetary and Space Science across the African continent, through cutting-edge collaborative research, capacity building and outreach.

My journey so far has brought me to a place where I carry out exciting and impactful scientific research, make an impact through teaching and mentoring, inspire and motivate the younger generation, and meet and collaborate with like-minded people from all over the world. Nothing good, however, comes easy and the road to success is often fraught with challenges. Getting to where I am today, came with focus, hard work, and determination. I have had to deal with cultural and traditional expectations of a girl or woman especially concerning starting a family by a certain age. Working in a male-dominated profession such as academia also comes with its own challenges. However, knowing who I am and what I want for myself has helped me to stay focused on my dreams. There are a few notable things that have helped me navigate my journey over the years that I would like to share here;

**1.** You have to work for the success you want to achieve.

**2.** Invest in mentors. Most times, there is no need for you to reinvent the wheel. People who have already been on the journey can better direct you through yours.

**3.** Create your own opportunities through preparedness.

**4.** Acknowledge and celebrate all your gains, especially the "small gains". Every success is worth celebrating. It serves as motivation.

**5.** Be humble and have a teachable spirit.

**6.** Speak up for your own rights and that of others.

**7.** It is true that you can do anything you put your mind to. It always seems impossible until it's done.

**8.** Challenges are a part of any worthwhile journey.

All that said, being a woman in STEM and particularly in my field of expertise is a privilege I do not take for granted. My journey so far has been worthwhile and I am hopeful for the future of Cosmochemistry and Astronomy in Africa.

**Awards**

1. 2023 West Africa Science Communication Awards

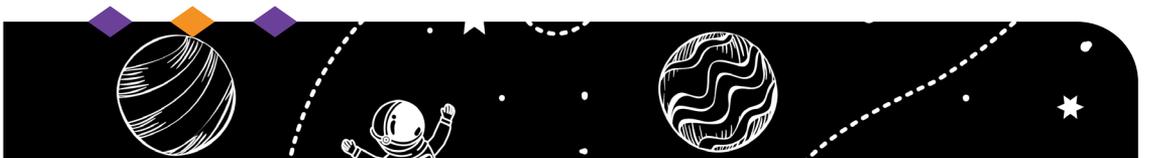

# Redait Basaznew

**Ethiopia**

**MSc student, Space Science and Geospatial Institute**

> *Science knows no boundaries or gender, so let us empower and inspire future generations of girls and women to reach for the stars and leave an indelible mark on the scientific world.*

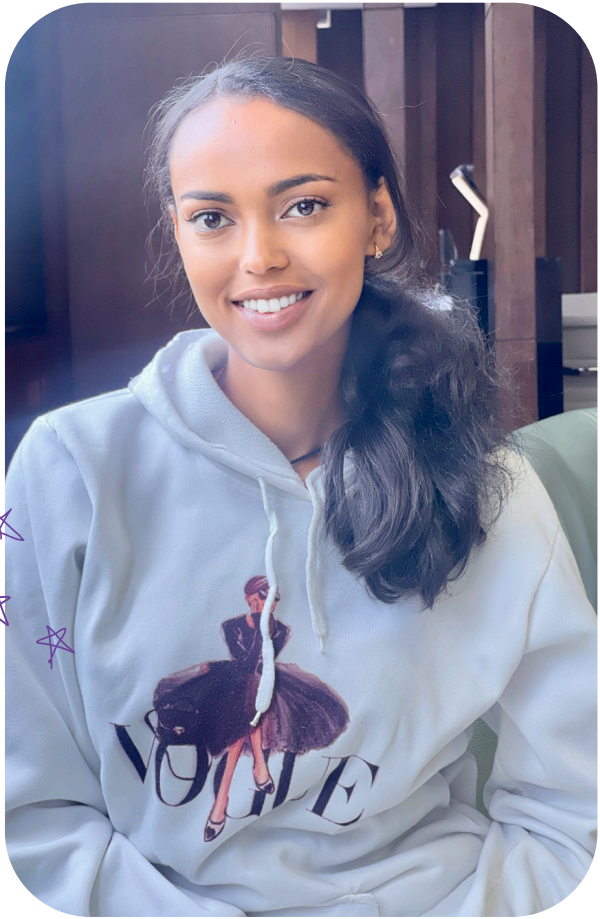

### A Journey Across the Cosmos: Unveiling the Secrets of the Stars

In the vast expanse of the universe, I found my calling as an astronomer specializing in stellar astronomy. From an early age, my fascination with the night sky ignited a spark within me, leading me to pursue scientific education and immerse myself in the wonders of astrophysics. Through years of dedication, I embarked on a celestial journey to unravel the mysteries of the stars, contributing to the ever-growing body of knowledge in the field of astronomy.

As an astronomer, I have been engaged in various research projects and observatory work, exploring the depths of the universe. My research focuses on the study of the eclipsing binary star system, where I investigate the internal stellar structure, orbital evolution, and the stability of mass transfer. Utilizing recent data and sophisticated analysis techniques, I aim to contribute to the understanding of these celestial phenomena and inspire future generations of scientists.

Science, to me, is not just a profession but a way of life. It is the guiding light that illuminates our understanding of the world, provides answers to our most profound questions, and drives innovation and progress. Science fuels my curiosity, encourages critical thinking, and empowers me to make a tangible impact on society. It is the cornerstone of human advancement, and its significance cannot be overstated.

To aspiring scientists, particularly girls, and women, I urge you to embrace your passion for science fearlessly. Believe in your abilities, break through barriers, and let your brilliance shine. Surround yourself with mentors and allies who support and encourage your dreams, for your unique perspective and contributions are invaluable to the scientific community. Be resilient, persevere, and explore the wonders of the universe with unwavering determination.

The path of a scientist is not without its challenges. I faced moments of self-doubt, encountered setbacks in research, and struggled to balance personal and professional responsibilities. However, I overcame these obstacles through perseverance, resilience, and the support of a strong network. Each challenge became an opportunity for growth and learning, strengthening my resolve to push the boundaries of knowledge
.
Throughout my career, I have had the privilege of contributing to groundbreaking research endeavors and fostering meaningful connections with fellow scientists. From publishing influential papers to presenting at conferences, each achievement has been a stepping stone towards my ultimate goals. Beyond my professional accomplishments, I have also advocated for greater inclusion and diversity in the field, aiming to create a more equitable scientific community.

My biggest dream is to witness humanity's first steps on a distant planet, to gaze upon a world never before touched by human hands. I yearn for the day when we unravel the secrets of interstellar travel, enabling us to explore the depths of the universe beyond our cosmic neighborhood. I aspire to mentor and inspire the next generation of scientists, fostering a diverse and inclusive scientific community. Additionally, I aim to collaborate with international space agencies to further our understanding of the cosmos and pave the way for future space exploration endeavors.

Looking ahead, I plan to expand my research, delve deeper into the mysteries of the universe, and contribute to scientific breakthroughs. As I continue to explore the frontiers of knowledge, I remain inspired by the wonders of science and the boundless possibilities that lie ahead. May my story ignite the spark of curiosity within others, propelling them toward their own astronomical odyssey. Let us embark on this incredible journey together as we traverse the cosmos and unlock the secrets of the universe.

**Awards**

1. The 2022 Ethiopian Physical Society North America (EPS-NA) Scholarship award at a graduate student level

2. The 2020 Ethiopian Physical Society North America (EPS-NA) Scholarship award at an undergraduate student level

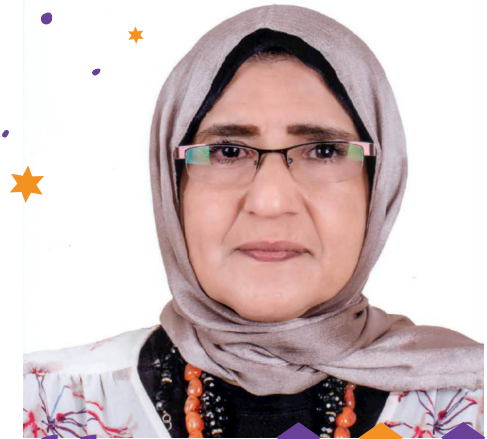

# Somaya Saad

**Egypt**

**Astronomy Dept.**, **National Research Institute of Astronomy and Geophysics**, **Helwan**, **Cairo**, **Egypt**.

*The sky is my home and life is wonderful with the stars. You Can Make the World Better Place."*

**M**y life was full of challenges and my dreams were always as far away as the stars, I reached almost all of them with trust, faith in God, and hard, continuous work. My passion for science and mathematics started at an early age where hands-on learning and interactive experiences in science classes played a major role in building my investigative personality. Physics and mathematics were among my favorite classes during high school.

The big leap for my dreams was to join the faculty of Science at Cairo University, where I studied various courses in physics and astronomy. After obtaining my Bachelor of Science in Astronomy, I was fortunate to work at the National Research Institute in Astronomy and Geophysics (NRIAG), one of the oldest astronomical institutes in the Middle East and North Africa. I started my Master's studies and became familiar with observational work (photography, photoelectric photometry, and spectroscopy) using the 2-meter telescope at Kottamia Observatory as well as skills in spectral and photometric reduction and analysis.

My master's study was on the classification of the peculiar Be stars (class of the variable stars). We were able to develop a method for classifying them avoiding the influence of the surrounding envelope which often misleads in their spectral classification. The study produced a Catalogue using the new classification of Be stars, and study of the line profile spectral variations of two B stars Leo and 17 Tau, which was published in the international journal of Astrophysics Space Science and this gave me confidence.

More than a big step towards my PhD study done in collaboration with senior professors Béla Szidle and Margit Paparo (Konkoly Observatory, the center of the variable stars in Europe), then I left to Budapest, Hungary to work in close touch with them. It was my first trip abroad and that meant a lot to me. Well, my first plane ride, it was time to be responsible and decide everything myself.

I couldn't forget the wonderful and fruitful discussion with Prof's Bella and Margit, it was a great time. The study dealt with the analysis of a huge amount of accumulated data ~ 100-year observations of the glob-ular cluster M15. We studied the double mode RR Lyrae stars within it. These stars are pulsating stars that have two modes: the fundamental and the first overtone. It was another step towards success as we reached completely new conclusions that were published in high-rank journals. Paparo taught me a lot of skills like how to write a paper and this was my first time to use Linux. Also, on weekends she would take me to different museums, parks, and many other beautiful places to introduce me to Hungarian culture. I cannot forget the role of professor Paparo in my lifestyle and scientific career.

After my PhD, I got two grants for a Postdoctoral position. The first was a KOSEF 'Korea Science and Engineering Foundation" for a year at the Department of Astronomy, Seoul National University, South Korea, where I met the great Professor Sang-Gak Lee. She was very kind and I learned a lot. We redetermined the distances to the globular clusters using the main-sequence technique with the high-accuracy parallax of HIPPARCOS astrometric mission. I used IRAF to reduce the observations of the globular clusters and also had ample observing opportunities with the 2-meter BOAO telescope and the 60-cm SOAO telescope.

The other postdoctoral position was two years working with Professor Jiří Kubat at the Ondrejov Observatory (B and Be stars center in Europe) in the physics of hot stars where I analyzed the spectra of binary systems containing a B star. It was also a very productive time and had ample spectroscopic observing opportunities with the 2- meter Onderjov telescope. I was lucky to meet Professor Petr Hadrava who was very kind and taught me how to use his programs FOTEL and KOREL for analyzing, modeling the light and spectral curves and the disentangling of the binary and multiple stars spectra. It was a wonderful stage in the development of my career, with us we published many excellent publications.

In 2004 I returned to my home institute (NRIAG). I assumed responsibility for supervising several masters and doctoral theses. It was one of the most enjoyable periods, as I was able to establish a scientific direction and work through research teams specialized in the field of stellar physics. I am very happy with these scientific teams that, through many observations of short-period eclipsing binaries they were able to discover the variation of some of the stars in the field, and then these systems were subjected to detailed analysis and study.

By working as a scientific team, we were able to collaborate in many scientific projects through international cooperation with Hungary, the Czech Republic, and Bulgaria, in addition to contributing to the establishment of the first Center for Scientific Excellence in Astronomy and Space Science to develop the work of the Kottamia Observatory through the fund of the Egyptian authority STDF Science Technology Development Foundation at the Academy of Scientific Research and Technology ASRT.

Then I held many positions, through which I contributed to developing scientific plans and policies for the Astronomy Department and the telescope at the Kottamia Astronomical Observatory, where I worked for a period as head of the stellar laboratory, deputy and head of the Astronomy Department. This gave me a lot of opportunities to contribute to institutional building in the field of astronomy, providing opportunities to conduct research and creating an encouraging environment for scientific research and capacity building in the Astronomy Department.

In 2005, I became a member of the IAU. This gives me the opportunity to work on the international level, and know about many grants. I became coordinator of the IAU offices of Astronomy Outreach, and Educa-tion. Another duty that I paid much attention to was simplifying science, astronomy education, and outreach for early stages where we found a lack of astronomy concepts in the school curricula.

With a wonderful team of young astronomers, we play a good role in our community to spread the basics and developments of astronomy and space science, through seminars, training courses, and workshops. Also, it was great to collaborate with the Network of Astronomy School Education (NASE) to train teachers on skills of how to teach astronomy in their classrooms in an interactive way, and even online.

During 2018-2021, I served as President of the Egyptian National Committee for Astronomy, which works in cooperation with the International Astronomical Union and I was one of the local organizers of the 40th ISYA which was held in Egypt in 2018. I met the fantastic professor Itziar Aretxaga, Director of ISYA who paid much attention and effort to teaching astronomy and giving hands to young astronomers.

At the beginning of 2022, I contributed to the establishment of the OAE-Egypt which was hosted by NRIAG, which is one of five centers in the world and operating under the supervision of the IAU-OAE. The role of OAE-Egypt expanded to the regional level and Arabic-speaking countries, and in order to be more effective a group of Arab scientists was created to communicate with each other and cooperate in spreading and teaching astronomy and contributing to improving the school curricula in our Arab world countries.

Recently, I was fortunate to be part of the African Astronomical Society AfAS and the African Network of Women in Astronomy (AfNWA). We are working together in the development of our continent and helping women and girls have more rights and visibility by offering them scholarships, training courses, supervisors, and awards to encourage more outstanding work. From my long experience, my main advice to young astronomers is to just believe in themselves, work hard to be qualified enough, and be good candidates for more opportunities.

Among the important role models, who have greatly influenced my professional life, are Professor Lotfia El-Nadi, founder of the LASER Institute at Cairo University, and Professor Sultana Nahar, a specialist in nuclear physics at Ohio University in the United States of America, and recently, while working with AfNWA, I met other wonderful professors: Mirjana Povic, Vanessa McBride, and the great late professor Carolina Odman. I knew from them the meaning of volunteering work to serve communities, they almost give all their time and attention without limits.

What I can say now, after working at the local, regional, and international levels, and working in close contact with many different societies and cultures, is that I have fully realized and learned to appreciate and achieve the concepts of inclusion, equality, and diversity, which will help in integrating the full potential of society and individuals together, which in turn will contribute to the integration of visions, beliefs, values, prosperity, and development of societies.

**Awards**

1. Award of community service and Science simplifying, the Scientific Forum of NRIAG in 2020.

2. Award of Professor Mahmoud Khairy Ali for Scientific Production, the Academy of Scientific Research and Technology, 2013.

3. Award of Professor Mahmoud Khairy Ali for Scientific Production, the Academy of Scientific Research and Technology, 2006.

# Hannah Worters

**UK**

**South African Astronomical Observatory**

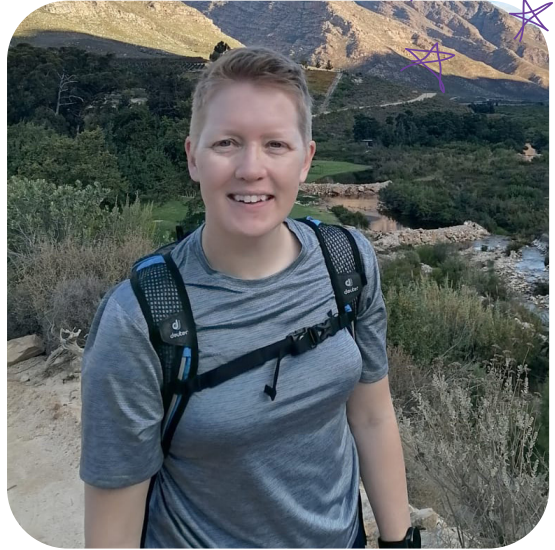

" *We all form a piece of the puzzle.*"

As a child, I thought that everything was known. All we newcomers had to do was absorb the world's wisdom from our elders. I asked a lot of questions, and gradually it became apparent that this wasn't quite the case; there was still so much that humanity had to learn about life, the Universe, and everything.

Have you ever looked at the moon through binoculars? It's mesmerising. You can climb mountains and ski into craters; cross vast, waterless seas and skip along the terminator. It was this that really drew me into astronomy. Before that, I'd had very little connection with space. At school, we were asked how long we thought it would take to travel to the moon. I had no idea, but all the boys had an answer. None of them was right, but that didn't matter: they had the confidence that suggested that this was their realm, not mine.

Wiring a plug-in science class was when I first realised that I was just as good as the boys, and that education was truly useful: not just for school but for everyday life. Soon enough I was fixing things around the house and designing and building things out of wood. Both of my parents seemed capable of absolutely anything, so this seemed pretty normal. There were comments that I wasn't very "ladylike", but no complaints about how useful I was.

I read more and more about astronomy and space travel. I lay outside at night and looked up at the stars, visited by foxes and hedgehogs while I compared the sky with the constellations mapped on a planisphere. My school wasn't used to teenagers wanting to be astronauts. The careers' advisor gave me a leaflet on electrical engineering and sent me on my way. Aged 15, I got a job in a camping shop and saved my wages to go to Space School, where I met people with the same curiosity and aspirations. At this point, I had to admit that astronautics was more of a dream than a goal for most of us, but there were related paths I could tread. This directed me towards a degree in Physics and Astronomy. I had to take a gap year after school – to fill a gap in my maths education – and I used that year for other things too: I restored an old VW Beetle, went backpacking around Eastern Europe, and joined an expedition to Arctic Norway, ice climbing and camping on glaciers (that part-time job paid off).

As an undergraduate, I was lucky enough to earn a placement to spend a year at an observatory in the Canary Islands. My wanderings had convinced the selection panel that I could cope alone, away from my lecturers and classmates, and I never looked back. Since first seeing photos of lightning strikes over Kitt Peak observatory (google it!) I knew that I wanted to work in one of those tin cans on a remote mountainside, and suddenly here I was. I couldn't believe my luck! But the excitement was mixed with quite some trepidation. I was thrown right into the deep end, expecting to train professional astronomers to use the telescopes that I was seeing for the first time. I had so much to learn - I observed with as many telescopes as I could and gatecrashed as much instrument maintenance and commissioning as possible that year. This paved the way for my PhD scholarship, where feedback on my application included the terms "unusual" and "special case", and not for the first time. I loved the path less travelled.

I chose my PhD placement based on the fact that the institute was a partner in the Southern African Large Telescope. I had done a project on liquid mirror telescopes and been captivated - the first examples in the early 1900s had encountered disturbances from passing carts, and the movement of livestock in a neighbouring barn! The fixed altitude axis of SALT had me really excited that it could have had a liquid mirror. I joined at the time when SALT had just been built and was being handed over to the commissioning team, of which I gleefully became a part. SALT has both an operator and an astronomer on duty each night, and I merged the two roles to become the first (the only) "operonomer", helping to characterise the First Light instruments and troubleshoot the integration of all the new systems.

My PhD research included spectroscopic observations of key stages in stellar evolution, using telescopes in South Africa, Chile, and Hawai'i. In other words, chasing rainbows in the light from a born-again red giant star, a catalogue of flare stars, and a recurrent nova: a nice variety of objects that go whizz, bang, bump in the night! I got to operate a variety of large telescopes at different observatories around the world and I loved the hands-on side of things. I became proficient in using – and training others to use – all the telescopes and instruments at the South African Astronomical Observatory, and as a postdoc I became the first Resident Astronomer at the Karoo base in Sutherland. Thus continued my life on the edge (of a cliff!), at the gateway to the Universe.

Working in the Telescope Operations division is a little different from the path most might take in astronomy, and it has been quite an adventure! I commissioned the first new telescope owned solely by South Africa for four decades. We ran a nationwide competition in which a learner from North West named it "Lesedi" (meaning light, or enlightenment) and gave a speech at the inauguration ceremony. When we have to renew the reflective coating on our telescope mirrors, I'm there in the lab. When the roof needs fixing, I'm up there in my harness. If an astronomer has a problem in the middle of the night, they phone me and I talk them through it. I've worked on the commissioning of a number of new instruments, and every time it's new and exciting. I have seen my shadow cast by the light of Venus, and marvelled at a moonbow. I've had a picnic on the edge of a volcano, watching lava flow into the sea by the light of the full moon. I've seen the exquisite detail of the rings of Saturn and the moons of Jupiter through a 1-metre telescope. I've visited six continents and met some inspirational characters along the way. I've encountered snakes, dassies, eagles, hares, springbok, rabbits, frogs, scorpions, porcupines...and rainbows. Let's not forget the rainbows.

# Victória Da Graça Gilberto Samboco

**Mozambique**

**PhD student**, Rhodes University, South Africa.

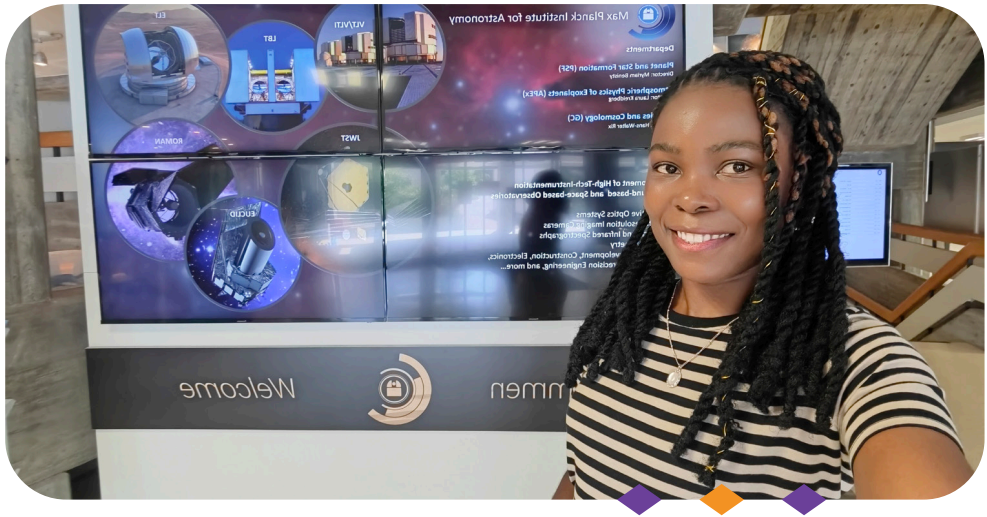

> *No matter what people say always follow your dreams."*

I am Victória da Graça Gilberto Samboco, a 26-year-old Mozambican woman from a simple family. Born in Maputo, the capital of Mozambique, I spent my childhood in Liberdade, a neighbourhood in Matola, the second province of Maputo. Despite the simplicity of our lives, my family instilled in me a deep love for learning.

In those early years, I did not have the opportunity to attend preschool, but my mother conducted home classes for me. Armed with a pencil, exercise books, and coloured pencils, she fostered an environment that fueled my curiosity. As a result, by age 4, I could already write, count, and read with proficiency.

At the age of 5, my parents took me to my grandmother's house in a rural region where life revolved around farming. The lack of electricity in this area meant that the night sky was an enchanting spectacle. This experience marked the beginning of my fascination with the sky, even though the thought of becoming an astronomer had not yet crossed my young mind. Returning home to Matola at the age of 7, I found myself in grade 3. Despite the contrast between the rural and urban environments, the night sky stood out as notably different. The absence of illumination in rural areas enhanced the beauty of the night sky, a stark difference from the luminosity in the region where I lived.

As I grew older, my interest in the sky increased. Scientific fiction movies became a source of delight, and I found myself frequently staring at the night sky in wonder. Excelling in all subjects throughout my academic journey, especially in challenging areas like math and physics, I began to consider a future in science.

The decision to pursue a scientific path was not without challenges. As I approached grade 11, a teacher from my neighbourhood intimidated me, saying that science was too difficult and that I would likely fail. This advice intimidated me initially but was fueled by an inner determination. I chose to follow my heart and enrolled in the science course. My resolve paid off, as I never failed a class or subject.

Upon completing high school, the time came to apply for university. Driven by a desire to become an engineer or scientist, I applied for admission exams in Civil Engineering and Meteorology, with the latter being my second choice. Despite my initial disappointment, I embraced meteorology after being admitted to Eduardo Mondlane University.

University life brought its own set of challenges. Questions and comments about my dedication to studying were constant, with people suggesting I should get married or focus on homemaking. These remarks didn't deter me. My commitment to learning was unwavering, and I found solace in the idea that, if possible, I would like to be paid to study Lol.

The turning point in my journey occurred during my second year of BSc in Meteorology in 2017 when I took an introductory Astronomy course. The class, the passion in the lecturer's eyes, and the fascinating aspects of the field, including the Square Kilometer Array (SKA), captivated me. In 2019, I volunteered for an outreach group led by the same professor, participating in night sky observations with telescopes and delivering speeches at schools.

Subsequent years saw my involvement in projects such as DARA big data and the DOPPLER project, furthering my skills in programming and radioastronomy. In 2021, I applied for DARA basic training, solidifying my interest in radioastronomy. This journey culminated in my admission to Rhodes University for a Master's by research project in Astrophysics. My project focused on developing SolarKAT, a solar imaging pipeline for solar interference mitigation in MeerKAT, under the guidance of Prof. Ian Heywood from Oxford University and Prof. Oleg Smirnov from RATT. Remarkably, I completed my master's thesis within 1 year and 8 months and achieved a distinction.

Simultaneously, I assumed the role of vice president in the Mozambican Astronomical Society. The joy of this achievement was heightened as it coincided with the launch event for the society. Always drawn to solving problems, my passion for science led me to address challenges such as Radio Frequency Interference (RFI) in the MeerKAT telescope.

I advise others, especially girls and women, to defy gender-based judgments and follow their dreams. I find immense joy in the continuous learning that science offers, and it is this joy that leads me forward. I am a PhD student at Rhodes University, working on "Pipelines for mining image-plane transients with MeerKAT." Collaborating with Prof. Oleg Smirnov and Dr. Cherry Ng, the goal is to develop fast imaging pipelines for transient search, potentially applied to technosignature search with MeerKAT.

My plans involve self-development to influence other girls and women in astronomy positively. I aspire to contribute to the development of astronomy in my country, Mozambique. While these dreams might seem distant, we have already taken a significant step by establishing the Mozambican Astronomical Society. The idea of solving problems and the innate joy I find in studying have always propelled me forward. I believe in science's importance because of its invaluable contributions to society – from groundbreaking discoveries to technological advancements and overall development. Scientists play a crucial role in solving real-world problems, and that's precisely what I aim to contribute.

Reflecting on my journey, my most cherished achievement is obtaining a master's with distinction. The joy of this accomplishment was magnified as it coincided with the launch event for the Mozambican Astronomical Society, a testament to the alignment of my personal and professional milestones.

As I continue my academic journey as a PhD student, I revel in the privilege of learning something new every day while working on "Pipelines for mining image-plane transients with MeerKAT." Collaborating with esteemed mentors like Prof. Oleg Smirnov and Dr. Cherry Ng, our goal is to develop efficient imaging pipelines for transient searches, potentially expanding the scope to technosignature searches with MeerKAT.

Looking to the future, my aspirations include influencing other girls and women positively and fostering their interest in astronomy. I also aim to contribute to the development of astronomy in Mozambique, shaping it into a prominent field in our country. While these ambitions may seem ambitious, the establishment of the Mozambican Astronomical Society serves as a testament to the collective dedication towards this shared dream.

In conclusion, my journey from a simple family in Mozambique to pursuing a PhD in Astrophysics at Rhodes University has been marked by determination, passion for learning, and a commitment to overcoming challenges. From the early days of improvised home classes to navigating through stereotypes and discouragements, each step has contributed to the person I am today.

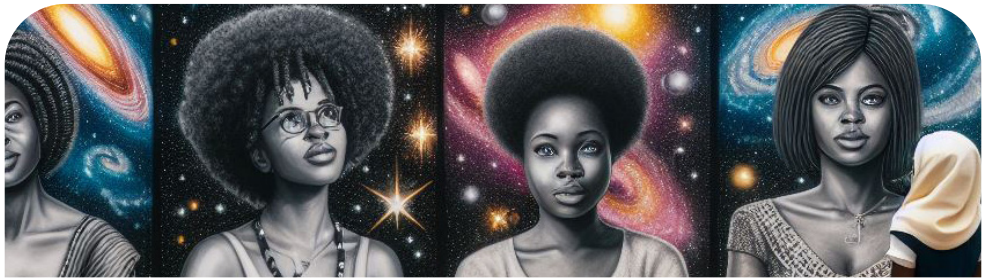

# Raïssa G. Ninon Amoussou

**Benin**

**Université Nationale des Sciences, Technologies, Ingénierie et Mathématiques (UNSTIM) d'Abomey/ Benin**

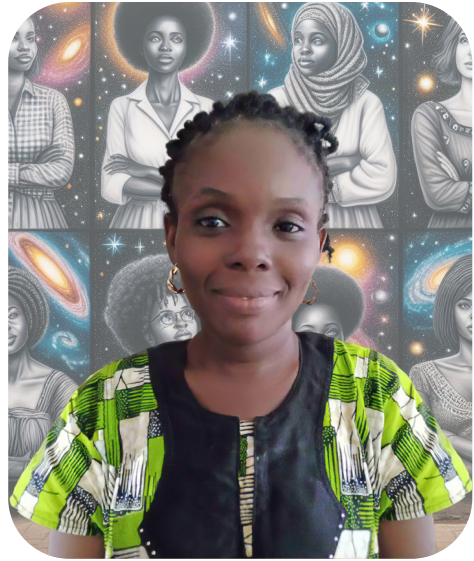

## " *Self-confidence - Trust in God - Perseverance"*

I obtained my CEP in 2002, and my BEPC in 2006. By wanting to choose the series in the second form class, I understood that science elevates minds and contributes to the development of technology and countries as well as the exploration of the universe. Thanks to science, people's living and working conditions are improved. Also, I appreciate rigor and precision in science. This is what motivated me to study science.

Thus, I obtained my Scientific Baccalaureate C series in 2009. After 3 years of study at university, I obtained a Bachelor's degree in Physics and Chemistry in 2012. And since 2014, I have been teaching Physics Chemistry in colleges in my country. I obtained my Master's degree in Theoretical Physics in 2019. And my passion for science and my curiosity led me to publish two articles in indexed scientific journals.

My research focused on the study of cosmological models in the f(T) theory of gravity and in Rastall's theory of gravity. Following a thesis defense on this work, I was promoted to the rank of Doctor in Theoretical Physics option Cosmology and Gravitation in December 2023; which makes me one of the rare women in this field in Benin.

Today, to reduce social inequalities in science, it is necessary for other women and girls to join the scientific community by standing out for work well done in science. Dear sisters, it is important to choose scientific fields and pursue a career in science for the betterment of society.

The biggest challenge is how to combine professional life, family life and research but with perseverance and remaining focused on these objectives, we achieve the result.

I hope to contribute to the development of Africa through my work; To be a reference in Cosmology in Africa and in the world to support young girls and women in science by being a source of inspiration for them.

# Feven Markos Hunde

**Ethiopia**

**Center for Theoretical Physics of the Polish Academy of Sciences**, Poland

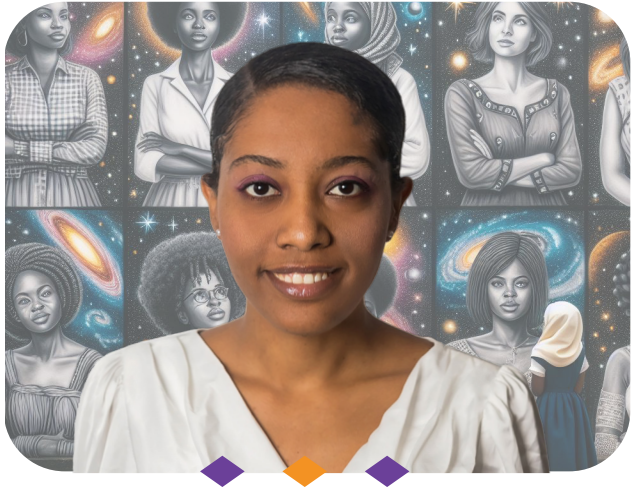

*"As African women, let us stand still and not give up. Together, we can create a beautiful future for our daughters to explore the universe."*

I grew up in Halaba Kulito, a small town in the southern part of Ethiopia. I grew up in a wonderful warm home filled with love and people. My family comes from a working-class background, and none of them had the opportunity to attend college. That's why my parents always stressed the importance of education for my siblings and me.

Since primary school, I started to be fond of stars and started staring at the moon every night. I enjoyed many aspects of school and was interested in every subject, but I was most drawn to physics and mathematics. I was inspired by reading the biography of Albert Einstein in my textbooks. As a child, I dreamed of becoming an astronomer after hearing about the late Ethiopian astronomer, Dr. Legesse Wetro. Starting from grade six, I started working and studying hard to fulfill this dream. I became obsessed with thinking about space and wondering about the magic up there. I developed a habit of reading at a young age, enjoying fiction, poetry, and anything else that came my way.

During my high school years, I relocated to Addis Ababa, the capital city. My enthusiasm for physics flourished as I received encouragement and support from my teachers. This also opened up numerous opportunities for me, including participation in stargazing and other outreach activities organized by the Ethiopian Space Science Society. Even though my family always encouraged me to follow my passion for computational science after high school, I didn't get much support from anyone in my immediate circle. I was under a lot of pressure to pursue a different career route, but I chose to pursue my passion. Since there was no Astronomy department at the undergraduate level in my country at that time, I chose Electrical and Computer Engineering and enrolled at Addis Ababa University. It wasn't about space, but I knew it would give me the knowledge I needed for my dream job. It provided me with valuable experience and knowledge in both physics and mathematics, which helped broaden my perspective for future scientific studies.

After completing my BSc in Electrical and Computer Engineering in July 2018, I was awarded a full scholarship to join the Ethiopian Space Science and Technology Institute in October 2018. This allowed me to pursue my dream field, starting a Master's degree program in Astronomy and Astrophysics. There were only limited opportunities, and few students joined Astronomy and other space science fields. I felt so lucky to get the chance. In my final year, I became increasingly passionate about astronomy and focused my research on white dwarf neutron star binaries. However, as a master's student interested in cosmology, it was challenging to find supervisors from outside the country.

At ESSTI, I had the opportunity to delve deep into astrophysics while pursuing my passion. Additionally, I was able to participate in outreach programs aimed at training high school students, attend continental and international conferences and workshops, connect with experts in the field, and build valuable connections. This experience enabled me to form a group of women interested in space and participate in the Space Governance Innovation Contest, where our group finished as one of the top three.

After finishing my master's study in November 2020, I knew that I wanted to do a PhD. However, it was not easy to get accepted to a PhD program at that time. The COVID-19 outbreak caused delays, making it difficult to pursue my interest in traveling abroad for my PhD. However, I remained active in the field of science by participating in various virtual summer schools and conferences and presenting my work. Moreover, I had the opportunity to supervise undergraduate students conducting research during their internship through the Society for Space Education Research and Development (SSERD). In November 2021, I ended up at the Center for Theoretical Physics of the Polish Academy of Sciences ready to search for the answers to nagging questions such as "How do dark matter halos behave?" and "How does the large-scale structure of the universe impact small-scale structures?" and others.

Transitioning to a new country with its own culture and weather, while also learning to be independent, presented significant challenges. Being a woman of color in the field of astronomy without easy access to others who shared my experiences added an extra layer of difficulty. However, the support from my mentor at Supernova Foundation as well as my PhD supervisors and colleagues helped make the transition much smoother. Currently, I'm doing my PhD. studying the mysterious dark matter using cosmological simulations on a small scale. The cosmic web is the large-scale structure where galaxies and dark matter halos reside, observed as a filamentary network of structures in galaxy surveys and simulations. It encompasses four environments: nodes, filaments, walls, and voids. My research focuses on understanding how these environments affect dark matter halos on small scales and determining which environment best serves to test different dark matter models such as CDM and WDM.

Astronomy is like a big adventure for me. What I love most is the feeling of wonder. Every new thing I learn is like discovering a hidden treasure. It's not just a job; it's like going on a journey to understand the big mysteries of our universe. There were tough times, like when people thought girls couldn't do science. But I didn't give up. I have my family, teachers, and colleagues who help me with my passion. Surrounding ourselves with encouraging people – particularly other women facing similar obstacles – fortifies our strength.

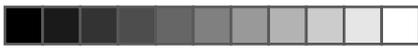

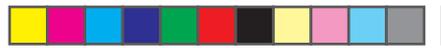

After finishing my PhD, I intend to continue my research and pursue a career in academia. However, my greatest dream is to see African women explore space beyond our planet. Creating a better environment for their daughters so that they do not experience the same challenges they did. Teach other Ethiopian girls that they, too, can reach for the stars. My story evolves during calm nights of study or exciting times in my research—the story of an Ethiopian dreamer who becomes a cosmic explorer. As a young African researcher, my goal is not just to gaze at the stars, but also to encourage other Ethiopian girls to look up, wonder, and dream big.

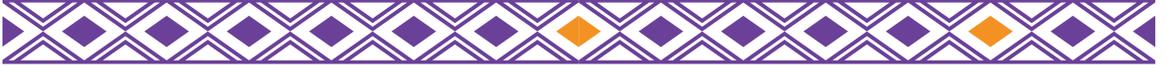

# Thobekile Sandra Ngwane

**Zimbabwe**

**MSc student, University of Cape Town and the South African Astronomical Observatory South Africa**

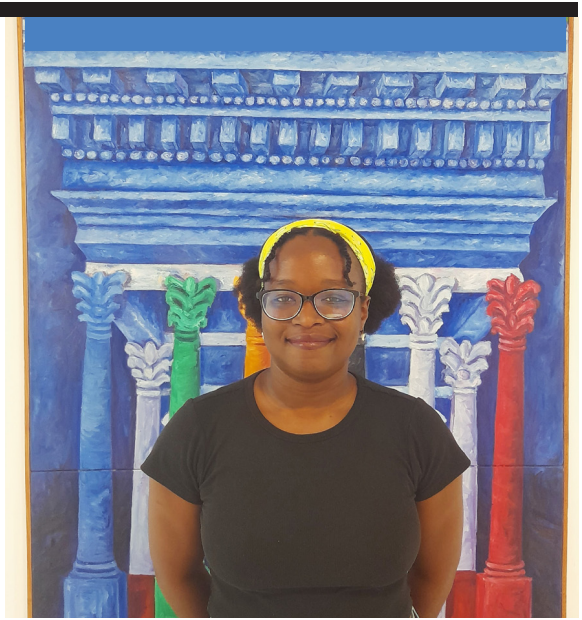

*I stand, on the sacrifices, of a million women before me, thinking, what can i do to make this mountain taller, so, the women after me can see farther"- Rupi Kaur*

y name is Thobekile, and I was born in Harare, Zimbabwe. My passion lies in astrophysics and astronomy, where I'm captivated by how the laws of physics help us understand the origins of our Solar system. I'm currently pursuing my master's degree in Astronomy, focusing my research on Near-Earth Asteroids (NEAs) and instrumentation.

Using the South African Astronomical Observatory's 1-meter telescope, Lesedi, I do rapid follow-up observations on newly discovered Near-Earth Asteroids (NEAs). I study small asteroids, a demographic that has been understudied and these rapidly fade as they move away from Earth. I have also had the opportunity to intern with the Office of Astronomy for Development (OAD), where I worked on assessing the impact of OAD-funded projects from a decade ago.

I should point out that my birthday actually falls on June 30, which is known as "Asteroid Day." On this day, we raise awareness of the important role asteroids have played in forming our solar system as well as the necessity of exercising caution when monitoring, identifying, and minimizing any possible impacts. We can create technology to identify and divert possible asteroid impacts by researching asteroids. We can monitor asteroids that might sometime in the future collide with Earth.

My interest in asteroids began during my undergraduate studies at the National University of Science and Technology in Bulawayo, Zimbabwe. My introduction to the asteroid search campaign by PACS e-Lab sparked my passion for working with asteroids and using specialized software to identify them. I lead the university's Citizen Scientist Asteroid Search Campaign team, and we've made two exciting provisional asteroid discoveries. I enjoyed being a part of that program and it was a great way to set the foundation for my future studies.

My path was not without challenges, particularly given the lack of promising prospects for an Astronomy career in Zimbabwe at the time. However, I persevered and secured a Mastercard Foundation Scholarship to further my studies on asteroids.

To all the young girls dreaming of STEM careers, my advice echoes a simple truth: never surrender your dreams. Never give up on that vision that you have for yourself. As someone who has had a couple of "gap years" due to financial constraints. I understand that it is hard to keep that flame inside you lit, but you have to try your best to keep it going and push through whatever hurdle that may come your way.

I struggled a lot with putting a timeline to my achievements. At times I felt like I had failed when I wasn't where I thought I would be. But life has no formula, no one template that we should all follow. I learned to appreciate that everyone's path is unique. By embracing my individual journey and staying open to new experiences, I was able to find opportunities that aligned with my interests and propel me forward.

Therefore, I urge you to be vigilant and seize every opportunity that aligns with your interests. Keeping your eyes open for chances to further your passion will not only enrich your journey but also open doors to unforeseen possibilities. Stay resilient, stay inspired, and remember that your unique path is part of the extraordinary story of success in the world of STEM.

Science is important to me. It serves as a powerful lens through which I can explore the complex world we live in. The process of scientific inquiry allows me to satisfy my curiosity, providing answers to questions that have fueled my fascination since childhood.

I plan to continue working on asteroids and pursue a career as an Instrumentation Scientist. I want to acquire the necessary skills on the new technologies and software used in the field. To be a part of the chain of knowledge that has been passed down for centuries and make my own contributions that will equip the next generation of women and men to outdo us in every field. This is the legacy I plan to leave behind.

# Amira Tawfeek

**Egypt**

**National Research Institute of Astronomy and Geophysics (NRIAG)**

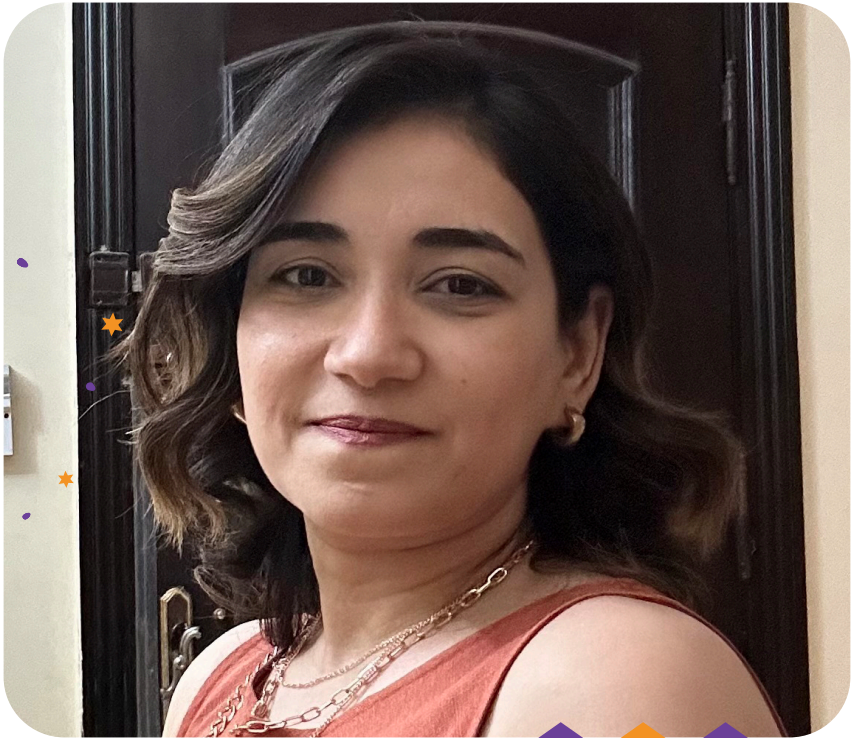

" *I'm a passionate researcher with a PhD in Astrophysics and a deep interest in Galaxy evolution and formation."*

**Unveiling the Mysteries of the Universe: A Journey into Astronomy**

From a young age, I gazed up at the night sky, captivated by the twinkling stars and the vastness of the cosmos. This fascination with the universe sparked a lifelong journey into the realm of astronomy, driven by a burning curiosity to understand the mysteries of celestial objects and their interactions. As a researcher in a prestigious institute of astronomy, I have dedicated my career to unraveling the complexities of gal-axies, from their formation to their evolution, using advanced analytical techniques and innovative algorithms.

My journey into astronomy began during my undergraduate studies, where I was drawn to the captivating beauty and intricate dynamics of galaxies. Delving deeper into the subject, I pursued a Master's degree in astrophysics, focusing my research on the effect of gravitational interactions on galaxy formation and evolution. This period of intensive study laid the groundwork for my subsequent doctoral research, where I embarked on a quest to develop a novel algorithm capable of identifying bar structures in galaxies across various environments.

Galaxies, with their spiraling arms and enigmatic structures, have long been a source of fascination for astronomers. One of the most prominent features found in many spiral galaxies is the bar—a long, elongated structure that extends across the galactic disk. These bars play a crucial role in shaping the morphology and dynamics of galaxies, influencing star formation, gas accretion, and the redistribution of stellar orbits. Understanding the prevalence and properties of bars in different galactic environments is essential for unraveling the underlying mechanisms driving galaxy evolution.

During my Master's degree studies, I focused on investigating the profound impact of gravitational interactions between galaxies on the symmetry of their morphology and structure. This research delved into the complex dynamics of galactic systems, where the gravitational forces exerted by neighboring galaxies can lead to significant distortions in their shape and organization.

Through a combination of theoretical modeling, numerical simulations, and observational analysis, I explored how gravitational interactions influence the overall symmetry of galaxies, including their spiral arms, bars, and overall disk morphology. By studying a diverse range of galactic systems, from interacting pairs to galaxy clusters, I aimed to uncover universal principles governing the interplay between gravitational forces and galactic structure.

This research not only deepened our understanding of the dynamic evolution of galaxies within the cosmic web but also shed light on the mechanisms driving galactic morphology and diversity. The findings from this study contribute to broader discussions in astrophysics regarding the formation and evolution of galaxies, offering valuable insights into the intricate dance of gravitational forces shaping the universe's rich tapestry.

During my doctoral studies, I conducted an in-depth investigation into galaxy triplet systems, focusing on morphological and photometric decomposition analysis. Galaxy triplets provide a unique laboratory for studying the interplay between gravitational interactions and galactic morphology within small-scale cosmic environments.

Using advanced imaging techniques and sophisticated decomposition algorithms, I meticulously analyzed the morphological and photometric properties of galaxies within triplet systems. By dissecting the light profiles of individual galaxies, I aimed to uncover the underlying structural components, such as bulges, disks, and bars, and examine how they are influenced by the gravitational interactions within the triplet.

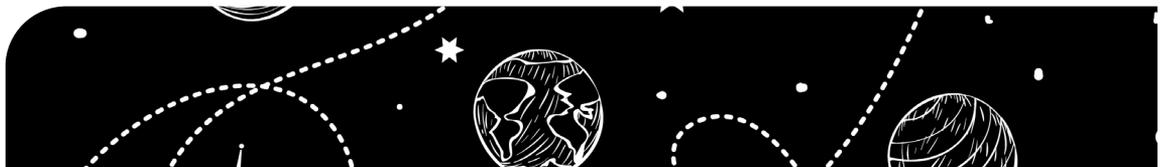

Through this research, I sought to elucidate the mechanisms driving the formation and evolution of galaxies in triplet systems, shedding light on the role of tidal interactions, mergers, and interactions in shaping their morphology and photometric properties. The findings from this study contribute to our broader understanding of galaxy evolution and provide valuable insights into the dynamics of small-scale galactic environments.

In the Post-Doctoral project, I was inspired to develop an automated algorithm capable of detecting and characterizing bar structures in galaxies with unprecedented accuracy and efficiency. Drawing upon my background in computational astrophysics and statistical analysis, I devised a methodology that leveraged machine learning techniques to identify bars in large-scale galaxy surveys. By analyzing diverse datasets spanning galaxy clusters, small groups, and isolated galaxies, my algorithm could discern subtle variations in bar properties across different environments.

The development of this automated algorithm represented a significant breakthrough in the field of galaxy morphology studies, offering astronomers a powerful tool for systematically cataloging and analyzing bar structures in galaxies. Through extensive testing and validation, I demonstrated the robustness and reliability of the algorithm, paving the way for its widespread adoption within the astronomical community.

The journey of a researcher is marked by countless hours of observation, analysis, and discovery. As I delved deeper into the intricate dynamics of galaxies, I found myself captivated by the beauty of the cosmos and the profound insights it holds about the nature of our universe. Each publication in prestigious international journals was not just a milestone in my career but a testament to the dedication and passion that drives scientific inquiry.

Throughout my career, I have been fortunate to collaborate with esteemed colleagues and mentors who share my enthusiasm for unraveling the mysteries of the universe. From attending conferences and workshops to engaging in spirited discussions with fellow astronomers, the pursuit of knowledge has been a collaborative endeavor fueled by curiosity and a thirst for understanding.

As I reflect on my journey into astronomy, I am reminded of the countless hours spent peering through telescopes, analyzing data, and pondering the mysteries of the cosmos. It is a journey filled with moments of wonder and awe, where each new discovery unveils a small piece of the vast tapestry of the universe. And yet, with each revelation comes a new set of questions, propelling us ever deeper into the realm of the unknown.

Ultimately, it is this sense of wonder and discovery that drives me forward as an astronomer. Whether studying the formation of galaxies or unraveling the mysteries of black holes, each day brings new challenges and opportunities to expand our understanding of the cosmos. As we continue to push the boundaries of scientific knowledge, I am honored to be part of a community dedicated to exploring the wonders of the universe and sharing its beauty with the world.

# Miriam M. Nyamai

**Kenya**

**South African Radio Astronomy Observatory (SARAO), South Africa**

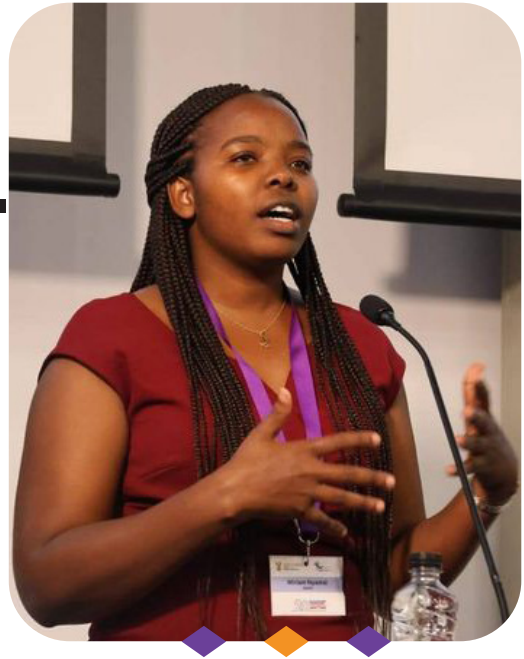

" *Observational radio astronomer with years of experience in conducting astronomy research. Adaptable and unwavering dedication to fostering a productive work environment and delivering high-quality results.* "

I am an observational radio astronomer with three years of experience in conducting astronomy research and operating the MeerKAT telescope located in Karoo, South Africa. As part of the telescope operations Science team, my work consists of three main parts: commissioning, telescope observations, and research using the SKA precursor radio telescope MeerKAT.

As part of my commissioning work, I set up observations for the telescope, reduced the data using various radio astronomy software, and wrote reports of the findings before finally making recommendations for observing strategies. My final role is to conduct independent research. My research consists of the study of radio transients using radio telescopes, in particular, MeerKAT and the Very Large Array in the USA. Radio transients that are of interest include cataclysmic variables, X-ray binaries, and flaring stars. I have authored a few papers published in high-impact journals, with others forthcoming.

Furthermore, I have extensive experience in tutoring postgraduate students in Physics and Astronomy. I have also been involved in an outreach program, the Development in Africa with Radio Astronomy (DARA; www.dara-project.org). This project has been delivering training in radio astronomy and related space-sector skills to over 300 young people across African Square Kilometre Array telescope partner countries. In these countries there are not many people who are experienced in radio astronomy;

hence, there is a need to train postgraduate students in this emerging area in order to make use of the facility.

My story of becoming an astronomer started in Kenya. I obtained a Bachelor of Education degree in Mathematics and Physics from Kenyatta University before embarking on postgraduate studies in South Africa. In 2014, I left my country Kenya to come to South Africa to attend the National Astrophysics and Space Science programme (NASSP; https://www.star.ac.za/) . Through the program, I completed an honors degree in Astrophysics and Space Science and a master's degree in astrophysics. The NASSP program offered sufficient bursaries to students. After completion, I applied for the South African Radio Astronomy Observatory (SARAO; https://www.sarao.ac.za/) Doctoral fellowship, which financed my PhD in Astronomy at the University of Cape Town. I became a professional astronomer in 2021 after obtaining a PhD in Astronomy from the University of Cape Town.

Science, especially astronomy, is important to me because studying the universe has led to amazing discoveries. For example, in 2015 astronomers using the Laser Interferometer Gravitational-Wave Observatory (LIGO) detected gravitational waves after Albert Einstein predicted their existence in 1916 in his general theory of relativity. Second, astronomy, among other things, teaches you how to handle large amounts of data and draw meaningful information or conclusions.

My advice to other girls and women is that, I know pursuing a career in science can be challenging but it is worth it in the end. Challenges make you strong and you get to experience being a researcher in the field and interact with like minded people. I enjoy being a scientist because it helps me think and come up with ideas that I can investigate using world-class telescopes such as MeerKAT. Therefore, I have the opportunity to hone my data analysis skills using large datasets. I also interact with and mentor students who share the same passion for science.

My biggest challenge as a woman in science is lack of immediate mentorship, sometimes you feel alone in this journey. However, I have been fortunate to have mentors who are far away physically but very close emotionally, and I will be forever grateful to them. They have taught me a lot about life in science and are always patient with me while offering valuable advice.

My main achievements at work and in life have been getting a doctorate degree from a foreign institution, far away from home, making new friends in South Africa, adjusting to the vibrant multiculturalism in Cape Town, and being awarded several prizes and scholarships during my academic journey. I have also gained experience by giving back to the community through outreach activities. I can honestly say that I have achieved my biggest dreams. In the future, I would like to continue in research and collaborate with astronomers around the world to continue pursuing science research.

**Awards**

1. African Astronomical Society (AFAS) prize for the best PhD thesis in Astronomy and Space Science Africa

2. 1st prize in poster presentation at South African Institute of Physics Annual Conference

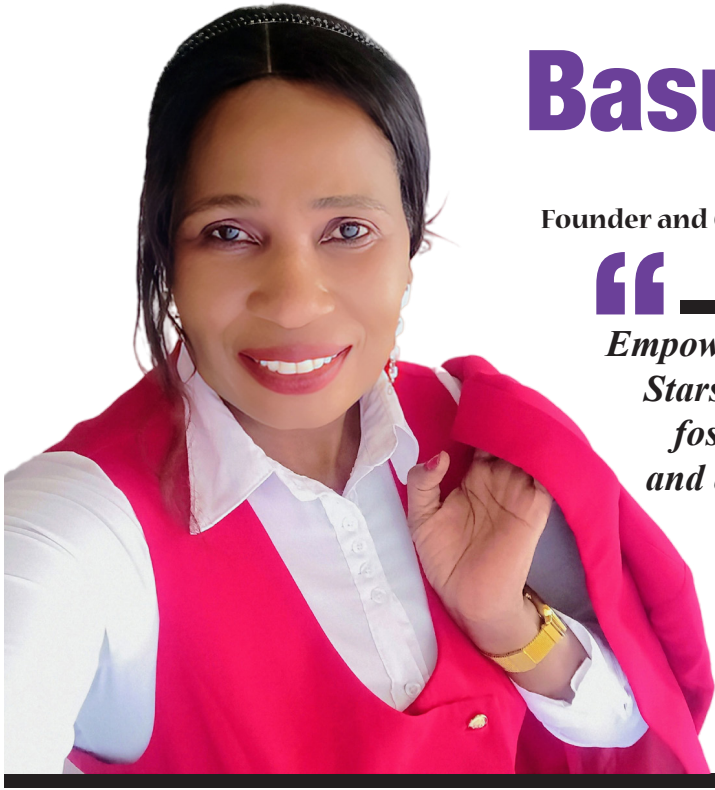

# Basuti Bolo

**Botswana**

**Founder and CEO, GoTospace, Botswana**

*Empowering Minds, Igniting Stars." Inspiring curiosity, fostering understanding, and embracing the infinite possibilities of space exploration. Together, we navigate the cosmos and illuminate the wonders of the universe."*

I am a multi-award winner, space scientist, mentor, advisor, public speaker, and founder and CEO of GoToSpace of Botswana.

My journey into space at first was a dream and later developed a passion for space exploration. I have been interested in STEM programs from a very young age. I used to be one of the best students in mathematics and science from primary to tertiary level. My first dream in space exploration was inspired by a picture of the Russian cosmonaut Yuri Gagarin, who became the first human to travel into space. This was when I was only 8 years old. "I used to look at the picture and vowed that, I too, will one day do the same".

Then later on, I wanted to be a pilot, at last studied Space and Atmospheric Science, and Space applications. My curiosity for Space exploration was sparked by an interest in knowing more about unexplained mysteries of things happening in space, such as the cause of some plane crashes. While studying for a Post Graduate Diploma in Space and Atmospheric Science, I carried out research on the Impact of Geomagnetic Storm (Space Weather) on Satellite Electronic devices and Navigation Systems. The research aimed to find out why Satellites crash or disappear in Space with an unknown failure.

My Dream turned into a career. I am one of the most influential women in STEM, Space, and Global Leadership as well as a gender equality champion. I am now a space scientist working as an Endowed Chair Educational Technologies; A multi-awards winning founder and CEO of GoToSpace, Botswana Country chair on Space exploration and aviation, International Academy of Space Law of Russia Ambassador to Botswana, United Nations Office for Outer Space Affairs Space for Women Network volunteer mentor, Africa Space Tourism Society Ambassador to Botswana, Space for Women Ambassador to Botswana. Women in technology and Global Women for Good ambassador.

I have been volunteering on numerous occasions in different organizations to assist in order to achieve targeted goals. I am one of the United Nations Space for Women Network mentors since 2020; I am a Jury for NASA Space Apps Challenge for the Countries of Philippines and Italy in 2021; and Africa50 Innovation Challenge to be one of the Africa50 Innovation Challenge Expert Panel in 2020.

When I began my career many years ago, I believed in myself, to drive my dream until I live it and see it happening. I made sacrifices to make my dream become a reality and I am not yet where I want to be. I am still at the beginning of it. Currently, I have obtained an admission to be trained on Human Space Flight; and Astronaut training programs certificates.

During my journey, I have experienced some challenges of gender equality and the belief that male students can do better than female students. I also experienced some challenges at work. There was insufficient support for sponsorship to further my studies. Despite doubts, discouragement and other misgivings from the academic wing, I proved them wrong by maintaining the highest mark consistently until I eventually won the trust and support. I followed my dreams and gave the best to all nations.

Despite doubts and discouragement, I focused on my dream and finally lived it. I felt that I could not do it alone, I had to encourage, motivate, mentor others, and be their role model. I founded the GoToSpace organization to educate, train, and mentor more than 1,000,000 girls and youth on Space and STEM programs to reduce the gender gap in STEM and Space sectors.

I believe that women and girls should be empowered and given access to quality education that is aligned with the 4th Industrial Revolution, Education System 5.0, and skills and knowledge of the 21st Century. The world is driven by technology and Innovation. STEM and Space are the current tools for sustainable development that can benefit all. Looking at the world population statistics, there are more women and girls than male, leaving them behind in the world cannot do much. Equality could reduce poverty, hunger, climate change, etc. because all will be equipped with quality education, skills and knowledge for sustainability.

I hold a Certification on Space Science and Technology Supervisor; a BSc Hons Geographical Information Systems; MSc in Information Systems and Data Management, Post graduate Diploma in Space and Atmospheric Science; and ongoing PhD. in Information Systems research on drones technology and applications.

**My advice to girls and women is**
"Live and follow your Dreams".
"Represent your country, culture, and nations and give the best to all".
"Never Quit, Follow your Dreams, The sky is not the limit but the beginning".
"Stand, Unite, Empower each other and bring solutions to the World".

**Awards**

1. Woman of Heart Global Awards: Outstanding leader in Technology, Research and Innovation Pioneer Award 2023.

2. Global Women Leadership 2023: Most Exemplary Leader in Space Science and Technology

3. Top100 Global Professional Women in Aerospace and Aviation 2022

# Tigist Simachew

**Ethiopia**

**MSc Student, Ethiopian Space Science and Geospatial Institute(SSGI), Ethiopia**

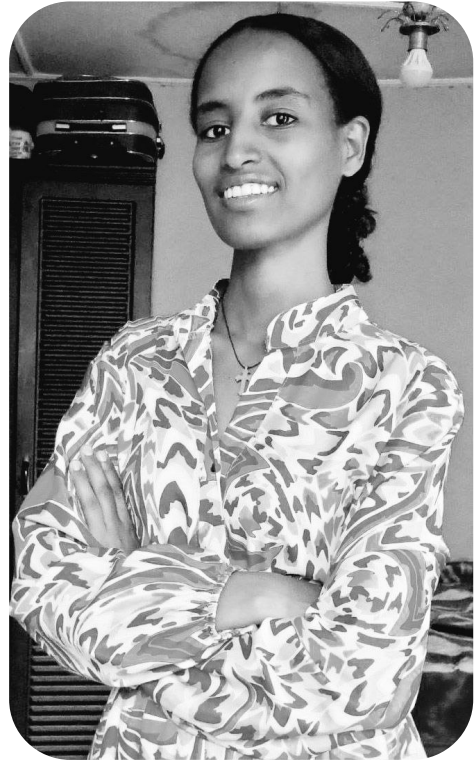

> *There is a plenty of research ahead for the next generation of astronomers to discover our universe."*

I worked as an assistant lecturer at Dire Dawa University after earning my bachelor's degree in applied physics there.

In my research, I mostly study low-mass star formation in molecular clouds that are subject to external pressure. I took astronomy courses at Dire Dawa University as an undergraduate, and those classes inspired me to learn more about the subject.

Science is important because it enables us to understand many things about the universe, such as how the universe was created, how life first began, where the universe is now, and what the universe will eventually be.

I genuinely value everyone's understanding of the importance of science in daily life. In particular, women should work more to make improvements in the world.

The most fascinating aspect of science is the big bang theory.

My objectives in science are to find solutions to unresolved issues in space exploration and to inspire female students to pursue science courses.

### Awards

1. Best female physics student of the year in 2020 by Ethiopian Physical Society (EPS)

# Zara Randriamanakoto

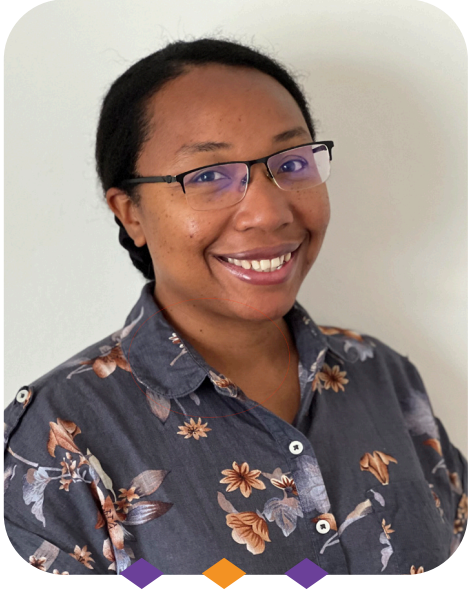

Madagascar

**South African Astronomical Observatory / University of Antananarivo**

*"Think big but start small".*

Zara Randriamanakoto is a 2023 National Geographic Explorer with extensive research experience on young massive star clusters. She is the founding president of the Malagasy Astronomical Society, the IAU adhering organization of Madagascar, and a strong advocate for women in STEM with a firm belief that science is for all.

Dr. Zara Randriamanakoto is a research astronomer at the South African Astronomical Observatory (SAAO), South Africa's leading astronomy research institute. Her main duties include conducting independent research with a focus on multiwavelength extragalactic astronomy and participating in postgraduate student training and science engagement.

Her research interests mainly revolve around the multiwavelength observations of the most massive star clusters in the Universe. She uses space-based (e.g. the Hubble Space Telescope and the Very Large Telescope) and ground-based South African telescopes (e.g. the Southern African Large Telescope) to help image local collisional ring galaxies and galaxy mergers that are ideal hosts of these massive star clusters. One of the key questions Zara aims to address in her work is whether the host galaxy environments play any major role in the properties of these cluster populations.

Noteworthy results from her work include establishing the first ever near-infrared relation between the brightest cluster magnitude and the host star formation rate. Such findings have helped to demonstrate that external factors such as the local environment partially regulate the different physical mechanisms that drive star formation. Zara also uses MeerKAT, the South African precursor of the Square Kilometre Array (SKA) which is the world's largest radio telescope, to study the lifecycle of (giant) radio galaxies for a better understanding of galaxy evolution.

Randriamanakoto is also a visiting research affiliate at the University of Antananarivo, Madagascar (her home country) with her main task being to supervise MSc students in Astrophysics. She is the founding president of the Malagasy Astronomical Society (MASS), the International Astronomical Union adhering organization of Madagascar. She was an executive committee member of the African Astronomical Society (AfAS) serving as its Early Career Representative Officer where she had the opportunity to implement meaningful initiatives for astronomy students and young researchers based in the continent. She also served as an external panel of the Hubble Space Telescope time allocation committee.

Zara completed a four-year postgraduate study in Physics Energetics at the Physics Department of the University of Antananarivo with the aim to specialize in renewable energy back then, given the ongoing electricity shortage in her home country. However, her academic journey saw its major turning point in 2008 when she was given the opportunity to pursue another BSc Honours in Astrophysics & Space Science through the University of Cape Town's (UCT) National Astrophysics & Space Science Programme.

Madagascar, along with seven other African countries, supported South Africa, through the South African Radio Astronomy Observatory (SARAO), in its bid to host the SKA and hence, the need to train young Malagasy students to become professionals in the field of Astronomy. She further pursued postgraduate studies in Astronomy at the same university where she earned her doctorate degree in 2015. Before joining SAAO as an observational astronomer in 2021, Dr. Randriamanakoto worked for three years as a SARAO Postdoctoral Research Fellow at UCT Astronomy Department followed by another three-year term Professional Development Program Postdoctoral Fellowship at SAAO.

She received noteworthy pan-African and international scientific awards for her scientific achievements in the field of Astronomy. She has recently been awarded a research grant by the National Geographic Society as a 2023 National Geographic Explorer and a laureate of the 2020 L'Oréal-UNESCO For Women in Science Sub-Saharan Africa Young Talents. She has also been recognized by the UN Economic Commission for Africa in 2022 as one of the 25 Outstanding African Women Scientists and by Canal+ International in 2020 with her research career profiled in a documentary film "Women in Science in Africa, a silent revolution". In recognition of her public engagement with Science, she was also selected as one of Mail & Guardian's 200 Young South Africans 2021.

Astronomy is not only Zara's field of research, it has become her passion and a powerful tool she uses to inspire girls and women to consider a career in science, technology, engineering, and mathematics (STEM). It is no wonder why in 2016 she co-founded Ikala STEM, a women-led community whose mission is to empower and raise the profile of Malagasy women in STEM. To date, there are more than 500 members spread across four continents and 20+ countries who work together to inspire and provide role models to the future generation. More than 40 activities including mentoring programs, talk series, skills transfer workshops, and the annual celebration of the International Day of Girls and Women in Science have been run since the creation of Ikala STEM. Zara and her team hope to reach out to girls and young women in Madagascar to study and engage with STEM topics while providing them role models that they could easily identify themselves with.

Zara's career path to become a professional astronomer was not straightforward. When she first arrived in South Africa, she had no prior background in Astronomy and she barely spoke English. As a result, she struggled with imposter syndrome. However, she persevered and stayed focused, and during her postgraduate studies, she won the most outstanding student presentation three times during the 2009, 2010, and 2013 Annual SARAO Postgraduate Bursa-ry Conference.

She hopes that sharing her story will help remind all girls and future scientists out there that science is for all, and one should not be scared to dare to venture off the beaten path. Here is her advice reflecting on her journey: "Think big but start small". There is nothing you cannot achieve with perseverance and self-discipline BUT you should also find mentors to guide you along the way".

**Awards**

1. 2023 National Geographic Explorer

2. 2021 Mail & Guardian Top 200 Young South Africans

3. 2020 Young Talent Sub-Saharan Africa L'Oreal-UNESCO For Women in Science

# Blessing Musiimenta

**Uganda**

**PhD student at University of Bologna, Italy.**

*Be kind to everyone and always do your best."*

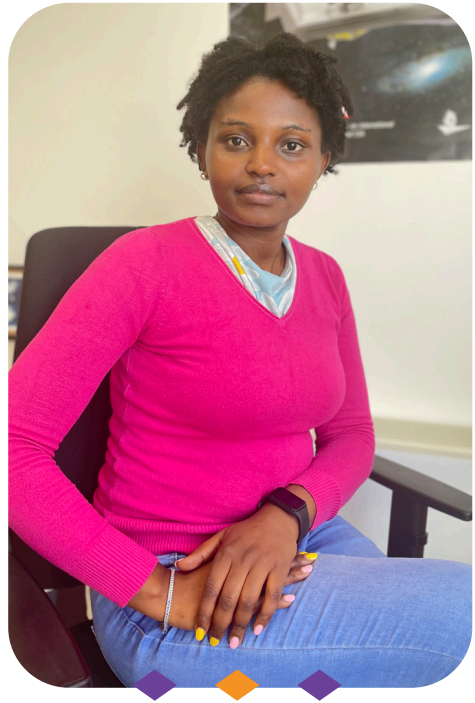

I am Blessing Musiimenta from Uganda. I am a Marie Curie fellow and PhD student at the University of Bologna, Italy.

How did I get interested in studying astrophysics? Well, until I joined the university,

I didn't know that doing research in astronomy is something one can do as a career. During high school, I always excelled in sciences, especially physics and mathematics. I therefore thought I would become an engineer, like most of the children I grew up with, the dream was to become a doctor, teacher, lawyer, or engineer as these were the professions always encouraged by our parents.

When the high school examination results came out to join University, I was offered electrical engineering on government sponsorship but at a diploma level. I didn't like that so I opted to apply for another course on private sponsorship. On my way to apply for a bachelor of business communication course, I met a friend who we chatted with and as I told him what I was applying for, he said to me "Don't put your physics to waste. Just apply for education (with physics) and you can do anything else after that". He went ahead to tell me how there are few women in physics and why I should not waste the good grades. Thanks to his advice, I applied for a Bachelor of Science in Education (physics and mathematics) at Mbarara University of Science and Technology (MUST) in Uganda.

My aim was to finish the course and get a job. Along the way during astrophysics courses, I developed an interest in doing a Master's degree. I was inspired by the teachers who taught me astrophysics for the first time. They were always travelling and the way they always taught it with passion reflecting on their faces. That made me want to do astrophysics.

I completed my Bachelor's and Master's degrees at MUST; thanks to the African Development Bank for higher education science and technology (ADB-HEST) and International Science Program (ISP) scholarships. While I did my Master's degree, I worked as a physics and mathematics teacher in secondary schools and a lecturer at Uganda Christian University. I also volunteered to teach a geophysics course at MUST for one semester.

After completing my Master's degree, I decided to apply for PhD positions abroad. This is because in Uganda, there are very few/no fully funded PhD, thus requiring me to work full-time jobs to support myself while I do PhD. So I thought, why not apply for a PhD where I get paid. I was also so eager to experience what life and education outside Uganda would be like. As a mother, this was a very tough decision to make, it came with huge sacrifice.

Thanks to Dr. Johan Knapen for his guidance in applying for PhD positions. I then won a prestigious 3-year Marie Skłodowska-Curie PhD fellowship at the University of Bologna as part of BiD4BEST (Big Data Applications for Black Hole Evolution Studies) project. This is a project that consists of 13 PhD students from all over the world pursuing PhDs in European Universities. I have been a member of the newest X-ray telescope aimed to survey the entire sky down to unprecedented depth (eROSITA).

In these two big projects, my work focuses on studying supermassive black holes when they feed (known as active galaxies). These black holes are found at the centres of galaxies and play a critical role in shaping how galaxies evolve, but there is still a lot we don't understand about them. I focus on understanding the strong winds of energy and matter released by these black holes, which can regulate the amount of material falling into them and influence the surrounding galaxy by preventing new stars from forming. It is difficult to detect these phenomena, so I use innovative multiwavelength and machine learning methods to find them and study their effect on their host galaxies. This helps us learn more about how galaxies grow and evolve.

Some of my biggest work achievements so far are having published 14 papers in peer-reviewed journals, my involvement in releasing important data sets and being granted telescope observing time (30 hours) with the European Southern Observatory, which is a rare opportunity at such a stage in this career. Outside of work, I have also achieved my dream of being a mother and raising my two daughters.

During my PhD I have had the opportunity to visit different science institutes as a research visitor such as Durham University (UK), Max Planck Institute of Extraterrestrial Physics (MPE, Germany), and Leibniz-Institut für Astrophysik Potsdam (AIP, Germany), collaborating with great scientists. I also had an opportunity to do an industrial internship with PreWarp, mainly focusing on developing forecasting models for fashion industries. This was a great opportunity to transfer skills learnt in academia to industry and vice versa and experience working outside academia.

I enjoy doing research in astronomy because it has offered me opportunities to travel to various countries across Europe and in China for conferences sharing my research, meeting new people from all over the world, and experiencing different cultures. I have had an opportunity to share my research and talk about black holes with the public in an outreach activity at the streets of Bologna during the European researcher's night. That was fun!

Of course, the journey was not smooth. Where do I start from? I remember when I was applying for PhD positions, it came with many rejections but also with acceptances. The rejections always hit differently, but I learnt that it's just part of this career of academia. Therefore, you keep trying, elsewhere again and again, until you get what you want.

My message to girls who want to join the science field: pursue your dreams, no matter how big they may seem. You should dream big and don't let anyone define your limitations. It is common to face cultural and social stereotypes, especially in high school, at home, and in the workplace, but you must push past these barriers. I am inspired by women who are able to balance their work and family life, you are super women!

My academic goal is to further my study of black holes and their influence on galaxy evolution, while also developing initiatives that promote participation of female students in STEM fields in Uganda and beyond.

In the future, I may decide to stay in academia or join industry. I am open to both options. If I stay in academia, I hope to find a permanent job, secure research funding, build my own research group, and make a meaningful impact in the astronomy community.

**Awards**

1. Marie Curie research fellowship

# Hasnaa Chennaoui Aoudjehane

**Morocco**

**Planetary Scientist, Hassan II University of Casablanca - Faculty of Science Ain Chock**

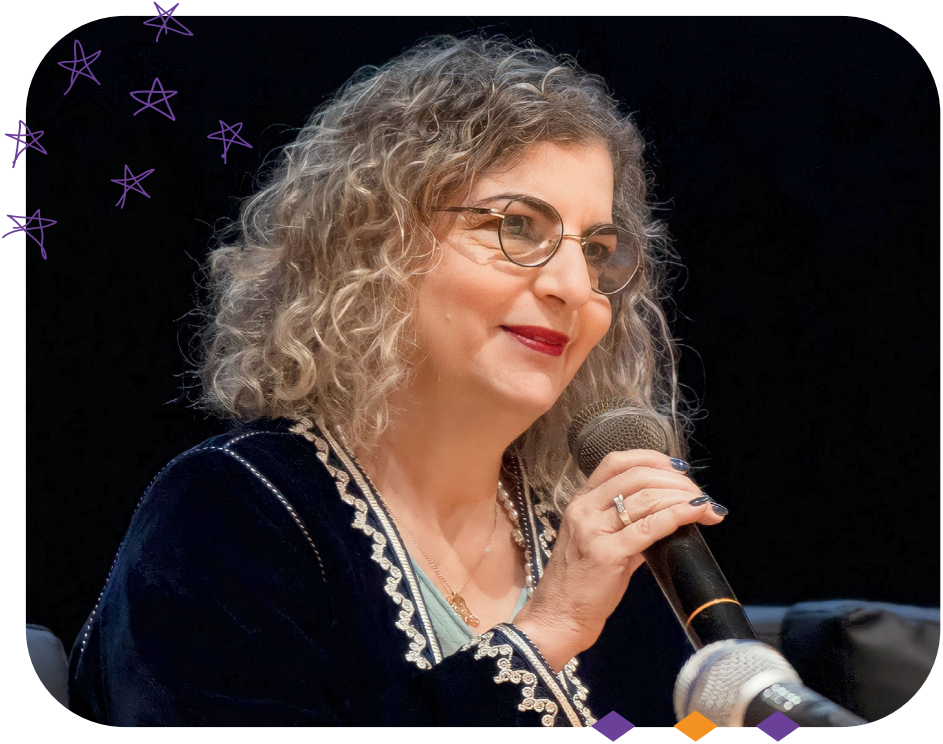

*Believe in your capacities and follow your dreams. Work hard, love the others."*

My journey with science dates back to my early childhood, where I was always very curious to learn. I asked a lot of questions and tried to understand everything around me. I enjoyed reading a lot and had a passion for scientific questions, which at that time were limited in my young mind to biology, especially human biology.

I naturally pursued a scientific high school diploma, which I obtained with top honors. This opened the doors to all higher education programs, including engineering and medicine. While my passion was for medicine, my sensitivity to the sight of blood (it made me faint!) led me to choose to study biology at the Faculty of Sciences to be close to the medical field.

During my first two years in the Biology-Geology program, I discovered the wonderful world of geology, which I didn't even know existed. I had inspiring teachers who made me love this discipline, in which I specialized. After earning my bachelor's degree in Geology, I went to Paris to pursue a DEA (Diplôme d'études approfondies) and then a thesis on an innovative topic in France, namely "noble gas geochemistry." With my PhD. in hand, I returned to Morocco and joined the same institution where I had studied, Ain Chock Faculty of Sciences, as a lecturer.

A few years later, in 2000, I decided to pursue a new PhD « Doctorat d'Etat ». I contacted my professor in Paris to ask if he was willing to supervise me again. He told me that the laboratory was still open to me and that he had shifted his focus to the study of meteorites. He said that I had a place in this field, given the beginning of the collection of several meteorites on Moroccan soil, and that local expertise on the subject would be welcome. Before this phone call with my professor, I didn't even know that rocks fell from the sky... That's how I immersed myself in the fascinating world of meteorites, impact craters, and planetary science. It became my research specialty, but also a passion. Similarly, it became a source of committed associative work on the preservation and valorization of Morocco's geoheritage, which is unparalleled in richness, as well as sharing knowledge through the creation of the ATTARIK Foundation for Meteoritics and Planetary Science.

My work involves teaching geochemistry and cosmochemistry at the university, as well as conducting research and supervising students. Thus, I have supervised several doctoral students who have defended their theses on various topics related to meteorites and planetary science. Other doctoral students are still in the process of preparing their theses. I also have research projects that need to be written up and published in suitable scientific journals. In addition to this academic work related to my role as a teacher-researcher, I have all the work associated with the associative aspect, namely mentoring mediators, giving lectures and workshops for students and the public, and disseminating knowledge through various means.

Science is vital to me; it enables us to address deep human questions in a factual, objective, demonstrated, and reasonable manner. I believe I have a Cartesian mind that needs to be convinced of the validity of things to accept them. Science opens minds and fights against obscurantism. It enables us to be rational and to promote reflection, exchange, and acceptance of others' opinions rather than closing ourselves off and thinking we possess knowledge. It is the best way to ask questions, try to answer them, and advance universal knowledge. It is a constant questioning for the true scientist who knows perfectly well that absolute truth is not held by anyone and that knowledge advances and develops with the contributions of all scientists. Science is a universe of discoveries and new questions. It also allows us to position ourselves well as human beings in the Universe and to grasp the scales of time and space, as well as to put much of the importance of daily life into perspective. It enables us to better appreciate the immensity of the creator behind these dimensions, many of which are imperceptible to humans.

My main advice to others, in particular other girls and women,is to follow your dreams, believe in your potential, work hard, love what you do, and love others.

One of the significant challenges I have had is related to society and the perception of gender roles in professions. Geology (like astronomy) is often seen as a male-dominated field because it requires fieldwork and frequent travel to remote regions. This perception also affects the professional and academic environment, which is a reflection of societal norms and evolution. Personally, I haven't faced this issue much in my family, where my parents raised us equally regardless of gender. However, I have encountered it later on, but I have been fortunate to share my life with a husband who provides unwavering support and accompanies me on field missions. He also handles all the logistics and household management during my frequent absences for analyses and conferences outside Morocco.

The second major challenge is the lack of analytical techniques adapted to meteorites in Morocco. This represents a constant challenge that I have always tried to overcome through international collaborations developed over years of research, which have led to trust in my work. This trust has built credibility that has opened doors to laboratories worldwide for my doctoral students. To achieve this, funding is also necessary to cover travel expenses, stays, and the often-costly analysis fees.

I am proud to have succeeded in 2012 in having a publication in the journal Science as the first author on the work done on the Martian meteorite "Tissint," a fall observed in Morocco. Another source of pride for me is having successfully classified and named all observed meteorite falls in Morocco since 2004 with Moroccan names. This Moroccan identity is very important to me. Representing my country, Morocco, in international scientific institutions dedicated to meteorites and planetary science. Successfully implementing a new science in Morocco within Moroccan universities by introducing a course on cosmochemistry into the academic curriculum. I am happy to have trained young researchers in Morocco to international standards and to have opened doors for them that they never imagined would be accessible to them one day.

The creation of the ATTARIK Foundation is a great achievement that allows us to share our knowledge with the public, especially with many children and young people, and to make them love science.

My most significant achievement and source of happiness are my children and my family. My dream is to have a research and exhibition center on meteorites and sciences in general in Morocco. I hope to continue to develop my scientific research with my doctoral students, write a book on the meteorites of Morocco based on the experience of the past twenty-four years, and continue to develop the activities of the ATTARIK Foundation to promote science in Morocco and disseminate knowledge.

I hope to succeed in getting the idea of a research and exhibition center modeled after the Museum of Natural History or Science Center accepted in my country and hometown, Casablanca.

**Awards**

1. Service Award of the Meteoritical Society 2023

2. Hypatia international award 2020

3. Prix Paul Doistaux Emile Blutet French Academy of Science 2009

# Ola Ali M. M. Saad

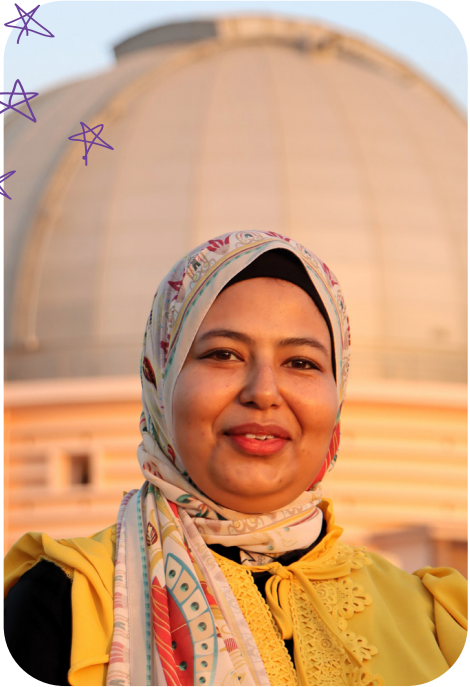

**Egypt**

**Assistant Researcher - National Research Institute of Astronomy and Geophysics – Egypt**

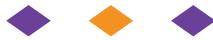

*Be sincere to your dream, time will respond to you, even if it takes a while."*

My name is Ola Ali, an assistant researcher at the National Research Institute of Astronomy and Geophysics. My fascination with astronomy began as a child, gazing out at the night sky from my balcony - my first observatory ever. Every new moon was a cause for celebration, a chance to marvel at the vast tapestry of stars.

A particularly inspirational figure was my science teacher in primary school, whose infectious love for science sparked a similar passion within us. She organized unforgettable trips, fueling my curiosity about the cosmos. These included:

· A trip to the Science Exploring Center, where I experienced the wonder of a planetarium show for the first time.
· A visit to the geology museum in Cairo, where I saw moon rocks and meteorites in person, realizing they were from beyond our world.
· Multiple trips to the National Research Institute of Astronomy and Geophysics - where I work now - to see telescopes and learn more about astronomy.

One unforgettable experience cemented my love for astronomy. During a high school trip to Karnak Temple, a historical site in Aswan, Egypt, I stumbled upon a meteor shower without knowing at that time what a meteor shower was. This awe-inspiring night also offered my first clear glimpse of the Orion constellation. With surprising clarity, I thought I could almost see the hunter himself! From that moment, I knew I wanted to dedicate myself to studying the sky and universe.

Fueled by this newfound passion, I embarked on a quest to become an astronomer. My high school teachers were incredibly supportive, particularly a biology teacher who left a lasting impression with these words: "Our body, through biology, is the small universe, while astronomy is the study of the bigger universe." This profound connection between ourselves and the cosmos continues to resonate with me to this day.

Cairo University's Faculty of Science made my dream come true, with its bachelor's degree program in astronomy. After graduating in 2011, I pursued a master's degree, focusing on the beautiful open star clusters, which I chose because of my love for the beautiful visible to the naked eye "Pleiades." My research delves into the secrets of these celestial families. Using powerful tools like observations from the Kottamia Observatory in Egypt, the 2MASS Catalog (which maps the infrared sky), and the Gaia Catalog (which tracks the motions of billions of stars), I uncover new details about open star clusters. Revealing previously unknown parameters, like their age or composition, brings me immense satisfaction. It's a constant process of peeling back the layers, expanding our understanding of these celestial wonders.

Science, for me, is more than just the pursuit of knowledge. It's a source of solace and inspiration. During challenging times, I find comfort in the vastness of the night sky, especially the breathtaking dance of countless stars on a clear night in the desert, far from the city lights. Sharing this passion through outreach programs, especially with children, brings me immense joy. Witnessing their wonder as they gaze at the stars for the first time is a truly magical experience.

Astronomy isn't just about celestial objects; it also connects me to something larger than myself. Contemplating the universe and its creation fills me with a sense of awe and deepens my spiritual connection with Allah. My journey as a female astronomer hasn't always been smooth sailing. There have been societal expectations and moments of doubt. However, the unwavering support of my family and friends has been my guiding light, as well as my professors at the university and institute, especially the guidance of my mentor and scientific role model, Professor Somaya Saad.

During this journey, I've learned and achieved so much. I operate small telescopes, conduct outreach events using them, and utilize the Kottamia Astronomical Observatory to take observations and publish them. I'm also a member of two scientific programs:

· "Observing and studying astronomical transient phenomena using the Kottamia Astronomical Telescope," STDF Egypt project No. 45779, from 1/8/2021 to 1/8/2024.
· "Poor Open Cluster in the GAIA Era," IM-HOTEP program No. 42088ZK between Egypt and France - 2019 to 2021.

In addition, I'm a member of Egypt's National Astronomy Education Coordinators (NAECs) and the Scientific Society of Astronomy & Space (SSASEgypt).

My dreams are vast. I want to continue exploring the universe, unraveling its secrets, and contributing to humanity's collective knowledge. Knowing that I am a part of something so vast and magnificent fills me with a sense of purpose. I also dream of creating a comprehensive outreach program to teach children astronomy.

Remember, never let societal norms dictate your passions. Pursue what ignites your curiosity, for the universe awaits!

**Awards**

1. Community Service and Science Communication Award from the National Research Institute of Astronomy and Geophysics December 2023

2. Best poster prize at Arabic Conference of Astronomy & Geophysics (ACAG 6) 2018 for a poster titled "2MASS study of Poole J1855+10.8, Ola Ali et. al.,"

# Lidia Dinsa Regassa

**Ethiopia**

**Msc student, Space Science and Geospatial Institute, Ethiopia**

My passion for Astronomy ignited during my elementary school years. Since then, I have been passionate about being an Astronomer. Even though I lived about 320 kilometers away from the capital of my country, which offered a better chance for Astronomy related pursuits, I was observing the activities being carried out only through some online means. When I was in high school, I found out that there was no undergraduate degree in Astronomy available in Ethiopia.

Then I decided to pursue Physics for my undergraduate studies, intending to obtain a master's degree in Astrophysics. It is with this clear picture that I joined the Physics department at Kotebe University of Education which is located in Addis Ababa. While enrolling there I was able to connect with people who shared my interests and passion. I completed my BSC with a good GPA. And I was also the female student president of the students' union of my university. Through that I developed leadership skills, I came across women facing challenges and managed to support them by being with the gender office of my university. The tight sched-

*On the way to Exploring the Cosmos fueled by Passion and Persistence, Ignoring Doubts, and Leading the Way Despite Challenges!"*

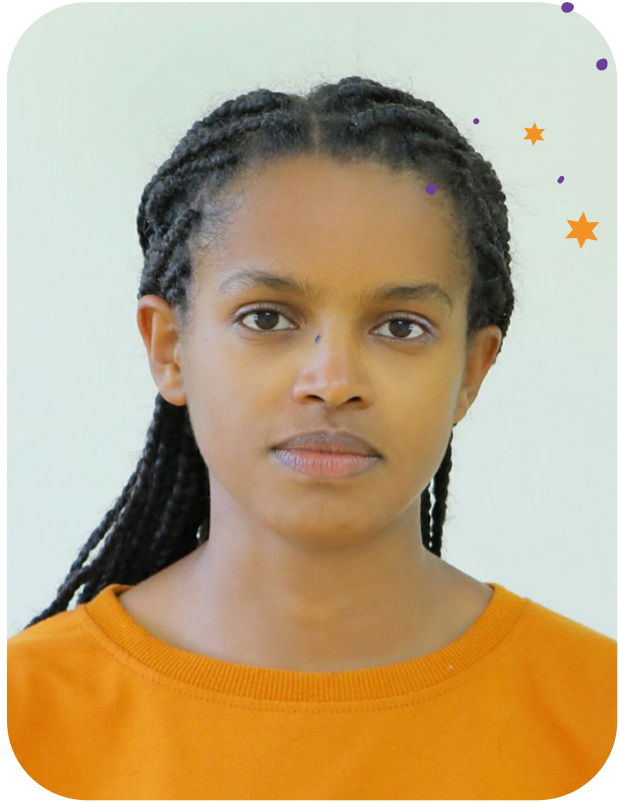

ules I had, the methods I used to balance my duties and my studies and my commitment has made my college years fruitful.

Being in Addis Ababa for BSc studies was also a milestone on my journey. Because it gave me the opportunity to network with like-minded people. Since I believed it would be a wise decision to actively engage with groups and events related to Astronomy, I joined the Ethiopian Space Science Society. I volunteered during a lot of events organized by the society. I was a dedicated volunteer during summer space trainings, public lectures, stargazing events, school outreaches, symposiums, and many other events. Together with the strong and passionate staff, I made a significant contribution to the success of such events and will continue to do so. Through this, I have gained invaluable experience while supporting and promoting society's initiatives.

I am a significant and active member of the "Ethio Asteroid Hunters", a citizen scientist group of the society where we search and report asteroids to the International Asteroid Search Collaboration representing the Ethiopian Space Science Society and our country. As of November 2023, we were able to make preliminary asteroid detections. And we also managed to put our country on the active citizen science group map which signifies the presence of an active citizen science group in the country.

I'm currently pursuing my MSc in Astronomy and Astrophysics at the Space Science and Geospatial Institute. And it feels like my childhood dreams are finally coming true. I'm now in my first year of my studies. To support myself financially I am working as a home Tutor. After completing my master's degree the plan is to continue my studies further and become a professional Astronomer.

My goal is to significantly impact both my country's and the world's Astronomy. I am eager to make a major contribution to science by carrying out groundbreaking studies. I see myself as an example for the younger generation who wish to pursue this path as their career. I picture myself among the few people who risk and dedicate their lives in order to contribute to the world of space science rather than to chase what seems to be easier and immediately rewarding.

My family, who was there for me despite my many challenges and believed in me despite the size of my dream, supportive friends, and teachers I came across had a positive impact on me. Even Though I am happy to be on the right path to my goal, I will not say that the journey was without challenges. The first and persistent challenge I encountered as someone pursuing Astronomy as her career was the pessimistic attitudes and thoughts of those around me who told me it would not be possible. From school peers to educated adults, I have received discouraging advice to let it go and follow a career that they considered "rewarding". Finally, as a woman committed to pursuing her aspirations despite challenges, I am confident that I will make it and I will also be the reason for others to make it.

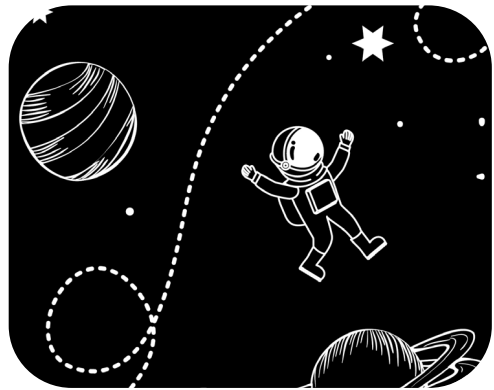

# Elizabeth Naluminsa

**Uganda**

**Makerere University**

*" My professional mission is to make science attainable, penetrable and relatable to everyone. De-mystification of science is the inspiration in all of my scientific endeavours."*

My name is Elizabeth Naluminsa. I am a professional astronomer/astrophysicist and a lecturer at Makerere University in Kampala, Uganda. My professional mission is to make science attainable, penetrable, and relatable to everyone. De-mystification of science is the inspiration in all of my scientific endeavours.

I am the daughter of Mr. James Mabaale and the late Mrs. Zeulia Nyanjura Mabaale. I grew up in our family home in a (then small) village called Kiwumu just a few miles outside Kampala city. Growing up, my parents were educationalists - my dad a teacher of chemistry, and mom an editor and linguistics consultant. As such, I was surrounded by reading material; story books, language books, travel books, novels, magazines, reader's digest, and science textbooks on physics, chemistry, geology, mathematics, name it. Except we did not have one single book on astronomy!

I picked an interest in astronomy during primary school. In one of the English textbooks, there was a story about Neil Armstrong and the moon landing. It fascinated me so much that I kept going back to that story over and over again. It ignited in me a curiosity about astronomy, but having no astronomy books at home, I pored over the geology books. With no light pollution at the time, our night sky provided the rest of the inspiration. By age 16, I was certain I wanted to pursue astronomy as a career. At university, I studied a Bachelor in Science (Physics major) with education, and then went to South Africa where I completed my postgraduate education in astronomy, astrophysics and space science. I got my PhD from the University of Cape Town in 2019.

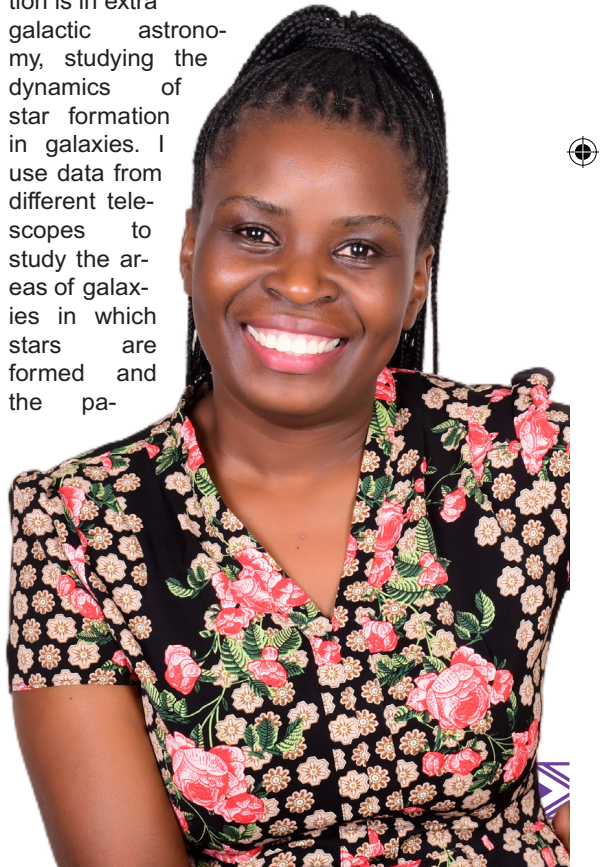

My research specialization is in extra galactic astronomy, studying the dynamics of star formation in galaxies. I use data from different telescopes to study the areas of galaxies in which stars are formed and the pa-rameters that precede that formation. This includes making computational models of the disks of galaxies to compare theory against observations.

In my teaching work, I teach both theoretical physics and astrophysics courses, and I also mentor young students who want to pursue STEM careers. Before I began my tenure at Makerere, I was an instrumentation postdoctoral fellow at the Southern African Large Telescope. My duties included writing software for calibration of a 3D spectroscopy instrument, carrying out observations on the SALT telescope (service astronomy), giving technical support to astronomers from across the world with regard to obtaining data from the telescope, etc.

I have experience in computational astrophysics, software, data science, instrumentation, service astronomy and telescope operations in addition to teaching, curriculum development, research, mentorship and outreach. What I enjoy the most about my work is that I get to learn new exciting things all the time. Whether it is learning to code a new algorithm on the computer, or discovering information in a data set, a new way to apply an old theory, a new way to teach an abstract concept, etc, there are always new things to discover everyday! I also like the fact that I get to inspire and encourage young people to take on STEM career paths.

My advice to young girls and women is that no dream is too big to be achieved. If science is your dream, go for it.

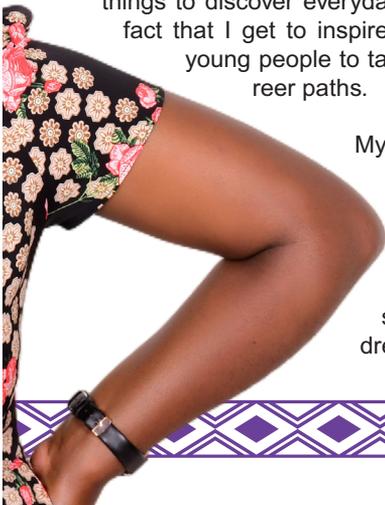

Your dreams might be one step of faith away. When I finished high school, there wasn't any university in my country that offered a course on astronomy, let alone a degree programme. But I spoke about my dream to study astrophysics with my lecturers, and they helped to direct me in the right direction so that I could manage to pursue my dream outside of my country. You will never know what your journey will look like if you don't take the first step. I ended up being the first Ugandan born woman to get my PhD in astronomy and the first black African woman to serve as SALT astronomer on the SALT telescope! And now I live to help my students pursue their STEM dreams just like my lecturers helped me pursue mine.

I do not take credit for any "achievements" in my life because these things wouldn't have been if I was not "standing on the shoulders of giants". So I prefer to not weigh my life in terms of achievements but rather to "collect" the moments and memories. To date, my most cherished moments are the moments when my students or mentees get a hold of their dreams or pick up the courage to pursue the dream where before there was fear. This always confirms to me that indeed together we can change the story of our continent as each one lifts another.

I have lived my dream ever since I began my journey into astronomy with the National Astrophysics and Space Science Programme back in 2011, and much as it hasn't been without challenges, it's been a dream come true nonetheless. Now that I am a professional astronomer, I intend to use my skills and talent to contribute to the growth of the field in my country through human capacity development, technological development as well as educational policy formulation. As is said in my language "The dance has only just begun – Gakyaali Mabaga!"

AfNWA call for Women in Astronomy in Africa | www.afnwa.org  83

# Sambatriniaina Hagiriche Aycha Rajohnson

Madagascar

**PhD student**, University of Cape Town

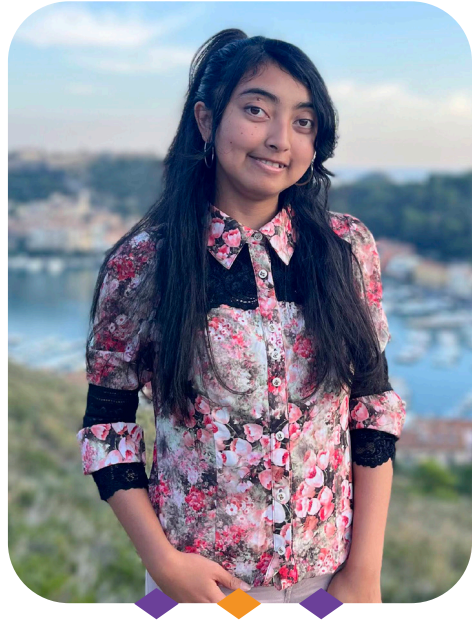

> *Fear not to target the skies nor the expanse of your dreams. Take small steps toward your goals, and always keep in mind, failures are but blips on the radar, with time as your ally.*"

**From Dreams to Charting the Sky**

My name is Sambatriniaina Rajohnson, but you can call me Sambatra. I know my first name can be a bit of a tongue-twister! I am currently in the final stages of my PhD journey at the University of Cape Town in South Africa. I originally come from the highlands of Madagascar, Antananarivo, also known as the "City of the Thousand".

I always had a deep love for scientific subjects during my school years, particularly Physics and Natural Sciences. With a vivid imagination and insatiable curiosity, I found myself drawn to science fiction and documentaries about Astronomy. Questions about the origins of our Universe and the possibility of extra-terrestrial life were never far from my mind. The sky, the stars, and their mysteries both fascinated and intrigued me. However, at that time, pursuing a career as an Astronomer seemed like nothing more than a distant dream. I simply believed that one day, I would love to be a scientific researcher, without a specific field in mind yet. I thought that decision would come when the time was right.

After obtaining my scientific high school diploma in 2012, I then decided to pursue my undergraduate studies in Physics at the University of Antananarivo, where I earned my Bachelor's degree in 2015. It was during my first year of Masters that students were required to select a specialization. News of a newly implemented program caught my attention: "Astrophysics"! Without hesitation, I seized the opportunity, recognizing it as the gateway to realizing my long-held aspiration. Enrolling in the program was a decision driven by both excitement and determination. I knew it would not be easy, but I was ready to take on the challenge.

Indeed, the path to achieving one's goals is not always straightforward; it is often filled with twists and turns. The astronomy curriculum at the University of Antananarivo only extends to the MSc level. To pursue further studies, one must either self-fund or secure a scholarship abroad. Recognizing this, I actively sought diverse opportunities to enrich my background in Astronomy.

One such opportunity arose when I was selected to attend the 39th International School for Young Astronomers (ISYA) in 2017, where I was exposed to the practical aspects of Astronomy for the first time. Later, towards the end of 2017 and the beginning of 2018, while preparing my Master's thesis, I had the opportunity to participate in the "Development in Africa with Radio Astronomy" (DARA) program. This program provided technical training in Radio Astronomy for students from eight African countries that are part of the African VLBI Network.

It was truly an eye-opener for me—I fell in love with Radio Astronomy! I believe it was this training that inspired and directed my future career paths. Moreover, it enabled me to find a supervisor to continue my doctoral studies at the University of Cape Town after completing my Master's degree in 2018 in Antananarivo. I am immensely grateful to Prof. Claude Carignan for providing me with this incredible opportunity, even just before his retirement. As such, never underestimate the power of networking!

Throughout all these years, and even after coming to Cape Town, things do not always go as planned. Projects can change, and data can encounter unforeseen problems. At times, I have experienced imposter syndrome, questioning whether I truly deserved such opportunities. Moreover, we all faced the global pandemic in 2020-2021, which significantly impacted my work. I could not present my research at conferences or travel as much as before, which is something I truly enjoy about Astronomy. However, do not let any of these challenges hinder your progress. Believe in yourself, as even small steps forward each day accumulate into significant progress in the end. When you reflect on how far you have come, you will feel a sense of pride in the journey you have undertaken. If you ever feel lost, do not hesitate to reach out and seek inspiring mentors or role models to guide you along your path.

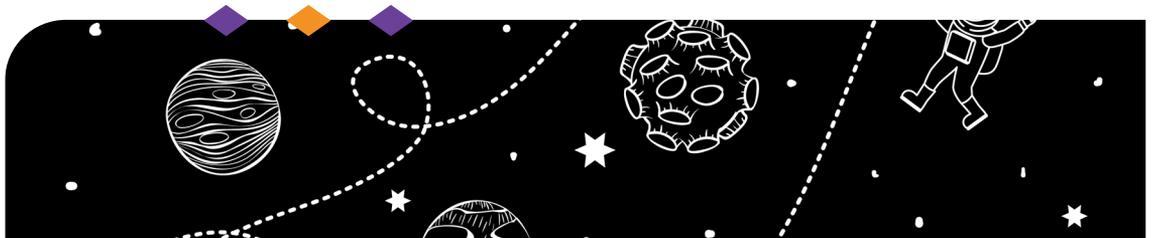

In terms of research, we are known as the Geographers of the Cosmos or Cosmographers, a term coined in Hélène Courtois' book "Le voyage sur les flots de galaxies" (Finding our place in the Universe) from 2016. Under the guidance of two inspiring astronomers, Em. Prof. Renée Kraan-Korteweg and Dr. Bradley Frank, my PhD project involves mapping the large-scale structures formed by galaxies hidden behind the Milky Way Plane.

This region is one of the most challenging parts of the sky for extragalactic astronomers to observe due to our galaxy's dense concentration of stars and dust, leaving vast portions of the sky unmapped. In regions with the highest obscuration, observations are only feasible through radio wavelengths, where we measure neutral hydrogen emissions from galaxies. Our aim is to complete the missing puzzle piece in the map of the Universe towards the constellation of Vela, out to 800 Million light-years away in cosmic time. This will help us identify new structures and understand their implications for the Universe's evolution. So far, my work has led to the discovery of over 1500 heavily obscured, gas-rich galaxies, shedding light on previously unexplored regions of the sky.

As such, I find science incredibly important as it enables societal advancement through continuous learning. Discovering new breakthroughs is always exhilarating, especially when you are the first to uncover something. Additionally, science serves as a wonderful platform for international collaboration, allowing one to connect with people from around the world. I will never forget the excitement I felt when I saw my name on my first co-authored paper through collaboration in 2021, and then when my first paper as the lead author about my work was accepted for publication in 2022.

Furthermore, being an astronomer is more than just conducting research; it is also about sharing our passions and discoveries with the public. Personally, I am passionate about inspiring the younger generation, especially girls, to cultivate a love not only for Astronomy but also for Science as a whole, encouraging them to pursue their dreams fearlessly. Currently, I serve as the regional coordinator for the Sub-Saharan Chapter of the Ikala STEM (Science, Technology, Engineering, and Mathematics) association, a non-profit organization dedicated to empowering Malagasy women scientists worldwide.

Looking ahead, I am eagerly excited about what the sky has in store for me, and I look forward to completing my PhD and soon being called "Doctor" in Astronomy! In conclusion, it is incredibly inspiring to see the increasing number of African women in science, offering hope for our continent's scientific advancement.

**Awards**

1. South African Research Chairs Initiative (SARCHI) Masters and Doctoral Scholarship awardee

# Ann Njeri

**Kenya**

**Astronomer, Newcastle University, UK**

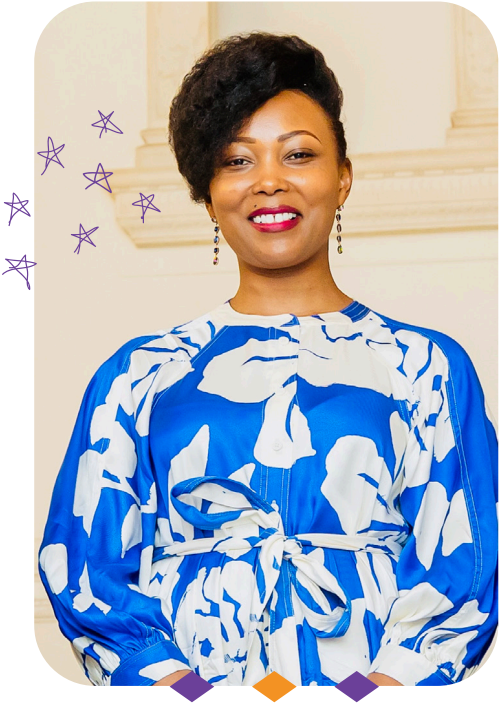

> *If you can't see it, then you can't be it!"*

I have always been fascinated by space and astronomy since as a little girl. That is the reason why I enrolled for a BSc in Astronomy and Astrophysics at the University of Nairobi. This was then followed by an MSc in Nuclear Science, with an Astronomy project for my MSc thesis.

I conducted my MSc research work at the Hartebeesthoek Radio Astronomy Observatory in South Africa under the supervision of Dr. Aletha de Witt and funded by the National Research Fund (NRF). I later secured a PhD scholarship offered by the Development in Africa with Radio Astronomy (DARA) project. This scholarship enabled me to pursue a PhD in Astrophysics, under the supervision of Prof. Rob Beswick, at the Jodrell Bank Centre for Astrophysics, University of Manchester. Currently, I am a postdoctoral research associate at Newcastle University working with Dr. Chris Harrison in the area of Extragalactic Astrophysics.

I have a strong passion for understanding our cosmic origins, with a particular desire to understand how supermassive black holes could influence the formation of stars and galaxies (including our own Milky Way galaxy).

As such, my current research at Newcastle involves the use of high-resolution radio imaging to map out large areas of the sky in probing the nature of the hidden supermassive blackholes. Supermassive black holes are thought to reside in the centre of each galaxy and hence influence the formation and evolution of these galaxies and stars in the Universe. In order to understand the mechanisms driving the symbiotic relationship between the supermassive black holes and their host galaxy, and to also separate accretion processes from star formation, a full census of supermassive black hole growth and star formation is required. My research involves separating these two processes in order to derive precise star formation rate measurements.

I have won several scholarship opportunities, grants and awards as a testimony to my resilience and hard work, including a PhD scholarship at a top institution (which is not easy for women from my background). I am also the project principal investigator for several science projects using major radio instruments across the globe.

Besides research work, I have founded and led an award-winning mentorship and outreach programme, https://www.elimishamsichana.org/. The programme is creating education opportunities and addressing the issue of gender disparity in education for girls in rural areas of Kenya and Uganda. The programme has already reached over 7,000 schoolgirls across 10s of schools and is mobilising local communities to keep their girls in school.

Additionally, in developing new research-focused teaching in rural parts of Kenya, I have established a collaboration between the Astronomy group at Newcastle University and Meru University of Science & Technology where lecturers and postgraduate students in the UK are offering weekly online seminars: "Introduction to Astronomy and Research". I also initiated the African Across Continents (Newcastle/Durham Universities, July 2024) which has recently fully sponsored ~40 early career researchers from Africa to attend the AGN Across Continents conference in the United Kingdom.

I have become sought-after as a world-expert in radio interferometry; an advanced technique of combining multiple radio dishes spread across a few metres to 10,000s of kilometres, to simultaneously observe the same astronomical objects. Consequently, I am an invited member of 5 international science consortia.

Using my technical expertise, and extensive computational methods, I have developed new techniques that revolutionise how to combine data from different radio interferometry facilities across the globe to produce unique images at unprecedented spatial details over large patches of the sky. This has previously been considered nearly impossible.

In the next five years, I aim to be an independent research fellow and a leading expert in multi-wavelength observational astronomy, supermassive black hole studies, very long baseline (VLBI) and other radio interferometric techniques. Eventually, I hope to return to Kenya, to secure a lectureship position, and contribute towards the establishment of a possible Square Kilometre (SKA) regional centre and a radio observatory.

Kenya is one of the partner-members of the proposed African VLBI Network, which aims to provide VLBI capabilities to the SKA in the near future, but local experts are required. Further, I aspire to be an international research leader and a leader in promoting equality, diversity and equity. I hope to initiate new projects within/outside my profession; including developing new research collaborations and education-initiatives in Africa and providing opportunities through STEM to the underserved-communities (particularly girls/women).

My advice to future women in Astronomy/STEM is Just go for it! Break societal barriers, stereotypes and expectations. Believe in yourself, seek out support and mentorship, and follow your heart. Science will always lead you home as there is still so much yet to be achieved and discovered!

**Awards**

1. L'Oreal UNESCO for Women in Science UK and Ireland Rising Talent Fellow 2024.

2. Making A Difference Award (Outstanding contribution to equality, diversity and inclusion), University of Manchester 2023.

# Brenda Namumba

> *By pursuing my research with confidence, passion, and conviction, I aspire to lead the way for other women."*

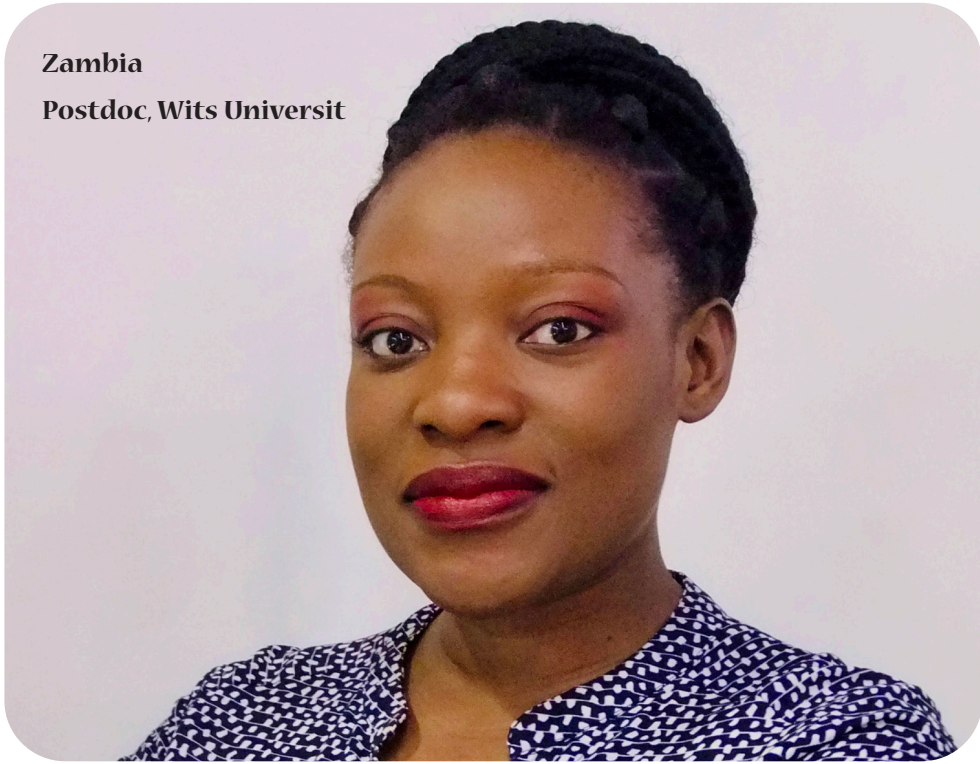

Zambia

Postdoc, Wits Universit

**M**y name is Brenda Namumba. I am currently a postdoctoral fellow at Wits University in South Africa, working on one of the most iconic projects for global cooperation implemented in Africa, the Square Kilometre Array. My research interests include neutral hydrogen in nearby galaxies, galaxy formation, evolution, and star formation.

I was born in Zambia. With an early strong affinity for science, I joined the School of Natural Sciences at the University of Zambia in 2005 and graduated with a BSc in Physics in 2010. In 2019, I attended my first academic workshop, the International Heliophysical Year SCINDA workshop in Zambia, thanks to a grant from the ICTP European institution. There, I gained my first in-depth knowledge of Astronomy and Space science. I became eager and curious to learn more, and my career journey as an astronomer began.

Coming from a low-income family, the only hope I had of continuing my studies was through a scholarship. The lack of infrastructure, such as computers, made me lack prerequisite courses, making it difficult to compete for scholarships. After being rejected for four applications over two years, I received the National Space Science and Astrophysics Programme (NASSP) scholarship in 2012 and moved to South Africa for postgraduate studies in Astrophysics. This opportunity changed my life, as NASSP focuses on African students from third-world countries, selecting based on motivation rather than merit.

In 2012, I obtained an Honours degree in Space Science and Astrophysics from Cape Town, where I chose to focus on radio astronomy. In 2015, I obtained my MSc in Astrophysics from the University of Kwazulu Natal, South Africa. My MSc thesis was on the evolution of neutral hydrogen (HI) in radio galaxies using KAT-7. In 2019, I completed my PhD at the University of Cape Town on the HI properties of dwarf galaxies in the Local Group.

Education has been an eye-opener for me. I have had opportunities to see the world, present my work at international conferences, and publish in scientific journals. Since June 2019, I have been a postdoctoral fellow at Rhodes University, sponsored by the South African Radio Astronomy Observatory. I am a recipient of competitive scholarships and grants. Being a woman in Africa comes with many challenges; still seen as a "housewife" in many societies, my story would have been different if I hadn't received these funds, for which I am appreciative. I was able to finish my studies, travel, and interact with international institutions, gaining scientific knowledge while broadening my cultural horizons, which I believe will benefit Zambia and Africa.

Through my research, I have won many awards, including the Women by Science "Mujeres por Africa," L'Oreal Women in Science, and the AFNWA Early Career Award for Women in Astronomy in Africa.

One thing I lacked growing up in terms of STEM career choices was mentors from my community. As such, I have taken it upon myself to be available as a mentor to young girls in Zambia and Africa who want to pursue STEM-related careers. I am part of Project Kuongoza, which empowers young girls across the Middle East, South Asia, and Africa.

I love astronomy research, and my ambition is to continue my research and apply the knowledge I gain to help develop Zambia and Africa at large, as well as to empower the girl child who is still hesitating to take up a STEM-related career for fear of failure. If I can make it, then you can do it as well.

**Awards**

1. Prof. Carolina Odman early career award for women in astronomy in Africa 2023

2. Women by Science 'Mujeres for Africa" 2022

3. L'Oreal Women by Science 2022

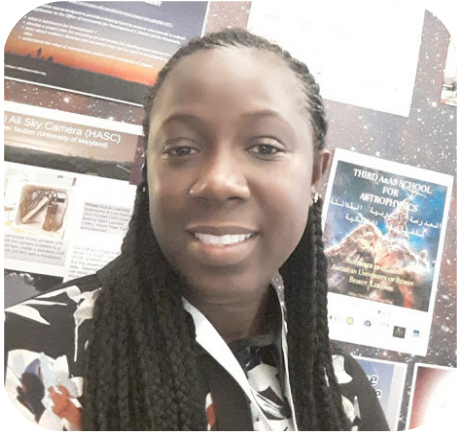

# Salma Sylla Mbaye

**Senegal**

**Institute of Applied Nuclear Technology/University Cheikh Anta Diop**

" *Patience, perseverance, persistent effort, passion, and networking are the keys to a scientific career. If you dream big, you have to be prepared to overcome big challenges.*"

My background is in atomic and nuclear physics with a post-graduate diploma, followed by professional training with an engineering diploma in telecommunications and computing. After my post-graduate studies, I taught physics and chemistry at secondary school for a few months before starting work as a Senior Technician and then as an IT Engineer.

During my five years working in IT, which consisted of installing applications and training users in IT tools, I also taught at university doing practical work and tutorials in physics for Undergraduates.

My work involves studying transient phenomena such as impact flashes on Jupiter and binary RR Lyrae variable stars using small telescopes.

Ever since I was very young I have been very interested in observing the sky. I lived in a region in central Senegal (the Kaffrine region) where there was almost no artificial lighting at the time, and as we went to bed in the open air because of the heat, we took advantage of the darkness to look at the stars and I even tried to count them from one part of the sky to another. When I got to secondary school, I was drawn to books on space science, even though there were not many of them in our libraries.

However, the desire to do research in astronomy/astrophysics was born during my postgraduate studies when I met the Belgian astrophysicist Pr. Katrien Kolenberg at a conference organised in Dakar, Senegal by the Director of our institute. After this meeting, I had the opportunity to take part in an astronomy conference in Burkina Faso in 2010, at which the first African Astronomy Society was set up. Then I became interested in and took part in the activities of amateur astronomers from the Senegalese Association for the Promotion of Astronomy (ASPA).

Following the meeting in Burkina Faso, I

found an internship in France at the Institute for Research into the Fundamental Laws of the Universe at the Commissariat of Atomic Energy (CEA) thanks to a grant from the International Science Program (ISP). I then followed two online training programmes in astronomy organised respectively by the Paris Observatory and the University of Lancashire. However, it was a few years later that I was able to start a PhD in astrophysics thanks to several collaborative projects, including the African Initiative for Planetary and Space Science (AFIPS) network.

Science is important for making new discoveries, improving existing methods or solving new problems. Overall, it helps to improve human living conditions. I recommend that young girls work hard at school, believe in their dreams, love STEM and be aware that science is at the heart of the development of any society, believe in themselves, science is for everyone (man or woman) and that there is still a gap of girls and women scientists. I invite them all to take up the challenge. I invite women to show solidarity and to work in networks to better defend the causes of women.

Everyone has potential, and women today play a major role in advancing of STEM. The involvement of more women in these fields would be useful for Africa to make up for its backwardness and be among the leaders in science and technology, such as space science with the advent of small satellites and astronomy.

What fascinates me about science is the constant quest for new results and innovations to meet new challenges. Science also gives you the opportunity to collaborate with scientists from all over the world, so you can make the most of your research and gain a broaderperspective.

The big challenge is to find the means to do my doctorate in astronomy in good conditions, as the discipline is not yet taught in our universities and there is also a lack of local infrastructure in the field, such as astronomical observatories. However, thanks to grant opportunities and a supportive team of supervisors, I was able to make my dream of a scientific career a reality.

The other challenge is to combine family life and research, taking long absences and leaving my very young children behind, but I have always been able to count on the support of my husband and parents in this respect.

One of my main achievements is to be able to do my doctorate in astrophysics and be the first to start it at my university and in my country. I have received four grants from different sources, one from the Father Louis Bryns Foundation for Belgium, one from the Organisation for Women Science for the Developing World (OWSD) for scientific visits in Morocco, and one from the French Embassy and the Ministry of Higher Education in Senegal for stays in France.

This scientific career has given me the opportunity to collaborate on several national and international projects and to represent the Astronomical Union in Senegal as the National Coordinator for Outreach (NOC-SENEGAL). This has enabled me to organise several basic astronomy training courses and to organise workshops in all areas of the educational cycle, from primary school to university, with the help of several partners and grants. I have also had the opportunity to belong to several networks such as AfNWA (African Network of Women in Astronomy), OWSD, and WFSW (World Federation of Scientific Workers).

My dream is to set up an astronomy and astrophysics laboratory, to help many young people find their way in this field and also to see more women in astronomy in Africa. As well as doing research, I want to develop my own structure to welcome my community, especially young people, and enable them to showcase their talent through training and capacity-building in astronomy, physics, and STEM in order to offer them opportunities in new technologies in the same way as young people in developed countries.

**Awards**

1. Organisation for Women and Science for the Developing World Doctoral fellowship

# Sinenhlanhla Precious Sikhosana

**South Africa**

**University of KwaZulu-Natal**, **South Africa**

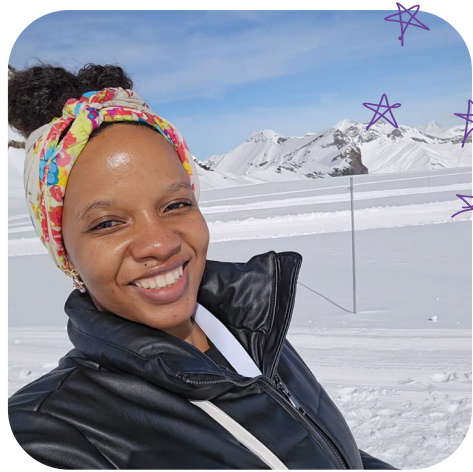

*I am a village girl who entered a field she couldn't even dare to dream. My PhD journey was populated with numerous awards, which have seen me share stages with ministers and lunch with Nobel Laureate Donna Strickland."*

I was born in eXambu, a village near Harding, South Africa. My primary and secondary alma mater are Harding Primary School and Durban Girls' Secondary School. I did my undergraduate and postgraduate studies at the University of KwaZulu-Natal (UKZN). My passion for science, in general, stems from the vast experience of village life in eXambu. The trips to the forest collecting wood and to the river, whether for a swim or laundry, always left me in awe of nature and inquisitive of how everything came to be. My love for astronomy, particularly, was first fuelled by my experience at the UKZN Astrophysics and Cosmology Research Unit's career week in 2010. I was fascinated by the talks that shared all the wonders of the universe and how one can explore the origins of it all as an astronomer. That marked the beginning of the journey of my astronomy career.

I am currently an astronomy lecturer at UKZN, South Africa. As a lecturer, I teach undergraduate astronomy, supervise MSc and PhD students, and conduct my own research. My research focuses on the largest gravitationally bound objects in the universe, known as galaxy clusters. These large-scale structures occupy an ideal position in the structure formation hierarchy; their massive scales make them perfect laboratories for studying astrophysical processes on large scales, and they can be used to constrain cosmological parameters.

When two galaxy clusters merge, a massive amount of energy is released into the intracluster medium, a plasma contained within galaxy clusters. This is the largest amount of energy released since the Big Bang. The deposited energy is dispersed through turbulence, shocks, and other astrophysical processes. Astronomers have found that in some galaxy clusters, these mergers result in the formation of faint cluster-scale diffuse radio sources known as radio halos and relics.

The formation mechanisms of these radio sources have puzzled astronomers for decades. The glaring conundrum is how the underlying population of electrons is transported to such large scales whilst maintaining relatively high energies, given that electrons' diffusion time scales are fairly short.

My research aims to understand the formation mechanisms of these radio sources and relate their evolution with the intrinsic properties of host clusters. My work with MeerKAT data led to the detection of the highest redshift radio halo, which challenged our current theoretical understanding of how these objects form. My ongoing work focuses on studying magnetic fields in galaxy clusters.

This field of study still has multiple open questions due to polarisation calibration and data analysis difficulties. I aim to shed light on magnetic field scales and strength by applying Faraday rotation analysis methods on MeerKAT observations.

In my academic career, I have been awarded numerous scholarships and grants, including the Department of Science & Technology TATA African Women in Science Doctoral Scholarship in 2018 and the Lo'real-UNESCO For Women In Science research grant in 2019. I was also amongst the top twenty young scientists selected to represent South Africa at the Nobel Laureate meeting in 2019. As of the beginning of 2024, I am now a PI of the Astrophysics Research Centre's (ARC) South African Radio Astronomy Observatory block grant. This was awarded to our research centre, which oversees the administration of three undergraduate astronomy scholarships.

I am currently a member of the African Astronomical Society's executive committee, ARC's outreach, diversity, and inclusion committees, and events coordinator at Mthethwamatic. This non-profit company seeks to alleviate the stigma of mathematics in South African communities. I firmly believe that it is our duty as academics to not only communicate science but also uplift the societies we live in. We cannot afford to be a secluded community, as this alienates the same people we ought to be engaging with if we are to transform the field.

Indeed, astronomy is one field that direly needs a paradigm shift. There needs to be more women who not only occupy this space but own it. This is not a simple task, as I have walked this path and faced multiple stumbling blocks. First, it was family members who couldn't comprehend my stubbornness; 'Who makes a living from studying mathematics and physics?' they said. Then it was my insecurities when I used to be the only female attending a physics class; 'You don't belong here', the voice in my head would say. To this day, as an accomplished astronomer, the imposter syndrome still lurks in the shadows; 'Sooner or later they will see that you are not capable', it says. But I have overcome all these hurdles with a strong-willed mind, which gently reminds me: that I am capable, I am intelligent, and all my achievements were because of my hard work. And this is the mindset that one needs to thrive in this field, especially young aspiring female astronomers.

**Awards**

1. Lo'real-UNESCO For Women In Science research grant in 2019.

2. Department of Science & Technology TATA African Women in Science Doctoral Scholarship in 2018

# Marie Korsaga

**Burkina Faso**

**Université Joseph KI-ZERBO**, Burkina Faso

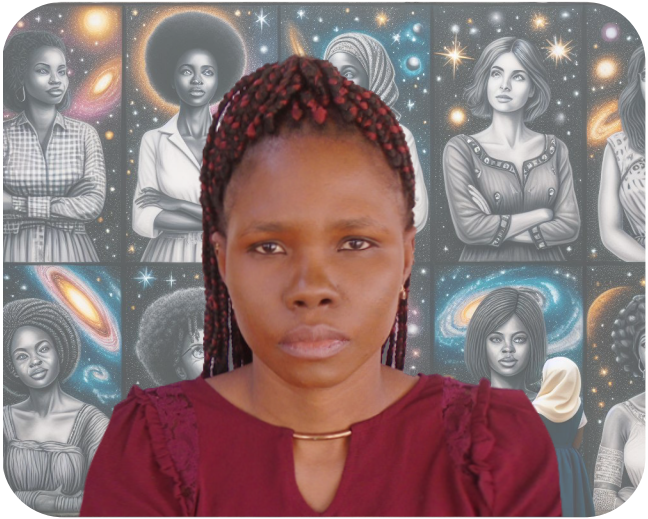

*"As the first female with a PhD in Astronomy in the West African region, I consider it my duty to break down social barriers surrounding women in sciences."*

My name is Marie Korsaga, I am an astrophysicist originally from Burkina Faso. I am currently an assistant professor at the University Joseph KI-ZERBO (the first and biggest university in Burkina Faso).

I carried out my undergraduate studies at this university until the completion of my Master's degree in Physics. After my Master's, I obtained a scholarship that allowed me to move to the Laboratoire d'Astrophysique de Marseille (LAM) in France, where, as an intern, I studied the kinematics of galaxies using Fabry-Perot observations.

Because I am always fascinated in studying the nature of dark matter which is a very challenging subject because it remains today the most mysterious matter in our universe that needs to be understood if we want to improve our understanding of the formation of galaxies including our home galaxy, the milky way, I therefore decided to start a PhD programme in 2014, jointly between LAM and the University of Cape Town in South Africa.

My PhD research focused on investigating the distribution of dark matter in nearby galaxies using state-of-the-art multi-wavelength spectral and photometric data – including infrared, optical and neutral Hydrogen observations. This study led me to understand the limitations of current approaches in constraining the distribution of dark matter in galaxies and to conceive strategies that would improve our knowledge of the formation and evolution of galaxies. The results of the study were published in several peer-reviewed papers.

After completing my PhD in 2019, I joined the IAU Office of Astronomy for Development (OAD) during which I researched the contribution of the science community in the fight against the COVID-19 pandemic. I have published different blog articles on the subject, which are featured on the IAU-OAD's official website. I also work as an associate researcher in astronomy at several institutions including the Instituto de Astrofísica de Andalucía (IAA) in Spain, the Observatoire Astronomique de Strasbourg (ObAS) in France where I study the scaling relation between the neutral hydrogen mass and the dark matter halo to better understand the nature of dark matter and global properties of galaxies..

As the first female with a PhD in Astronomy in the West African region, I consider it my duty to break down social barriers surrounding women in sciences. As a result, I am a junior member of the International Astronomical Union (IAU), a founding member and sits on the editorial board of "L'Astronomie Afrique", an online and free outreach magazine in French for the lovers of Astronomy, Space sciences, and Africa.

Similarly to many African countries, Burkina Faso has a very young population. In fact, 57% of its population are under the age of 20 years. Although the country offers free education until the age of 16, science education still needs a lot of effort. However, science class enrolments only count for 14% of all enrolled students. One way to get that number high is through science promotion, especially among young children. Because the enthusiasm for science has to be stimulated early on in the minds.

In this direction, I initiated the creation of a non-profit association FeBSTIM (Femmes du Burkina en Sciences, Technologie, Ingénierie et Mathématiques) which brings professionals, students, etc. to improve the status of women in science and to inspire more girls to do STEM. We have therefore led actions, seeking to promote science in schools in Burkina Faso, dispelling stereotypes, and highlighting women's achievements in science.

As I used to say, change cannot take place without the strong commitment of women who already have a career in science. It is necessary that these women take action with of course the support of governments and the private sector, scientific organizations, etc. to make their voices heard, to convey the message towards the development of measures promoting parity, and to make the scientific research environment more inclusive and conducive to women.

I am therefore certain that if the right policies are put in place and that the opportunity is given to girls from an early age, the continent will have many more women in science, and gender parity in this area which is so important for the development will be achieved in the relatively near future.

The message that I would like to address to young girls is: believe in yourself, work hard to reach your goals, and never doubt your abilities.

**Awards**

1. African Network of Women in Astronomy Early Career Award 2021

# Sthabile Kolwa

**South Africa & Zambia**

**University of Johannesburg**

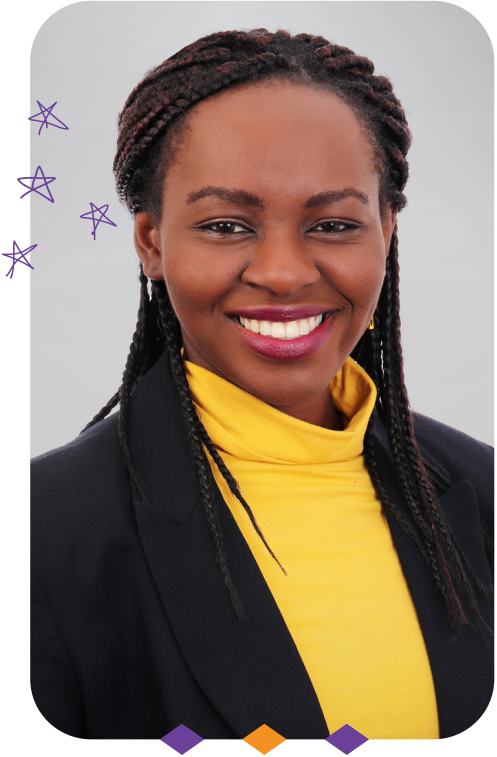

> *Try again, fail again, fail better" is one of my most-loved mantras; I even tattooed it onto the shin of my right leg. It reminds me that there is no harm in trying and you miss 100% of the shots you do not take".*

Sthabile is South African raised and Zambian by heritage. Growing up in Johannesburg, Sthabile developed an early interest in both the sciences and the arts. While curious about the natural world and the physical principles that underlie it, Sthabile also drove into the world of theatre and music. This interest was evident in their outstanding academic achievements through primary and high school. It also showed in their keen involvement in school stage productions such as the high-school production, Bugsy Malone, and a rendition of For Colored Girls for which Sthabile was the director and her friends, the actresses.

Toward the end of their high school years, Sthabile had a major decision to make. Whether they would pursue an artistic life or a scientific one. Being adept at both made the decision difficult, but they eventually chose the sciences. Sthabile's reading of books such as A Brief History of Time and the biography of Richard Feynman galvanised a deeper interest in physics and cosmology.

Sthabile's pursuit of life as an astronomer began with undergraduate studies at the University of Cape Town where they met many of their future colleagues and mentors. During this time, Sthabile's studies were funded by the National Research Foundation's Square Kilometer Array (SKA) student bursary initiative. They completed an M.Sc. degree at the University of the Western Cape. After submitting an excellent thesis and were recruited into the 2016-2019 class for the International Max Planck Research School in Munich.

Spending three and half years in Munich, pursuing a PhD., learning basic German, and exploring Europe, Sthabile considers these years to be some of the most gruelling, challenging but also exciting, and mind-expanding years of their young life. While always knowing they were queer for most of their lives, Sthabile found the courage to embrace their gender identity while living in Germany. Through the process of self-acceptance, Sthabile became even more fearless and confident than ever and was eager to continue pursuing the sciences as a proudly African, queer, and non-binary femme.

After defending her PhD. in 2019, Sthabile earned a postdoctoral fellowship from the South African Radio Astronomy Observatory (SARAO). In 2021, they were appointed as a lecturer at the Physics Department of the University of Johannesburg. Here, they have been teaching and conducting research since.

Sthabile's research examines how active galactic nuclei impact galaxy evolution through cosmic time. They have published rather widely on this topic and continue to do so. Realising the idea of being on the cusp of great discoveries about the known Universe is the most enlivening part of the research process. They have scientific collaborators in the U.S., Germany, and the U.K. and are one of a few African women astronomers participating in Astrophysics research. They are currently a member of the African Astronomical Society's Executive Committee and early-career science sub-committee.

Sthabile is also actively involved in public science communication and outreach. They have given invited talks at several European Astronomical Society meetings and universities in the U.K. and featured in Nature magazine in efforts to spark meaningful dialogue on the cultural diversity issue in Physics and Astronomy. Sthabile hopes that by dispelling false ideologies about which cultural groups or 'races' are more competent at deep scientific and intellectual thought, Astrophysics will become a more welcome space for all who wish to pursue it.

If Sthabile had one wish, it would be that every individual in the world could look up at the clear and unpolluted (by terrestrial light) night sky and be told, in their language, about the vastness of the cosmos and their place in it. Sthabile plans to continue pursuing a career in Astronomy while also becoming an expert on the application of artificial intelligence and machine-learning within Astronomy.

# Priscilla Muheki

Uganda

Mbarara University of Science and Technology
Founding AfNWA board member

*Believing in oneself is the secret of unleashing the potential that is within."*

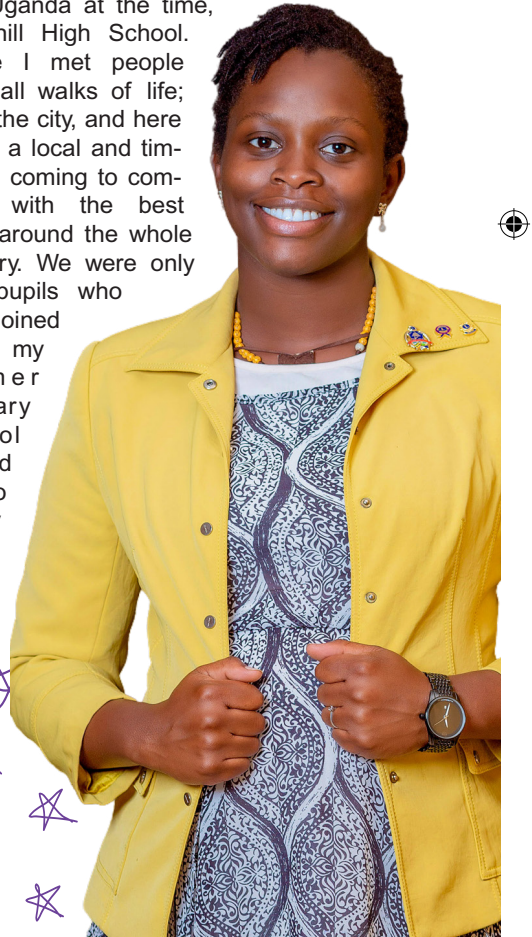

I am a lecturer in the Department of Physics at Mbarara University of Science and Technology (MUST) where I teach Physics and astronomy courses to undergraduate and postgraduate students.

Growing up, in a typical African family with a very humble background, influenced by culture and societal norms and beliefs at the time in which it was believed that sciences were for the boys and arts for the girls, there was very little inspiration around science. We were rather fond of role plays of the things we saw around us, particularly family and community life.

I was born in a family of 8, of whom 6 are girls and two are boys born much later after me. As children, we cherished games that revolved around building homes since we understood the future of growing and becoming wives. The only biggest curiosity I had at the time was how I could navigate the clouds so that I reach heaven since we were taught in our religious classes that "heaven is up above". And so life was such. At school, on the other hand, I preferred to study Mathematics and Science and they were always my best-performed subjects. My parents were both from the arts; Dad being a civil servant and mum in secretarial works as an editor of the biggest Newspaper in the country at the time; "The New Vision". And so I wouldn't say I drew any inspiration towards the science from them.

Onward I moved to secondary school after performing quite well and joined one of the best girls' schools in Western Uganda at the time, Maryhill High School. There I met people from all walks of life; from the city, and here I was a local and timid girl coming to compete with the best from around the whole country. We were only two pupils who had joined from my former Primary school and not so many from those that were ahead of us. Then I made a decision that I was going to excel at what I was doing and shouldn't let any of that intimidate me.

I must say we had a very smart, beautiful, and intelligent female Physics teacher and this made me fall in love with the subject. I was always looking forward to the Physics lessons compared to all subjects. And my love for the subject grew even more.

It is now that I can say that mentors and role models are very important in nurturing our passions and dreams. From that time, from my love for Physics and its application, I was sure I wanted to be an engineer in the future, and for which I worked hard. On the other hand, my father really hoped I could offer Medicine and Surgery at the University and so kept encouraging me to study Biology. However, innately, that was never my dream, because then (actually till now) the phobia I had for injections was a world apart. Fast forward, I joined the same school for my advanced level to study Physics, Chemistry, Biology, and Mathematics.

Unfortunately, at the end of my advanced level, I didn't score the points to get Engineering on a government scholarship but was rather offered a Bachelor of Science (Physics and Mathematics) with education at MUST. I knew for sure that I didn't want this course and on top of that be a teacher. Nevertheless, my father insisted that I had to take it since it had been given on a government scholarship. I applied for the East African scholarship to pursue civil engineering at the University of Dar es Salaam in Tanzania but was unsuccessful. I reluctantly had to accept and commence with the course for lack of options. Little did I know that this would be the beginning of a whole new captivating adventure!

While pursuing this course, we had an "Introduction to Astrophysics" course unit which was taught by a very interesting professor; Assoc. Prof. Simon Anguma. Astronomy/astrophysics was something that was new to me. And I believe till now, there is still a big number of people who can't imagine you can have an experience of it in Uganda. Prof. Simon who always taught it with so much passion and humor interested us to explore the subject further.

At times we would hold lectures at night as we viewed the night sky, mapping out the different constellations, tracking the phases of the moon, etc (obviously we didn't have any telescope). This is when I realised that whatever I had always seen e.g. the dark spots on the moon (which by the way in my culture is believed to be a woman with a hoe going to dig … funny!) had a scientific explanation that I could know if I delved into learning astronomy. This prompted me to offer all the other astrophysics course units which were electives in the following years. But with all that, my dream of becoming an engineer had never vanished. I was hopeful that after completing the degree I could still pursue my childhood dream of becoming an engineer.

After completing my bachelor's degree, I was encouraged by Prof Jurua Edward to enroll for a Master's degree specializing in astronomy under the sponsorship of the International Science Programme (ISP) which I initially declined but after being counseled by my mother to go further, I decided to enroll. This was the beginning point of my career in astronomy.

At this time I was already married with a 2 months old baby. I was soon bound to realize how far I could stretch myself; spending almost half of the night awake with the baby, having to wake up early to work on some assignments, do some reading, and then go for classes. It was so draining, I won't lie. I wondered if that was the right decision I had taken. But I testify that I am so glad I didn't give in to my weakness. And so it is very important to be resilient! As I completed my dissertation, I had my second baby. At the time, I had several undergraduate cours-es to teach as well. It was now a battle of whether I should proceed with a PhD or just let go and concentrate on what was on my plate. As if that was not enough, astronomy was still in its initial stages of development and so the expertise was very limited. This didn't offer me any motivation to proceed as I felt it would be very challenging. And so I decided to take a one-year break as I re-organised my life. I tried to consider applying for PhD positions abroad but was specifically looking for sandwich opportunities since I still had a young family to look after. Unfortunately, these were very scarce and the one I found needed me to have an affiliation at the desired institute which wasn't possible as I had no connections at all.

Eventually, I decided to settle and still pursue my PhD at MUST, but this wasn't easy for the first year as I figured out a research problem. Along the way, I met an amazing person who later turned out to be one of my PhD advisors, Dr Eike Guenther from the Thuringer Landessternwarte (TLS), Tautenburg in Germany who after a chat offered to work with me for my PhD. Don't hesitate to look out for help, there are people who are always willing to support you to realise your dream.

My research work focused on long-term spectroscopic monitoring of two very active M dwarfs in search of strong and very energetic magnetic events known as flares and coronal mass ejections that play a vital role in the dynamics of stellar and planetary systems. This type of star (much cooler than our own Sun) is the most abundant in our galaxy and have been found to commonly host planets. We study the activity rates on M stars and try to model how they affect the evolution of planetary atmospheres and life on those planets. Interesting, huh? The evolution of planetary atmospheres is an interesting but also complicated process. Understanding this process is vital for comprehending the evolution of the Earth, and the emergence of life. Additionally, it also sheds more light on whether life is common in the universe or rare. In other words, were we extraordinarily lucky to exist, or is the formation of habitable planets that have the potential to have life a common process?

I was able to realize my dream of doing a sandwich PhD through the sponsorship of ISP and TLS and of course, my advisors who were more than just supervisors but friends. At the start of my PhD, I also had an opportunity to attend the 39th International School for Young Astronomers in Addis Ababa, Ethiopia. As they say in my local language, "abagyenda bareeba", literally translates as "those who travel get to discover", this school in so many ways changed my life. The people I met there, that have mentored me till this time including Prof. Mirjana Povic. She always encouraged me to keep moving even when the times were quite tight and I am so grateful to the heavens that we crossed paths. Of course, I can't forget my family, for whose support I am forever indebted to. They believed so much in me and that made the heavy burden a bit lighter to carry. I am so grateful for all the lessons learned on this journey. It is always vital to look out for people who encourage you rather than pull you down, those who build you and encourage you to follow your dream, those who believe in you and stand by you. These beautiful souls still exist. In the end, I can say it was so rewarding and fulfilling to attain a PhD alongside my 4 little munchkins!

Looking back at how I grew up with little mentorship, I decided to be that which I didn't get. Maybe if I had found mentors and role models much earlier, my story would have been different. Nevertheless, they say life may not always be straightforward for everyone but what matters is to reach your goal. I have since committed myself to doing as much outreach as possible especially to girls from primary level up until postgraduate level. I share my experiences and encourage them to not let the light inside them go off. I was never extraordinary, and neither did I have an extraordinary upbringing. I was just an ordinary girl from an ordinary Ugandan home but I believe that I have now done extraordinary things. I have superseded the norms and beliefs of my society: a girl can't excel at science like the boy, you can't attain a higher degree when you are married especially higher than your husband, you can't have a family and maintain it if you always travel for research visits, you can't have a career and still afford to have children, and so much more. To all the girls and women that will have the chance to read my story, I want to reecho the words of Eleanor Roosevelt, **"A woman is like a tea bag - you can't tell how strong she is until you put her in hot water".**

Don't fear to go for that which may seem impossible because it can only be possible after you do it. Know that you can do everything you set your mind to, don't give up rather work up!

I wish you all the best as you aspire to be the great scientists of our times, joining us to explore the cosmos and proving to the world that we can do it… "… oo oh, we got our feet on the ground and we are burning it down oh oh got our head in the clouds and we're not coming down ..." Alicia Keys.

### Awards

1. 2023 OWSD Early Career fellowship

# Jacinta Delhaize

**Australia**

**University of Cape Town, South Africa**

*Fascinated by the mysteries of galaxy formation and evolution. Co-founder and co-host of The Cosmic Savannah podcast, showcasing Afrocentric astronomy to the world."*

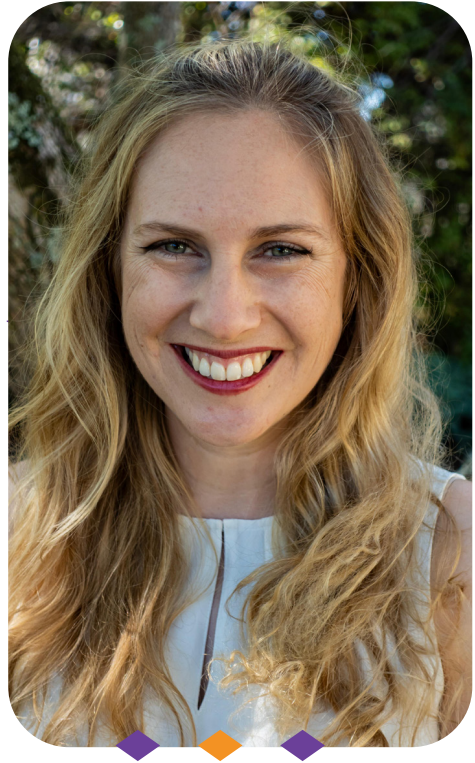

I am Dr. Jacinta Delhaize, a research astronomer, lecturer, and science communicator. I am originally from Australia but have lived in South Africa and worked at the University of Cape Town (UCT) since 2018.

My journey into astronomy started when I was 7 years old and I wanted to be an astronaut so that I could float. My teacher told me that I couldn't be an astronaut because I was Australian and a girl! My parents were so angry with my teacher and they told me I could be absolutely anything I wanted to be. I am very grateful to them for always encouraging me to pursue my passions.

From that age onwards, I was totally captivated by books showing beautiful photos from the Hubble Space Telescope of galaxies, planets, and nebulae. This fascination with space has shaped my personal identity and driven my academic pursuits. During high school, I realized that mathematics and physics are just the languages of the Universe. Once you start unlocking those languages, you gain deeper layers of understanding about the Universe's mysteries and an even greater appreciation and awe of its beauty and complexity.

I knew I wanted to pursue a career related to space and astronomy, but I didn't really know how. There was no astronomy degree offered, and almost no astronomy research being done in my state of Western Australia (WA) at the time. So I decided to go to university to study a general Physics degree. As luck would have it, this turned out to be the right choice and placed me in exactly the right place at exactly the right time.

While I was in the second year of my Physics degree, two senior radio astronomers moved to the University of Western Australia in Perth, where I was studying. They told me they were there to help set up WA's bid for a giant radio telescope to be built in the state. This sounded too good to be true! I made sure to stay in touch with them as much as possible and offered to do little research projects with them.

At the end of the 3rd year of my degree, I applied for and received a summer internship at the Gemini South Observatory in Chile. I had applied unsuccessfully the previous year, but am so glad I tried again because it was one of the best experiences of my life! I spent 3 months in Chile working on a research project with a professional astronomer and helping with observing on the 8-metre Gemini South optical telescope at the top of the Andes mountains. I'll never forget watching the sunset through the vent gates of the telescope dome, then later that night staring straight into the centre of the Milky Way and being completely dazzled by the sheer number of stars I could see from up there.

From that moment on, I was totally hooked. I was certain I wanted a career in astronomy research. I returned to WA and completed my Honours research on a radio astronomy project. Then I was nominated for the Western Australian Science Student of the Year Award, and to my utter surprise…I won! This led to many opportunities to be an ambassador for science and astronomy. I travelled to rural areas all around the (very large!) state of WA talking to school kids and the public about how incredible astronomy and science are, and how exciting it would be to host the SKA telescope. This is when I realised how much I love science communication.

In further excellent timing, just when I was ready to start my PhD, the International Centre for Radio Astronomy Research (ICRAR) opened in Perth. So I was again in the right place at the right time to receive world-class supervision on radio astronomy research. I spent 9 months a year doing my PhD at ICRAR, and 3 months a year at the University of Oxford in the UK working with my co-supervisor. I had the opportunity to travel to so many conferences all around

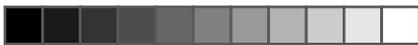

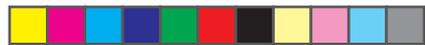

the world and discovered that I absolutely loved travel and networking and talking about science.

So when I finished my PhD, I applied for postdoctoral research positions outside of Australia because I knew I wanted to live abroad. At first, it was quite disheartening to receive a lot of rejections from my applications, but finally, I received a job offer from the University of Zagreb in Croatia. My first thought was "Where is Croatia?" So I googled it – it's in central/Eastern Europe, by the way!

I wanted an adventure so I accepted the job and moved to Croatia all on my own. It certainly was an adventure - full of many ups and downs. The country was beautiful, and the culture fascinating, but the language was very difficult to learn even though I tried. It was hard to feel cut off from society with that language barrier, which I am sure is a feeling that many people reading this have experienced before.

I lived in Croatia for 4 years and didn't know what to do after that. I thought maybe I should leave astronomy and go do something else. But first I gave an astronomy career one last try and applied for a Postdoctoral Research Fellowship offered by the South African Radio Astronomy Observatory (SARAO). To my surprise (again!) I got the job! So I moved from Croatia to Cape Town, again by myself, and took up my Fellowship at the University of Cape Town (UCT) in 2018.

I instantly loved South Africa and felt so at home. It somehow felt completely natural to me, almost like I had always lived there. I love the different people and cultures, food, landscape, and the incredible, invigorating astronomy community.

But the second move did come with its challenges. I was very burned out and felt very sad and lonely. But I felt guilty for feeling bad because I could see so many people around me with far worse struggles, so for a long time I hid my feelings. This of course only made things worse, and I started having panic attacks. Eventually, I sought help, was diagnosed with depression and anxiety, and started treatment. Unfortunately, it was a long recovery process and I was off work for some time. But my bosses were extremely understanding and told me to take as much time as I needed. I am still so grateful to them for their support and for giving me the chance to fully recover.

A few years later, at the age of 35, I was also diagnosed with ADHD. I was happy about the diagnosis because I finally felt validated in my struggles. I find it very difficult to concentrate while listening to lectures or seminars, and while reading papers. This led me to believe I was lazy and not smart and was part of the reason why I had burned out and considered leaving my astronomy career in the past. But now I am learning how to harness my neurodiversity as a super power! I love my brain! It lets me come up with very unique ideas and solutions to problems that no one else has thought of.

When it came to the end of my 3-year Fellowship, it was time to apply for another position. I decided to put in an application for a permanent Lecturer position that had been advertised in my department at UCT. I "knew" I wouldn't get the job because I was too junior for the role, so I was just hoping to get some job interview experience. But to my utter shock (yet again) I was offered the job!

So now I am a permanent member of academic staff in the Department of Astronomy at the University of Cape Town. I can honestly say I love my job and have never felt better, physically, mentally, and emotionally, in my whole life. I lecture a third-year undergraduate course on Galactic and Extragalactic Astrophysics, and I love training up new "proto-astronomers!"

I also supervise a group of postgraduate students doing their Masters and PhDs in astronomy. We call ourselves the "RADHIANCE" research group. We try to unravel the mysteries of how and why galaxies have changed over the history of the Universe. We want to understand the role that cosmic gas plays in allowing stars to form in galaxies, and in feeding the supermassive black holes that lurk in their centres! We also want to understand how these "burping" black holes blow gas out of the galaxies and prevent new stars from forming.

To do this research, we mostly use South Africa's amazing new MeerKAT radio telescope. Just as MeerKAT is a "precursor" to the upcoming SKA telescope, our research acts as a pathfinder for science that will be done with the SKA in the future. In this way, I hope to help empower the next generation of African astronomers to fully harness the immense opportunities afforded by the SKA.

The achievement that I am most proud of in my career so far is co-founding The Cosmic Savannah podcast. This freely-available science communication initiative showcases Afro-centric astronomy to both the African and international public. My colleague and I launched this podcast in 2019 and six seasons later, our team has grown and the podcast is still going strong! To the best of our knowledge, this is the first astronomy podcast by professional astronomers coming from the African continent.

Through The Cosmic Savannah podcast, we hope to raise awareness for, and pride in, the incredible, world-leading science and technology emerging from African astronomy. We also aim to provide a platform for under-represented voices and provide diverse and positive role models for young people. By engaging the public with African astronomy, especially in the lead-up to the SKA, we hope to empower people with a general interest in science, general critical thinking skills and an appreciation for the scientific method.

My first piece of advice to others, especially young women interested in astronomy, science (or other) career is to believe in yourself and ignore those thoughts that you're just an "imposter". Imposter Syndrome has plagued me for most of my life and almost led to me quitting my career. I felt that none of my achievements were real because I had just "tricked" people into believing I deserved it and that at any moment they were going to realise I was just a "fake". This is why I was so surprised every single time I was actually successful at something, and almost felt guilty!

Imposter Syndrome is a very, very common feeling, especially among women, and can sometimes be paralysing. So my advice is to ignore the demon in your head telling you that you're not good enough. Tell that stupid demon that you don't care what it says, you're going to go right ahead and do the thing anyway. And in the moments when you really are having trouble believing in yourself, believe others who believe in you.

My second piece of advice is to take very good care of your mental health. The one essential tool you need to do your job, especially in science and academia, is your mind. So value it and respect it enough to make sure it's as healthy as possible.

Oh and lastly, listen to The Cosmic Savannah! You can find it on any podcast platform or on our website www.thecosmicsavannah.com ;-)

Good luck! I know you can do whatever you choose to pursue. I believe in you and I'm cheering for you!

xox Jacinta

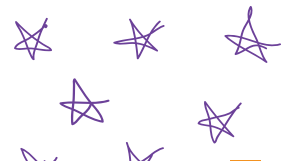

**Awards**

1. Finalists for the Science Communication Award: The Cosmic Savannah. South African NSTF-South32 "Science Oscars" 2021.

2. Western Australian Science Student of the Year 2008.

# Olayinka Fagbemiro

**Nigeria**

**Nigerian Space Agency/ Astronomers Without Borders (AWB) Nigeria**.

> *I believe it's time for young people, especially girls in Africa, to pursue their dreams in STEM, breaking barriers and stereotypes that have held them back. Let's rise and show the world that nothing is impossible!"*

## Journey to the Stars: My Astronomy Story

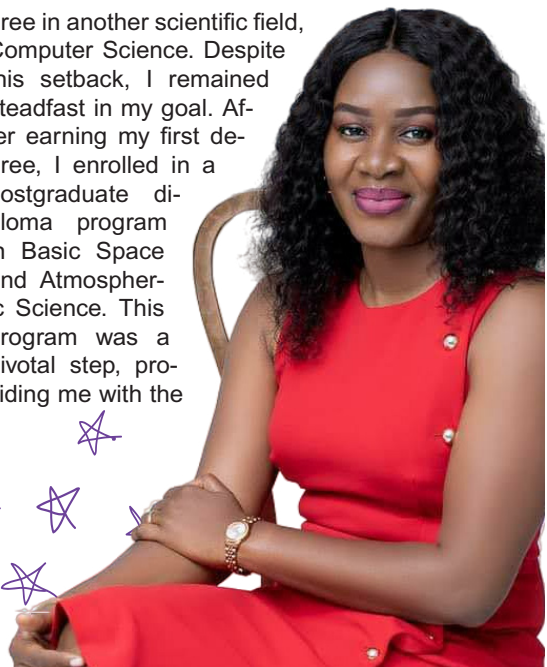

### Early Fascination and Inspiration

Growing up in a rural African town, the night sky was my playground. As a young girl, I was mesmerized by the stars, their twinkling lights painting a beautiful canvas in the dark sky. My fascination with the cosmos was boundless, and I dreamed of one day uncovering its mysteries. My father, a geographer, was my biggest cheerleader and mentor. He introduced me to astronomy, teaching me about constellations and the movements of celestial bodies. His enthusiasm and knowledge fueled my curiosity and solidified my ambition to become a scientist.

### Academic Pursuits and Initial Challenges.

My love for the sciences flourished throughout my schooling. I excelled in STEM subjects and dreamt of becoming an astronomer. However, my university did not offer an astronomy program, so I pursued a degree in another scientific field, Computer Science. Despite this setback, I remained steadfast in my goal. After earning my first degree, I enrolled in a postgraduate diploma program in Basic Space and Atmospheric Science. This program was a pivotal step, providing me with the foundational knowledge to delve deeper into space science.

**Professional Journey and Experience**
Seventeen years ago, I joined the Nigerian Space Agency, embarking on a fulfilling career in space science. My journey began with Space Education Outreach at the United Nations African Regional Centre for Space Science and Technology Education in English (ARCSSTEE). This role allowed me to merge my passion for astronomy with my love for education, reaching out to students and educators across Nigeria.

At ARCSSTEE, my responsibilities included developing educational programs, conducting workshops, and engaging with communities to promote space science. I aimed to make astronomy accessible and exciting, particularly for young girls, and to dismantle the barriers that often discourage them from pursuing STEM careers.
The Importance of Science

To me, science is the bedrock of all developmental abilities. It drives innovation, fosters critical thinking, and is essential for addressing global challenges. Science is not only crucial for societal advancement but also incredibly fun and rewarding. Considering the significant gender gaps and inequalities in STEM, I passionately advocate for every girl to pursue STEM and reach the zenith of their careers.

**Advice to Aspiring Astronomers**
To young girls and women aspiring to enter the field of astronomy, I offer this advice: aim for the stars, literally. Break every gender barrier and strive to reach the peak of your abilities. Dream big, work hard, and shatter every seen and unseen glass ceiling. Science is not gender-sensitive; anyone with the right attitude, interest, and effort can excel.

**The Joy of Science**
What I enjoy most about science is the creativity, the endless possibilities, and the collaborative spirit it fosters. Science allows us to explore the unknown, to innovate, and to solve complex problems. The sense of discovery and the ability to contribute to our understanding of the universe is incredibly fulfilling.

**Overcoming Challenges**
Working in Africa, I have faced significant challenges, particularly the lack of adequate infrastructural and financial support. Often, improvisation is necessary to achieve my goals, and institutional backing for larger projects can be scarce. However, these challenges have taught me resilience and ingenuity. I have learned to make the most of available resources and to seek out collaborations and partnerships that can amplify my impact.

**Achievements and Recognition**
One of my most significant achievements has been coordinating Astronomers Without Borders (AWB) in Nigeria. Over the past decade, we have used this platform to popularize STEM among young people, especially girls and children from disadvantaged backgrounds. Through AWB, we have reached thousands of young people, creating awareness about astronomy and inspiring a new generation of scientists.

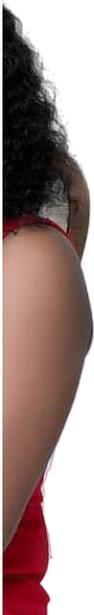
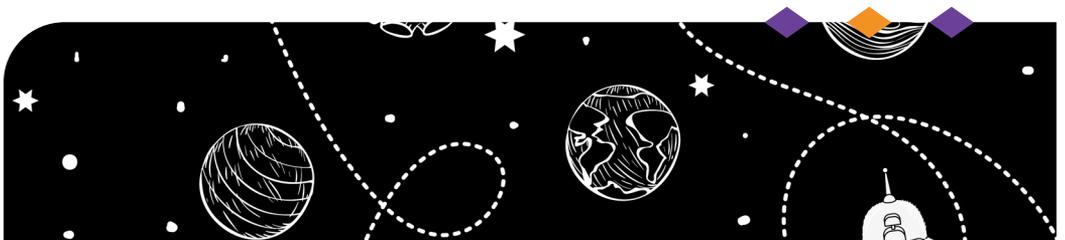

I am particularly proud of our work with kids from Internally Displaced Persons (IDP) camps and rural areas. We have provided opportunities for these children to learn about space science, fostering their curiosity and encouraging them to pursue STEM careers. Mentoring thousands of girls and women through AWB has been incredibly rewarding, and I am proud of the impact we have made.

### Dreams and Future Plans

One of my biggest dreams is to see more out-of-school children in Nigeria return to the classroom. I also aspire to mentor many more young girls, building their confidence and helping them excel in STEM careers. In the future, I plan to create more awareness about astronomy and space science, raising the next generation of STEM experts across Africa.

My vision is to establish a scholarship fund to support young girls from underprivileged backgrounds who aspire to study astronomy. I also hope to develop more educational programs and initiatives that make space science accessible and exciting for all.

### Conclusion

My journey in astronomy has been a testament to the power of passion, perseverance, and the pursuit of knowledge. It has taught me that no dream is too big and no star is too far. To all the young girls out there looking up at the night sky with wonder, I say this: the universe is vast and full of possibilities, and so are you. Reach for the stars, and you will find your place among them.

This is my story, and I hope it inspires others to pursue their dreams and contribute to the ever-expanding field of science. The night sky that once fascinated me as a young girl continues to inspire me, and I am committed to unlocking its mysteries and sharing its wonders with the world.

**Awards**

1. Fellow, 2024 Karman Fellowship

2. 2020-21 TechWomen Fellow, an initiative by the U.S. Department of State for Educational and Cultural Affairs.

3. International Aeronautical Federation (IAF) Young Professional Award

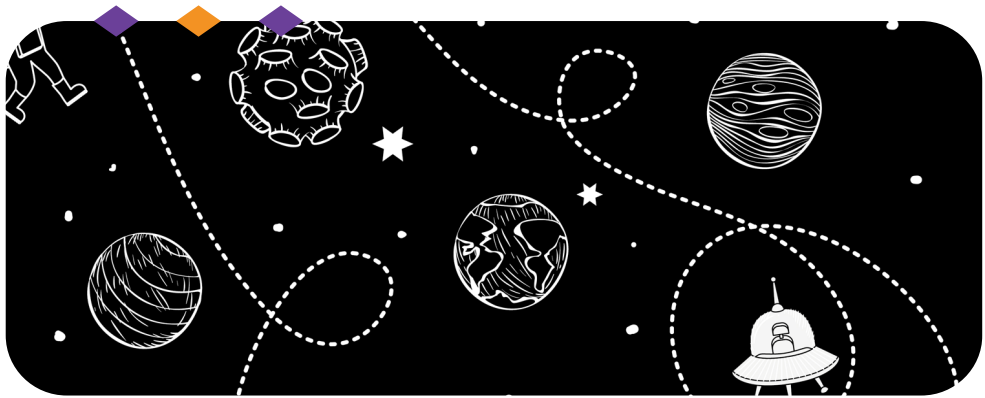

# Ilani Loubser

**South Africa**

**North-West University**, **South Africa**

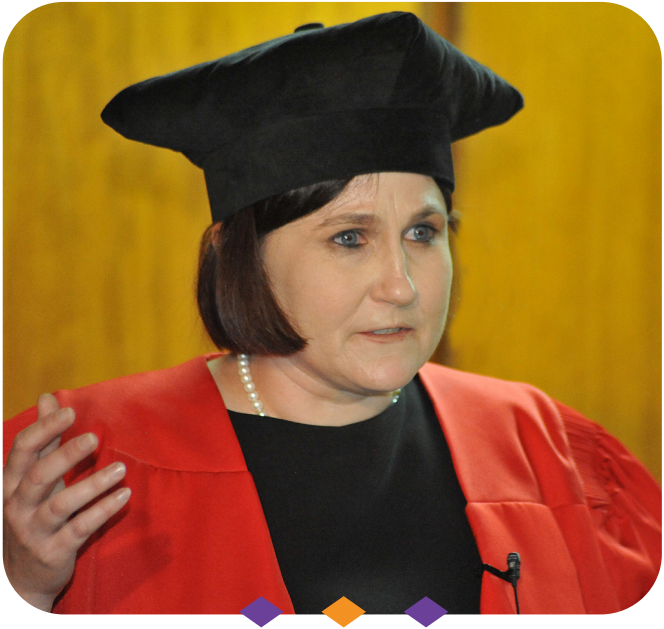

" *Perseverance, curiosity, and support from a community can transform dreams into reality, no matter where you start from. With determination, it's possible to achieve great things and inspire others along the way."*

My name is Ilani Loubser. I'm a professor of physics and an astronomy researcher at the Centre for Space Research (CSR) on the Potchefstroom Campus of North-West University (NWU) in South Africa.

I grew up in Potchefstroom, where I never would have imagined that I would one day become a professional astronomer and work in my (small) hometown. As a child, I wasn't interested in academics or anything resembling schoolwork, but I was always curious about how things worked. In high school, thanks to my parents' constant persistence, I started reading science books. I was particularly drawn to astronomy and dreamed of becoming a scientist, though it didn't seem like a realistic career choice in a small town like ours.

One day, I read an article in our local newspaper, the Herald, about a local astrophysicist, Prof. Okkie de Jager, who worked at the university. It was a revelation to me that astronomy was a viable career, not just something done at NASA. When I was 17, my mother, who worked at the university, arranged for me to visit the physics department and meet Prof. Harm Moraal, the head of the department. He looked like the quintessential professor with his shorts and long grey beard. He showed me pictures of his work, mostly from trips to Antarctica, and from that moment, I knew I wanted to pursue astrophysics.

After my matriculation, the Herald published a story about me and my goal of becoming an astrophysicist, complete with a picture of me next to my telescope. This caught the attention of another professor, Johan van der Walt, who invited me to visit the university's observatory. In January 2001, I began my B.Sc in Physics and Applied Math, confident I was on the right path. My father was skeptical about my career choice but supportive. My older sister's announcement that she wanted to be a philosopher made me seem like a normal child in comparison (she is now a philosopher at NWU).

I completed my B.Sc and enrolled in an Honours degree at NWU in 2004. Thanks to my mother's employment at the university, my tuition was free. That year, I participated in a radio astronomy project at the Hartebeesthoek Radio Observatory. One evening, while walking from the control room to the hostel, I told Prof. van der Walt that I wanted to focus on optical astronomy. They were nearly finished with the construction of the Southern African Large Telescope (SALT) in Sutherland at the time. He offered me an exploratory M.Sc project in optical spectroscopy on galaxies, which remains my specialty today.

During this time, my mother became ill, and my father resigned to care for her. In September 2005, my mother passed away, and shortly after, I submitted my M.Sc dissertation. I received a SALT-Stobie Scholarship to pursue my PhD at the University of Central Lancashire, a partner in SALT. In January 2006, I moved to Preston, England, for a fresh start, knowing I would return to South Africa eventually.

In April 2009, I returned to take up a South African Radio Astronomy Observatory (SARAO) post-doctoral fellowship at the University of the Western Cape, the first post-doc in their new astronomy group. This was a transformative time for astronomy in South Africa, with preparations for the SKA and the appointment of new SKA research chairs. One sunny Sunday morning, my father called to tell me about a lecturer position at NWU he saw advertised in (again!) the Herald.

In April 2010, I came full circle back to Potchefstroom. There were no optical astronomers and very few female staff members at the department. I had the opportunity to build an extragalactic optical astronomy research group from scratch. It started with one Honours student and it has grown to where I've graduated 13 Honours students, 12 M.Sc students, and 3 PhD students as of 2024. Currently, I have three M.Sc students, one PhD student, and three post-doctoral fellows in my Galaxies research group. Establishing a group in the traditional physics department was challenging but rewarding. I faced opposition, both conscious and subconscious, but also received plenty of support. My research focuses on using optical spectroscopy, in combination with radio, X-ray, and other observations, to understand galaxy evolution, particularly in groups and clusters. Teaching is also a significant part of my job, including undergraduate physics courses and postgraduate astronomy courses.

In January 2020, I became the first female professor of physics at NWU, and in 2021, I received a B-rating from the National Research Foundation as an internationally recognized researcher. Despite the progress, there are still only three female academic staff members in physics at NWU across both the Potchefstroom and Mahikeng campuses, so our work, especially in addressing gender imbalance, is far from done.

What I love most about my work is that no two days are the same. There are always new opportunities for projects, collaborations, discoveries, and learning skills. There is no shortage of challenging problems to solve. Even though Potchefstroom can feel isolated, I enjoy close collaborations with astronomers across the country and world-wide. It's incredibly rewarding to be part of the South African astronomy community that has grown so much over the past two decades, and I am sure that the best is yet to come!

Lastly, it takes a village (preferably one with a local newspaper) to raise and shape a female astronomer. Acts of kindness, even small ones, from professional astronomers can change the course of an aspiring young astronomer's career.

**Awards**

1. The first female professor in physics at North-West University.

2. Holds a B-rating from the National Research Foundation as an internationally recognized researcher.

3. Winner of the 2018 South African Women in Science Award (for Distinguished Young Woman Scientist: Astronomy).

# Dorcus Mulumba Khamala

**Kenya**

**PhD. student at Technical University of Kenya.**

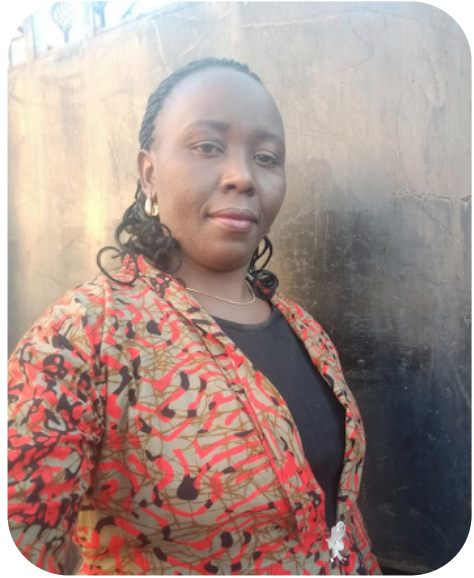

*Director of WEGO Publishers; EduTech Company."*

I am Mrs. Dorcus Mulumba Khamala, in my early forties and a mother of four girls. I am currently pursuing a PhD. in astrophysics at the Technical University of Kenya. I studied Bachelors in Education in Mathematics and Physics and taught in high school for some years.

I started my master's while teaching and would attend classes during school holidays. During one of the sessions, one of the lab technicians asked me in a casual talk; Have you ever heard about SKA? , there is a BIG telescope coming, please go read about it! That is where my interest in astronomy started as I read about many endless possibilities and areas of study in Astronomy!

I am currently studying about galaxies and specifically AGNs. It is very interesting as I discover new worlds unknown to many.

As I taught, I realised how many students have a negative attitude towards sciences and mathematics; this is usually passed to new high school students immediately they report to high school.

I am currently a director in WEGO publishers an eduTech company that delivers science and Mathematics lessons in a way easier to understand using animations to make abstract concepts more real. This is really making learning in high school more fun! I am passionate about outreach and talking to students about science.

# Michelle Lochner

South Africa

**University of the Western Cape/ South African Radio Astronomy Observatory - South Africa**

*When combined with human ingenuity, artificial intelligence has the power to make new discoveries about our incredible universe."*

When I was very young, around 6 years old, my father pointed out a beautiful grouping of stars known as the Pleiades. He explained that there weren't just seven stars but hundreds of stars loosely bound together in something known as an open cluster. Some people think that when you understand something, it removes the mystery and beauty but I found the opposite. Seeing something beautiful like the Pleiades and then understanding how it worked started my lifelong love of science.

I was very lucky to be able to read many books about astronomy as a child but I never knew this was a possible career. After all, you don't see many astronomers on TV! I decided to study physics and maths at Rhodes University simply because those were my favourite subjects. Along the way, I also picked up computer science and developed a taste for programming, something that has been incredibly beneficial throughout my career.

While at university I heard about the Square Kilometre Array, the largest radio telescope ever to be built, and somehow managed to convince a few people at the Johannesburg office to take me on as an intern during my vacations. School and university tend to do a poor job exposing people to real careers and this vacation work, simple as it was, had a lasting impact on my decision to become an astronomer. I learned that it was a real job and, more importantly, I could stay in the country I loved (South Africa) and have an exciting career.

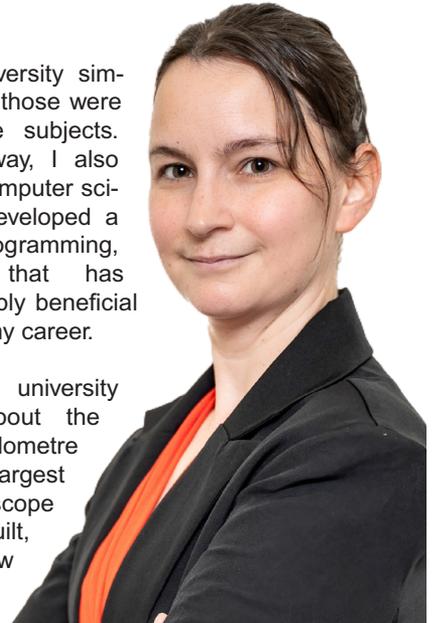

During my third year of study, while still deciding which physics field I wanted to pursue for postgraduate, I came across an advert for an international competition run by the European Space Agency. It involved analysing some real data from their space telescope INTEGRAL and writing a report. It was my first taste of data analysis and I loved every minute. I won my category of the competition and went on a trip to visit ESAC in Madrid, Spain. It was an exhilarating experience that finally cemented my decision to become an astronomer.

I decided to pursue my honours with the National Astrophysics and Space Science Programme at the University of Cape Town. It was an intense but exciting experience. After honours, I started my master's, later upgrading it to PhD, with Prof. Bruce Bassett. I was entranced by cosmology, the study of the formation, evolution and fate of the universe as a whole, but I also deeply enjoyed problem-solving and developing new techniques. Bruce combined both, allowing me to work on novel statistical methods for cosmology and laying the groundwork for my future career. I was extremely lucky that the SKA South Africa (now the South African Radio Astronomy Observatory) funded my studies from undergraduate to the end of my PhD, which also enabled me to travel extensively. Throughout my career, I have been to dozens of countries and taken hundreds of flights, having incredible experiences and making lasting friends at conferences all over the world.

A PhD is not for the faint of heart and there were certainly moments when I was ready to give up the career entirely. I always recommend to students to only start a PhD if you are fairly confident this is the career you want. However, I'm incredibly grateful I was able to push through and get my degree. After PhD, it is standard to take up a postdoctoral position which is a two or three-year research post. I originally planned to remain in South Africa so I only applied to one position at University College London (UCL). Usually, students apply for dozens of positions so this was somewhat risky but I was offered the position and moved to London for two years. It was during my postdoc that I first became interested in machine learning, a branch of artificial intelligence, and I wrote what is currently my most highly cited paper applying machine learning to classify supernovae (exploding stars used as cosmological probes).

At this point, I encountered an issue faced by almost every astronomer in the world dubbed the "two-body problem". Named after a famous physics problem, it refers to the fact that astronomers frequently have to move countries every few years which can put incredible strain on long-term relationships. My husband wanted to stay in South Africa so we had a long-distance relationship for two years, which is not something I would recommend to anyone. I was lucky to have the support of an incredible group of friends in the other postdocs and students at UCL, as well as my two sisters who live in London. As special as the experience in London was, it was with joy and relief that I received the offer for a more permanent position jointly between the African Institute for Mathematical Sciences (AIMS) and SARAO.

I worked at AIMS/SARAO for four years, focusing on applications of machine learning to various problems in astronomy and cosmology. Throughout my career up to that point, I had (like most young scientists in temporary positions) considered a permanent position to be the end goal. So when I started this post, I suddenly realised I had no real idea what to do. I was overwhelmed by opportunities and demands on my time, being unsure of what to say yes to. I now recognise this as a fairly natural aspect of that career stage and my main advice to anyone starting out with this career is to figure out your priorities and focus on them because you can't do everything. Academics have considerably more freedom than most careers. In some ways, we are more like entrepreneurs than a standard employee. I realised I had the power to build my own career profile, choosing whether to focus more on research, supervision, outreach, teaching, service work, or different combinations of them. But no one can tell you what the right path is for you, you have to try and figure it out for yourself.

While I had a great experience at AIMS, the group was focusing more on machine learning whereas I have always considered myself an astronomer. So I was delighted to receive a job offer from the University of the Western Cape, where I was hired as a Senior Lecturer, with a joint Staff Scientist position at SARAO. The group at UWC is large, diverse and vibrant and, barring the untimely interruption from the pandemic, I've had a wonderful experience since 2020.

My research has continued to focus on applications of statistics and machine learning. With new telescopes like the SKA and the Vera C. Rubin Observatory (under construction in Chile) soon to be delivering petabytes of data, there is an incredible opportunity to make new scientific discoveries. But with millions of sources to look at, there aren't enough students in the world to manually go through them and make these discoveries. My research is currently focused on combining machine learning and human expertise to automate scientific discovery. Bruce and I released a software framework called Astronomaly (which is publicly available) to search for rare and anomalous sources in almost any astronomical dataset.

While applying Astronomaly to data from MeerKAT, a beautiful radio telescope in South Africa, I discovered a new type of radio source which we called SAURON (a Steep And Uneven Ring of Non-thermal Radiation), which could be the remnant of a merger of two supermassive black holes. This was the most exciting and fun scientific project I've ever worked on, especially when our silly acronym got picked up by the media and several popular science YouTubers. It showed that artificial intelligence has the power to make new discoveries and can lead to exciting new science.

My day-to-day consists of a substantial amount of time programming, analysing data, making plots and thinking through the scientific implications. Through my research projects, I've had the opportunity to collaborate with researchers from all around the world: the USA, UK, France, Canada, Chile and Australia, just to name a few. Working with incredible people from many different cultures, and often travelling to meet with them, is an amazing perk of this career that is not often found in other careers. I've been invited to speak at over 20 international conferences, 6 of them as a keynote speaker.

I also find supervising postgraduate students to be by far the most challenging and the most rewarding aspect of this job. I meet with them regularly, guiding them through their research but also trying to build up their skills as scientists.

Teaching is also an important part of any academic career, although I do not focus on it as some academics prefer to do. I predominantly teach computational physics, which is a core component of any physics degree. I particularly find the opportunity to work with students from less privileged backgrounds fulfilling. Many of my students have had few interactions with computers and seeing some of them take to programming once they're exposed to it is a wonderful feeling. I know that even if they choose not to pursue physics (many don't), they have learnt an incredibly valuable skill that prepares them well for industry.

Although I had wonderful friends and I met the love of my life and future husband at university, studying physics as a woman can be isolating. Most women I encountered were studying psychology or other humanities subjects and I struggled to relate to them. Later in life, when I finally encountered more women interested in the same field during my postdoc, I realised how hard it actually had been to be the only woman in a class of forty. Gender equality in physics is something I actively work towards today to try to improve conditions for future generations.

Following a great idea from Bruce Bassett, I founded the Supernova Foundation Mentoring Programme (www.supernovafoundation.org) which is an international mentoring and networking programme aimed at women and gender minorities in physics. We pair senior women physicists with students for personal mentoring, as well as host webinars on topics ranging from CV writing and work-life balance to imposter syndrome and harassment. The Supernova Foundation has over 300 members from over 50 countries and has become an invaluable resource for many women who find themselves isolated without senior role models to support them. Running this Foundation, and especially watching my own mentees develop from undergraduates to becoming independent scientists, has been one of the highlights of my career.

It is easy for girls and women to lack confidence when they find themselves interested in maths, engineering and science, which are often socially considered male-dominated disciplines. Early on in my career, I often found myself talked over, ignored or disrespected. These are often fairly unconscious and subtle behaviours towards women. Although distressing, I learnt to trust myself regardless of how I thought others perceived me. I relaxed and tried to just be myself as far as possible, speaking out when I felt I had something to add. Slowly, other scientists began to realise I knew what I was talking about and started to defer to me and respect my opinion. The field has already improved significantly for women and my hope is that with each generation, it gets a little easier.

My advice to anyone who is interested in a career in science is not to let anyone tell you that you can't do it because of your gender, race, nationality or anything else. Pursue what you love, as many others have before you. An incredible universe is out there, waiting for you to unravel its mysteries.

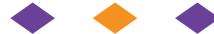

**Awards**

1. NSTF-South32 Award finalist, 2021 & 2024

2. 2nd runner-up South African Women in Science Awards, Distinguished Young Woman category, 2023

3. NRF P-rated (awarded to the most promising young researchers in South Africa), 2020

# Rosalind Skelton

**South Africa**

**South African Astronomical Observatory**

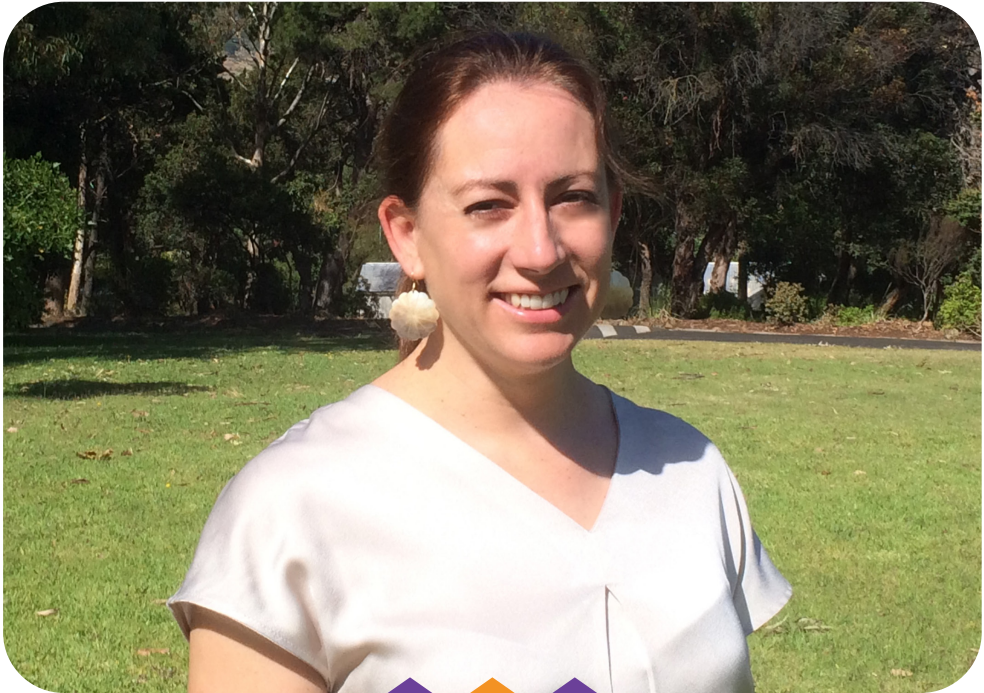

> *Acting managing director of SAAO.*

**M**y love for science and mathematics grew with me from a young age. I grew up in small towns in the North West and Northern provinces of South Africa, and my family loved to go camping in the beautiful wild bushveld around us. I remember many happy evenings of stargazing, scanning the skies for satellites (which were more rare in those days), and making wishes when we saw a shooting star. In high school I did a memorable science project on galaxies - they had already captured my imagination - and read many popular science books on cosmology and the forces of nature.

I remember seriously considering the answers to a long list of questions that I'dm asked someone who was studying cosmology on the difference between cosmology and astronomy, where and how one studied, and what the career prospects were (very challenging and competitive in his view; he later went into finance). Despite my strong interest in astronomy, I wasn't yet convinced what kind of scientist I wanted to be - I was fascinated by many things, including physics, archaeology, and ancient civilisations, and how the brain works.

After a gap year, I decided that a BSc in Physics and Applied Mathematics at the University of Cape Town was the place to start my university studies. My first and second-year astronomy classes and wonderful afternoon tutorials in the planetarium with the late Prof. Tony Fairall fuelled my love of astronomy further.

I had a romantic picture in my head that I wanted to study overseas and was thrilled when I was able to make that happen through a semester exchange at the University of British Columbia in Canada in my Honours year. There I took a fascinating course on galaxy dynamics, and vinteracted with the astronomy graduate students, which cemented my desire to continue studying and gave me an idea of what I wanted to pursue for my master's degree.

The National Astrophysics and Space Science Programme (NASSP) had recently started so I joined the Master's programme at UCT in 2005. The astronomy department was excited to welcome Renee Kraan-Korteweg as head of the department that semester. She became my Master's supervisor and a great role model, who revi alised the department and set up the astronomy major course - the only such undergraduate degree in the country for many years.

In the holiday before I started my MSc I spent my first night in Sutherland visiting the observatory. It would become a place I'd spend many many nights at in future. There I met my other MSc supervisor, Patrick Woudt, and was awed by the vast Milky Way overhead and the slightest smudge of the Andromeda galaxy in the far north, barely visible to the naked eye.

My desire to broaden my horizons and continue my studies elsewhere, and perhaps also to use the German I'd learned at high school, spurred me to apply for the International Max Planck Research School in Germany for my PhD. I was very excited when my application to Heidelberg University was successful! In the busyness of preparing to move overseas, I didn't manage to finish my thesis before leaving, as I should have! Luckily I had very supportive supervisors in both places and was able to get it done soon after arriving in Germany, though I did need to burn the candle at both ends, working on my new PhD project at the same time. My master's work led to my first publication on the Norma cluster of galaxies, hidden behind our Milky Way in the so-called Zone of Avoidance.

For my PhD, I was blessed to work with an incredibly inspirational young research group leader at the Max Planck Institute for Astronomy, Eric Bell, an expert on galaxy evolution who introduced me to large, deep, multi-wavelength galaxy surveys. Alongside him, I benefited from Rachel Somerville's deep theoretical understanding and was able to use her state-of-the-art semi-analytic models of galaxy formation to explore how galaxy mergers affect the galaxy population over time.

This theme continued into my postdoctoral work at Yale University in the USA. The timing of my postdoc worked out excellently - I was in the right place at the right time to join the 3D-HST survey project led by one of the most influential scientists in my field, Pieter van Dokkum. 3D-HST uses the low-resolution grism spectrograph on the Hubble Space Telescope to measure the redshifts of many galaxies at the same time. This gives us their distances, the essential 3rd dimension required to work out their intrinsic properties from the amount of light we measure at different wavelengths.

My work involved doing photometric measurements on images from many different telescopes around the world. It gives me great satisfaction that the catalogues we produced have been used by many people, both within and outside the collaboration and all around the world, to study how galaxies have evolved across cosmic time. This has been my most useful contribution to the field.

In 2013 I returned home to South Africa to take up a postdoctoral position at the South African Astronomical Observatory (SAAO). The postdoc years felt very uncertain and without many jobs available in academia in South Africa, I was even starting to consider my prospects as a Lindy Hop swing dancing teacher (I wouldn't have been the first to make that transition!). Luckily a job as an astronomer on the Southern African Large Telescope (SALT) team opened up in 2016 and I joined the team that operates SALT at SAAO, where I've been based ever since.

My time as a SALT astronomer was split between operations, research, and service to the astronomical community. I observed about one week out of every six, doing night shift at the telescope in Sutherland until Covid-19 hit. During lock-down we started observing remotely from Cape Town; we were one of the few large telescopes in the world to get back to observing quickly. We operate SALT in queue-scheduled mode - the team of SALT astronomers does all the observations on behalf of our users around the world, sending them their data the morning after their targets are observed. We are also responsible for monitoring the data quality and checking the health of the instruments, liaising with SALT's users to help them propose and set up their observing blocks, and analyse their data by providing data processing pipelines.

I have lectured the NASSP Honours spectroscopy course for the past five years and have been supervising NASSP Honours and Master's students since my post-doc, gaining experience over time through co-supervision. I am the chair of the NASSP Partnership, the network of partner institutes involved in the programme, and so my involvement in NASSP has grown as I moved from being a student to being a supervisor, lecturer, and leader.

My research group has grown over the years and my focus has shifted slightly towards the more nearby universe and the effects that the environment has on galaxies. I work closely with Assoc. Prof. Sarah Blyth at UCT, who co-supervises a number of students with me, and through her and the LADUMA project I have developed closer links to radio astronomy.

I have recently moved out of SALT operations to focus on management - I am the acting managing director of SAAO. I have enjoyed this challenge and learnt a tremendous amount in a short time. It is difficult to make enough time to support my students and keep up some research through my group's work, but I hope to find a way to balance all the competing priorities. I consider balance to be essential in all aspects of my life!

When I said in my interview for the SALT astronomer position that I hoped to make a difference to South African astronomy in future, I would never have guessed that I would be in this position of leadership. I am excited about what the next few years hold for me in my career - still a great unknown. I encourage those starting out their careers to work hard, persevere, take opportunities, and be positive - you never know where life might take you, may it be a fulfilling journey!

# Mirjana Povic

Serbia (old Yugoslavia) and Spain

Professor and researcher in astrophysics at the Space Science and Geospatial Institute in Ethiopia

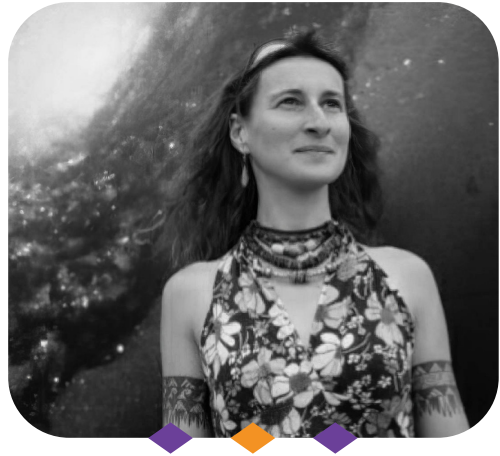

" *Education and the peaceful use of science, including astronomy, are the most powerful tools for transforming our society and combating poverty in the long term, making this world a better place for all, regardless of where children are born.* "

I was born in 1981 in the country that no longer exists, the former Yugoslavia. The Balkan wars marked most of my childhood. The civil war started when I was 8-9 years old, and when it ended, after the NATO bombings, I was almost 20 years old. Growing up in the former Yugoslavia with all that was going on around us, when the consequences of conflict enter into every aspect of our daily lives, when hearing about violence becomes our normality, and when the empathy and basic moral and ethical values by which we should all be governed are compromised, was not an easy thing to do. In those conditions, projecting my childhood dreams about life, the world and humanity was not easy either. I grew up in a very small household of eight, without a father, but with three mothers: my grandmother Ruza, the person who has had the biggest impact on my life, my mother Jasmina and my aunt Vukica. Being guided and raised by women and their life struggles significantly shaped the way I was. Books, studies and losing myself in nature and its beauty were ways to escape the madness of our reality, and contemplating the beauty of the night sky (a perfectly dark sky due to the long and regular power cuts we suffered) and its infinity was the safe space to project dreams: dreams about life, the world, humanity and who I want to be. Contemplating the beauty of the night sky also triggered my curiosity about the Universe, which grew over time.

To imagine at that time, in the circumstances that existed, that one day I would become an astronomer and scientist was like an impossible dream. However, after a long journey and many people who played a role in shaping it, here I am as an astrophysicist, trying to understand how galaxies form and evolve through cosmic time. After a long journey I learned that no matter how impossible things may seem, in most cases with our determination and the help of others if we keep our long-term vision as our guiding star it will be easier to go through the challenges without giving up and fulfill our dreams. Education transformed and changed my life, but if I had not had access to free education - from primary and secondary school, through university in Serbia, to a full PhD scholarship in Spain - I would never have been able to become a scientist, no matter how good student I was.

This personal experience and having been all over Africa during the last 20 years of my work, many children who, no matter how bright they are and how hard they work, will never be able to change their lives because they were born in a poor family or grew up in difficult conditions, significantly shaped my dissatisfaction with the world we live in. Africa has always been my passion. Since I was a child I was attracted by the infinite beauty of Africa's diversity and knowledge, but also by the unfair history and unfair current reality of the African continent and inequalities that we have in our world.

The last twenty years of my life were years of a journey between exploring the Universe outside our Earth, and Africa, our Universe on Earth. It was a journey of learning, understanding and finding ways to contribute to the betterment of our society through our life and work, and passion for astronomy, science and education. It was a journey full of diverse challenges and valuable life experiences that taught me that with determination, long-term vision and working together, the changes we would like to see in our world and society are possible. We shall always keep in mind our responability as citizens of planet Earth toward people who had less opportunities.

Awards:

1. 2021 – European Astronomical Society (EAS) inaugural Jocelyn Bell Burnell Inspiration Medal

2. 2019 and 2020 – Ethiopian Space Science and Technology Institute (ESSTI) and Ethiopian Space Science Society recognitions for development of astronomy in Ethiopia

3. 2018 - Inaugural Nature Research Award for Inspiring Science

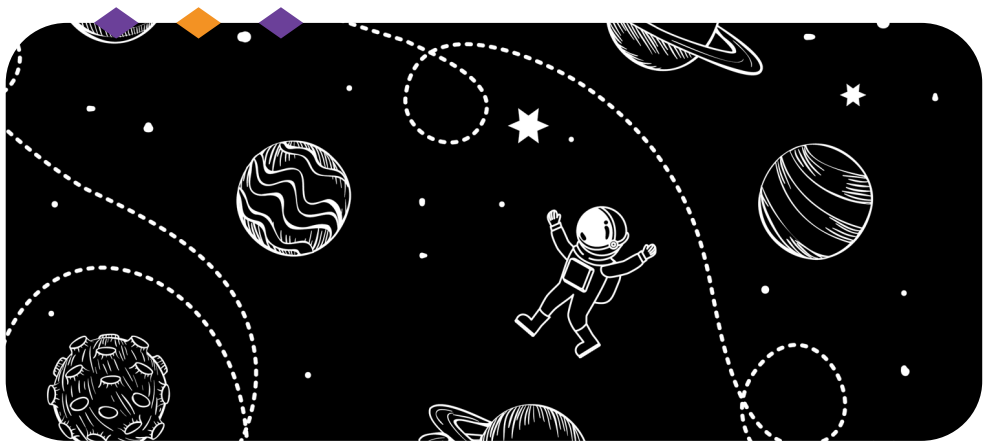

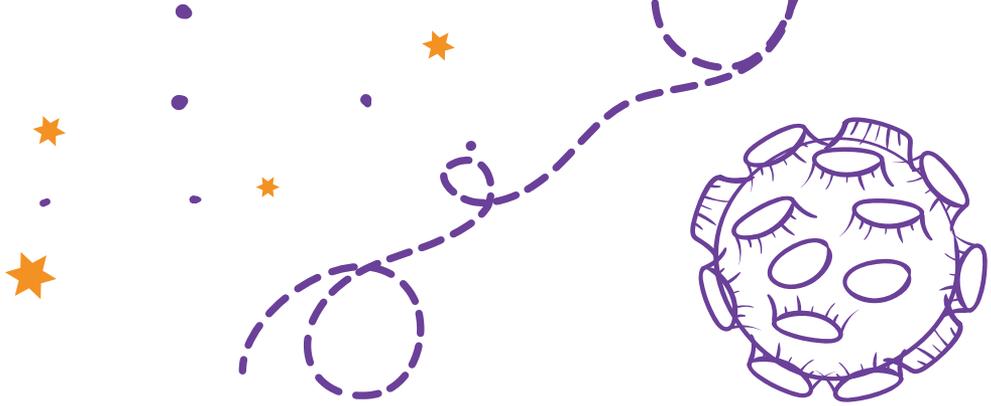

*The AfNWA Board is grateful to all women who shared their stories, and all people who showed their interest in this book and will use it to inspire many others. We acknowledge the continuous support of the African Astronomical Society (AfAS) and its Secretariat, the South African Department of Science and Innovation (DSI) through the support given to AfAS, the Nature Research and Estee Lauder through the inaugural Inspiring Science Award given to Prof. Mirjana Pović, the International Science Program (ISP) and Uppsala University for their continuous support, and African Science Stars for their amazing guidance and help with book editing and design.*

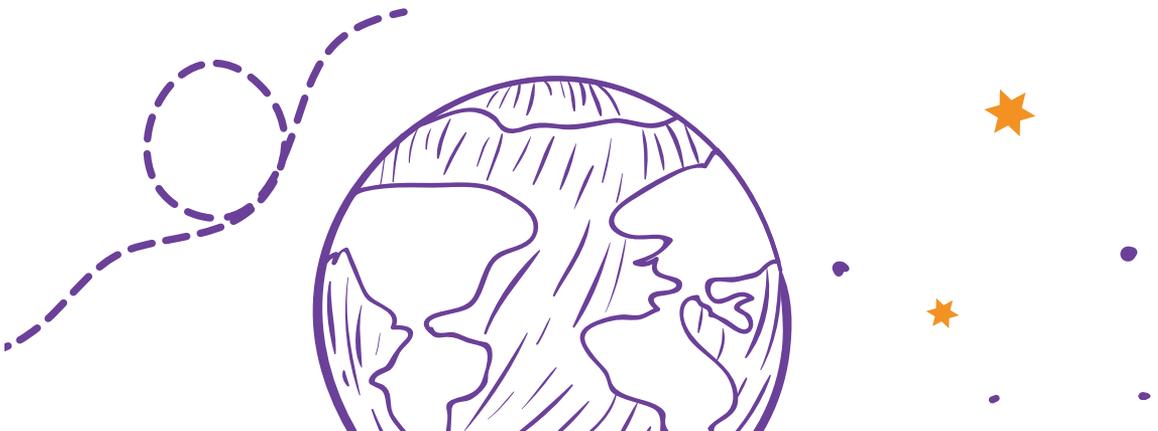

*Thank you very much for your reading!*

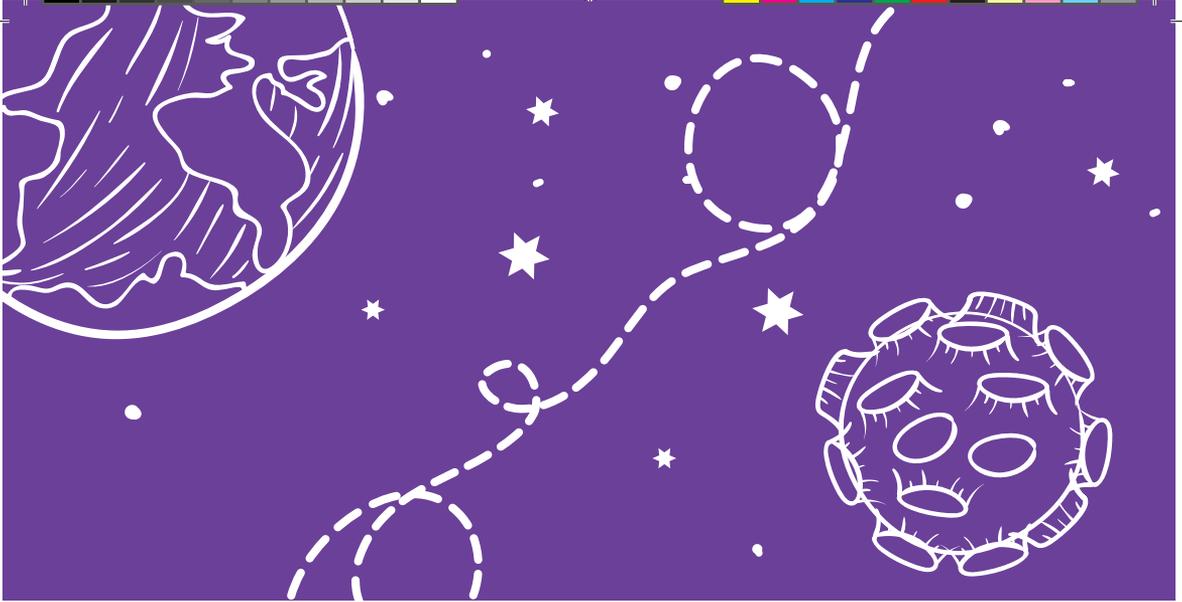

*All funds raised from the sale of this book will go towards AfNWA and AfAS activities focused on promoting STEM through astronomy to empower girls across Africa who are studying and living in difficult circumstances. For more information, please contact us.*

*Thank you very much for your contribution and support.*

www.afnwa.org

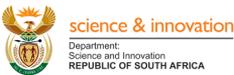 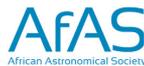 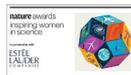 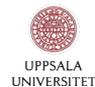 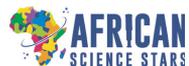